%% file: main.tex
\documentclass[prd,twocolumn,preprintnumbers,floatfix,letterpaper,showpacs,showkeys,superscriptaddress,nofootinbib]{revtex4-2}
\usepackage{natbib}
\usepackage[utf8]{inputenc}
\usepackage{graphicx}  %
\usepackage{rotating, graphicx}
\usepackage{bm}        %
\usepackage{xcolor}
\usepackage{amsmath}

\usepackage[caption=false]{subfig}
\usepackage[nomargin,inline,marginclue,draft]{fixme}
\usepackage[colorlinks=true, allcolors=blue]{hyperref}
\usepackage{wrapfig}

\begin{document}

\title{High-Statistics Measurement of Antineutrino Quasielastic-like scattering at $\enu \sim$~6~GeV on a Hydrocarbon Target}

\newcommand{\Rutgers}{Rutgers, The State University of New Jersey, Piscataway, New Jersey 08854, USA}
\newcommand{\Hampton}{Hampton University, Dept. of Physics, Hampton, VA 23668, USA}
\newcommand{\Dortmund}{Institute of Physics, Dortmund University, 44221, Germany }
\newcommand{\Otterbein}{Department of Physics, Otterbein University, 1 South Grove Street, Westerville, OH, 43081 USA}
\newcommand{\JMU}{James Madison University, Harrisonburg, Virginia 22807, USA}
\newcommand{\Florida}{University of Florida, Department of Physics, Gainesville, FL 32611}
\newcommand{\UCIrvine}{Department of Physics and Astronomy, University of California, Irvine, Irvine, California 92697-4575, USA}
\newcommand{\CBPF}{Centro Brasileiro de Pesquisas F\'{i}sicas, Rua Dr. Xavier Sigaud 150, Urca, Rio de Janeiro, Rio de Janeiro, 22290-180, Brazil}
\newcommand{\PUCP}{Secci\'{o}n F\'{i}sica, Departamento de Ciencias, Pontificia Universidad Cat\'{o}lica del Per\'{u}, Apartado 1761, Lima, Per\'{u}}
\newcommand{\INRM}{Institute for Nuclear Research of the Russian Academy of Sciences, 117312 Moscow, Russia}
\newcommand{\Jlab}{Jefferson Lab, 12000 Jefferson Avenue, Newport News, VA 23606, USA}
\newcommand{\Pittsburgh}{Department of Physics and Astronomy, University of Pittsburgh, Pittsburgh, Pennsylvania 15260, USA}
\newcommand{\Guanajuato}{Campus Le\'{o}n y Campus Guanajuato, Universidad de Guanajuato, Lascurain de Retana No. 5, Colonia Centro, Guanajuato 36000, Guanajuato M\'{e}xico.}
\newcommand{\Athens}{Department of Physics, University of Athens, GR-15771 Athens, Greece}
\newcommand{\Tufts}{Physics Department, Tufts University, Medford, Massachusetts 02155, USA}
\newcommand{\WM}{Department of Physics, William \& Mary, Williamsburg, Virginia 23187, USA}
\newcommand{\FNAL}{Fermi National Accelerator Laboratory, Batavia, Illinois 60510, USA}
\newcommand{\Purdue}{Department of Chemistry and Physics, Purdue University Calumet, Hammond, Indiana 46323, USA}
\newcommand{\MCLA}{Massachusetts College of Liberal Arts, 375 Church Street, North Adams, MA 01247}
\newcommand{\UMD}{Department of Physics, University of Minnesota -- Duluth, Duluth, Minnesota 55812, USA}
\newcommand{\Northwestern}{Northwestern University, Evanston, Illinois 60208}
\newcommand{\UNI}{Facultad de Ciencias, Universidad Nacional de Ingenier\'{i}a, Apartado 31139, Lima, Per\'{u}}
\newcommand{\Rochester}{Department of Physics and Astronomy, University of Rochester, Rochester, New York 14627 USA}
\newcommand{\Austin}{Department of Physics, University of Texas, 1 University Station, Austin, Texas 78712, USA}
\newcommand{\USM}{Departamento de F\'{i}sica, Universidad T\'{e}cnica Federico Santa Mar\'{i}a, Avenida Espa\~{n}a 1680 Casilla 110-V, Valpara\'{i}so, Chile}
\newcommand{\Geneva}{University of Geneva, 1211 Geneva 4, Switzerland}
\newcommand{\Chicago}{Enrico Fermi Institute, University of Chicago, Chicago, IL 60637 USA}
\newcommand{\hired}{}
\newcommand{\OregonState}{Department of Physics, Oregon State University, Corvallis, Oregon 97331, USA}
\newcommand{\oxford}{Oxford University, Department of Physics, Oxford, OX1 3PJ United Kingdom}
\newcommand{\umiss}{University of Mississippi, Oxford, Mississippi 38677, USA}
\newcommand{\upenn}{Department of Physics and Astronomy, University of Pennsylvania, Philadelphia, PA 19104}
\newcommand{\AMU}{Department of Physics, Aligarh Muslim University, Aligarh, Uttar Pradesh 202002, India}
\newcommand{\wroclaw}{University of Wroclaw, plac Uniwersytecki 1, 50-137 Wroa\l{}aw, Poland}
\newcommand{\Mohali}{Department of Physical Sciences, IISER Mohali, Knowledge City, SAS Nagar, Mohali - 140306, Punjab, India}
\newcommand{\CINVESTAV}{Departamento de Fisica Col. San Pedro Zacatenco, 07360 Mexico, DF, Av. Instituto Polit\'ecnico Nacional, Mexico}
\newcommand{\york}{York University, Department of Physics and Astronomy, Toronto, Ontario, M3J 1P3 Canada}
\newcommand{\ND}{Department of Physics, University of Notre Dame, Notre Dame, Indiana 46556, USA}
\newcommand{\ICL}{The Blackett Laboratory,  Imperial College London,  London SW7 2BW, United Kingdom}
\newcommand{\warwick}{Department of Physics, University of Warwick, Coventry, CV4 7AL, UK}
\newcommand{\qmul}{G O Jones Building, Queen Mary University of London, 327 Mile End Road, London E1 4NS, UK}

\newcommand{\mascencioThanks}{Now at Iowa State University, Ames, IA 50011, USA}
\newcommand{\amitbashyalThanks}{Now at High Energy Physics Department, Argonne National Lab, 9700 S Cass Ave, Lemont, IL 60439}
\newcommand{\ricfregianThanks}{Now at Department of Physics and Astronomy, University of California at Davis, Davis, CA 95616, USA}
\newcommand{\mateusfcarneiroThanks}{Now at Brookhaven National Laboratory, Upton, New York 11973-5000, USA}
\newcommand{\anfilkinsThanks}{now at Syracuse University, Syracuse, NY 13244, USA}
\newcommand{\finerThanks}{Now at Los Alamos National Laboratory, Los Alamos, New Mexico 87545, USA}
\newcommand{\kleykampThanks}{Now at Department of Physics and Astronomy, University of Mississippi, Oxford, MS 38677}
\newcommand{\bamThanks}{Now at University of Minnesota, Minneapolis, Minnesota 55455, USA}
\newcommand{\byaeggyThanks}{Now at Department of Physics, University of Cincinnati,  Cincinnati, Ohio 45221, USA}

\author{A.~Bashyal}\thanks{\amitbashyalThanks}  \affiliation{\OregonState}
\author{S.~Akhter}                        \affiliation{\AMU}
\author{Z.~~Ahmad~Dar}                    \affiliation{\WM}  \affiliation{\AMU}
\author{F.~Akbar}                         \affiliation{\AMU}
\author{V.~Ansari}                        \affiliation{\AMU}
\author{M.~V.~Ascencio}\thanks{\mascencioThanks}  \affiliation{\PUCP}
\author{M.~Sajjad~Athar}                  \affiliation{\AMU}
\author{A.~Bercellie}                     \affiliation{\Rochester}
\author{M.~Betancourt}                    \affiliation{\FNAL}
\author{A.~Bodek}                         \affiliation{\Rochester}
\author{J.~L.~Bonilla}                    \affiliation{\Guanajuato}
\author{A.~Bravar}                        \affiliation{\Geneva}
\author{H.~Budd}                          \affiliation{\Rochester}
\author{G.~Caceres}\thanks{\ricfregianThanks}  \affiliation{\CBPF}
\author{T.~Cai}                           \affiliation{\york}  \affiliation{\Rochester}
\author{M.F.~Carneiro}\thanks{\mateusfcarneiroThanks}  \affiliation{\OregonState}  \affiliation{\CBPF}
\author{G.A.~D\'{i}az~}                   \affiliation{\Rochester}
\author{J.~Felix}                         \affiliation{\Guanajuato}
\author{L.~Fields}                        \affiliation{\ND}
\author{A.~Filkins}\thanks{\anfilkinsThanks}  \affiliation{\WM}
\author{R.~Fine}\thanks{\finerThanks}     \affiliation{\Rochester}
\author{A.M.~Gago}                        \affiliation{\PUCP}
\author{H.~Gallagher}                     \affiliation{\Tufts}
\author{P.K.Gaur}                         \affiliation{\AMU}
\author{S.M.~Gilligan}                    \affiliation{\OregonState}
\author{R.~Gran}                          \affiliation{\UMD}
\author{E.Granados}                       \affiliation{\Guanajuato}
\author{D.A.~Harris}                      \affiliation{\york}  \affiliation{\FNAL}
\author{S.~Henry}                         \affiliation{\Rochester}
\author{D.~Jena}                          \affiliation{\FNAL}
\author{S.~Jena}                          \affiliation{\Mohali}
\author{J.~Kleykamp}\thanks{\kleykampThanks}  \affiliation{\Rochester}
\author{A.~Klustov\'{a}}                  \affiliation{\ICL}
\author{M.~Kordosky}                      \affiliation{\WM}
\author{D.~Last}                          \affiliation{\upenn}
\author{T.~Le}                            \affiliation{\Tufts}  \affiliation{\Rutgers}
\author{A.~Lozano}                        \affiliation{\CBPF}
\author{X.-G.~Lu}                         \affiliation{\warwick}  \affiliation{\oxford}
\author{E.~Maher}                         \affiliation{\MCLA}
\author{S.~Manly}                         \affiliation{\Rochester}
\author{W.A.~Mann}                        \affiliation{\Tufts}
\author{C.~Mauger}                        \affiliation{\upenn}
\author{K.S.~McFarland}                   \affiliation{\Rochester}
\author{A.M.~McGowan}                     \affiliation{\Rochester}
\author{B.~Messerly}\thanks{\bamThanks}   \affiliation{\Pittsburgh}
\author{J.~Miller}                        \affiliation{\USM}
\author{O.~Moreno}                        \affiliation{\WM}  \affiliation{\Guanajuato}
\author{J.G.~Morf\'{i}n}                  \affiliation{\FNAL}
\author{D.~Naples}                        \affiliation{\Pittsburgh}
\author{J.K.~Nelson}                      \affiliation{\WM}
\author{C.~Nguyen}                        \affiliation{\Florida}
\author{A.~Olivier}                       \affiliation{\Rochester}
\author{V.~Paolone}                       \affiliation{\Pittsburgh}
\author{G.N.~Perdue}                      \affiliation{\FNAL}  \affiliation{\Rochester}
\author{K.-J.~Plows}                      \affiliation{\oxford}
\author{M.A.~Ram\'{i}rez}                 \affiliation{\upenn}  \affiliation{\Guanajuato}
\author{R.D.~Ransome}                     \affiliation{\Rutgers}
\author{H.~Ray}                           \affiliation{\Florida}
\author{D.~Ruterbories}                   \affiliation{\Rochester}
\author{H.~Schellman}                     \affiliation{\OregonState}
\author{C.J.~Solano~Salinas}              \affiliation{\UNI}
\author{H.~Su}                            \affiliation{\Pittsburgh}
\author{M.~Sultana}                       \affiliation{\Rochester}
\author{V.S.~Syrotenko}                   \affiliation{\Tufts}
\author{E.~Valencia}                      \affiliation{\WM}  \affiliation{\Guanajuato}
\author{N.H.~Vaughan}                     \affiliation{\OregonState}
\author{A.V.~Waldron}                     \affiliation{\qmul}  \affiliation{\ICL}
\author{C.~Wret}                          \affiliation{\Rochester}
\author{B.~Yaeggy}%
\author{L.~Zazueta}                       \affiliation{\WM}

\collaboration{The MINER$\nu$A Collaboration}
\noaffiliation
\date{\today}

\newcommand{\minerva}{MINERvA }
\newcommand{\numi}{NuMI }
\newcommand{\numu}{\nu_{\mu} }
\newcommand{\numubar}{\bar{\nu}_{\mu} }
\newcommand{\pperp}{p_{T}}
\newcommand{\ppar}{p_{||}}
\newcommand{\degrees}{\circ}
\newcommand{\enuqe}{E_{\overline{\nu}}^{(QE)}}
\newcommand{\enu}{E_{\overline{\nu}}}

\begin{abstract}
    We present measurements of the cross section for antineutrino charged-current quasielastic-like scattering on hydrocarbon using the medium energy (ME) \numi wide-band neutrino beam peaking at antineutrino energy $<\enu>\sim 6$ GeV. The measurements are presented as a function of the longitudinal momentum ($p_{||}$) and transverse momentum ($p_{T}$) of the final state muon. This work complements our previously reported high statistics measurement in the neutrino channel and extends the previous antineutrino measurement made in a low energy (LE) beam at $<\enu>\sim 3.5$ GeV out to $p_{T}$ of  2.5 GeV/c. 
     Current theoretical models  do not completely describe the data in this previously unexplored  high $p_{T}$ region. The single differential cross section as a function of four momentum transfer ($Q^{2}_{QE}$) now extends to 4 GeV$^2$ with high statistics.   
     The cross section as a function of $Q^{2}_{QE}$ shows that the tuned simulations developed by the \minerva collaboration that agreed well with the low  energy beam measurements do not agree as well with the medium energy beam measurements. Newer neutrino interaction models such as the GENIE v3 tunes are better able to simulate the high $Q^{2}_{QE}$ region. 
\end{abstract}

\maketitle

\section{Introduction}
Recent results from the neutrino oscillation experiments  NOvA\,\cite{NOvA:2021nfi} and  T2K\,\cite{Abe:2019vii} hint  that  charge-parity  (CP) symmetry is violated in the lepton sector and favor a normal mass ordering of neutrino mass-states. Precise determination of the PMNS\,\cite{Workman:2022ynf} CP violating parameter ($\delta_{CP}$) requires new measurements with larger statistics and significantly smaller systematic uncertainties. The DUNE\,\cite{DUNE:2020jqi} and HyperK\,\cite{Yokoyama:2017mnt}  experiments aim to measure $\delta_{CP}\ne 0$ with greater than 5 $\sigma$ sensitivity at maximal $\delta_{CP}$, which require that less than 2\% of interaction rate uncertainties come from cross-section models\,\cite{DUNE:2020jqi}. This 2\% uncertainty can be achieved with improved cross section measurements. The data presented here overlap the energy range of the DUNE experiment,
although on a hydrocarbon target instead of argon. 

We present results on the   charge-current quasi-elastic (CCQE) process,  $\bar{\nu}_{\mu}p\rightarrow \mu^{+}n$, which is  a significant component of  the charged-current interactions rate\,\cite{RevModPhys.84.1307} in the few GeV energy range. 
To achieve high fiducial mass, present and future neutrino experiments employ detectors made of heavier nuclei (argon in the case of DUNE and water in the case of HyperK) where   nuclear processes and final state interactions (FSI) will affect the interpretation of CCQE interactions.  As the primary neutrino interaction occurs within a nucleus, the pure CCQE process itself is not experimentally accessible.   We instead use a CCQE-like signal definition (charged current event with no pions in the final state) based on the event topology observable outside the nucleus. Our  CCQE-like definition is  similar to the CC0$\pi$ definition used by other experiments \,\cite{T2K:2020jav},\,\cite{MiniBooNE:2013qnd}, \,\cite{Papadopoulou:2022aoo}. 
The advantage of concentrating on CCQE-like processes  is that, as a 2-body process, the full kinematics of the interaction are approximately determined from the outgoing charged lepton kinematics alone. This provides an estimate of the incoming neutrino energy, as needed for oscillation measurements,  and can be applied in a consistent fashion to both neutrino and antineutrino interactions, despite final state differences.

This work improves upon our previous measurement in the antineutrino channel\,\cite{Patrick:2018gvi} by utilizing 20$\times$ more statistics, improved background subtraction methods, and access to an extended kinematic region due to the higher energy beam.  It complements our previous muon-neutrino cross section measurement at similar beam energy\,\cite{Carneiro:2019jds}. 
 We first present the double differential cross section as a function of muon momenta ($\pperp$ and $\ppar$) as it is largely
model-independent and allows stringent tests of interaction models.  We also present cross section measurements as a function of the estimated four-momentum-transfer squared variable, $Q^{2}_{QE}$, using the CCQE hypothesis. Assuming the nucleus is at rest, the $Q^{2}_{QE}$ is given as:
\begin{equation}\label{QE based variables-2}
    \begin{aligned}
    Q^{2}_{QE} = 2\enuqe(E_{\mu}-p_{\mu}\cos\theta_{\mu}) - m^{2}_{\mu}
    \end{aligned}
\end{equation}
where $\enuqe$ (neutrino energy based on CCQE hypothesis) is given as:
\begin{equation}
    \begin{aligned}
            \enuqe = \frac{m^{2}_{n}-(m_{p}-E_{b})^{2}-m^{2}_{\mu}+2(m_{p}-E_{B})E_{\mu}}{2(m_{p}-E_{b}-E_{\mu}+p_{\mu}\cos\theta_{\mu})}\\
    \end{aligned}
\end{equation}
To provide a useful comparison with global neutrino energy cross sections, we also 
present the total antineutrino CCQE-like cross section $\sigma(\enu)$ corrected from $\enuqe$ to the $\enu$ (true neutrino energy).  The correction from the observable $\enuqe$  to $\enu$ is sensitive to nuclear effects and introduces additional  model uncertainties. %
The cross section as a function of $\enu$ is then the corrected number of events as a function of $\enu$ normalized by the flux. Since the correction is generated using the simulation, the correction from $\enuqe$ to $\enu$ introduces model dependencies.  %

\section{The MINERvA experiment}
The MINERvA experiment was located on-axis in the NuMI neutrino beam\,\cite{Adamson:2015dkw}, which serves as the neutrino source. In the \numi beamline, a 120 GeV proton beam impinges on a 1.2 meter long target to produce pions and kaons. Negatively charged mesons are then focused by two magnetic horns. The focused beam decays within a 675 meter long decay pipe to produce leptons and antineutrinos. Data were taken between  June 2016 and  February 2019. 

 The \minerva detector consisted of planes of scintillating strips interleaved with nuclear targets, a central tracker region consisting only of scintillator, an electromagnetic calorimeter formed by adding lead planes and a hadronic calorimeter
formed from iron plates. Muon charge and momentum measurements were provided for muons with momenta above $\sim 1.5$ GeV/c by the MINOS near detector\,\cite{WOJCICKI1999182}, which was located directly behind the MINERvA detector. 
The data presented here are from the central tracking region of \minerva which consisted of 108 tracking planes of scintillator composed of 88.5\% Carbon, 8.2\% Hydrogen and 2.5\% Oxygen. The tracker could reconstruct charged tracks with a kinetic energy ($T_p$) threshold of $\sim 120$ MeV for protons.  

\section{Event selection}

\subsection{Experimental selection}
The definition of CCQE-like process is given in our previous lower energy (LE) results\,\cite{Patrick:2018gvi}. 
Reconstructed antineutrino  candidates are required to have one positively charged muon,  no other reconstructed charged tracks ($N_{extra}$ = 0)and small recoil energy. This is intended to select  the classic CCQE signature $\numubar +p \rightarrow \mu^{+}+n$ where the neutron deposits little energy in our detector. CCQE-like candidate events are further defined by limits on the recoil energy deposited by the outgoing hadrons. Reconstructed non-muon recoil energy outside a 100 mm radius around the neutrino interaction vertex is required to be less than a value that varies with $Q^{2}_{QE}$ and is similar to that of the low energy analysis\,\cite{Patrick:2018gvi} but has been loosened by 50 MeV due to increased instrumental backgrounds in the higher intensity ME beam.
The final state muon is required to enter the downstream, magnetized MINOS near detector for charge determination and full momentum reconstruction. 
To match the forward MINOS acceptance, and assure that the muon charge is positive in the presence of neutrinos in the antineutrino beam, we require that the reconstructed $\theta_\mu < 20^\degrees$ and that the muon have positive charge and $\ppar$ between 1.5 and 15 GeV/c.

\subsection{Signal definition}
To match our detection capabilities, the 
CCQE-like process is defined at generator level by requiring a final state with a $\mu^{+}$ with polar angle 
$\theta_\mu < 20^\degrees$ 
with respect to the beam and $\ppar$ in the range 1.5 to 15 GeV/c,  no final state protons with kinetic energy $T_p$ above the proton reconstruction threshold (120 MeV) and no mesons or heavier baryons. Interactions that include nuclear excitation photons below 10 MeV are allowed.  The proton energy requirement is motivated by the requirement that there be no additional charged tracks in the interaction.  For $T_p \ge 120$ MeV, the efficiency for reconstructing a proton as an additional track rises quickly, making the efficiency of the $N_{extra}=0$ selection difficult to model.  As a result, we report our results for the restricted kinematic regions where we understand our efficiency the best. The reconstruction efficiency as a function of maximum kinetic energy of  final state protons is provided in the supplementary material to show the effect of this threshold.

 This CCQE-like definition is designed to exclude non-elastic interactions such as resonances ($\bar{\nu}_{\mu}+N\rightarrow \mu^{+}+\Delta$) and deep-inelastic scattering  (DIS, neutrino scattering off of quarks inside the nucleons), but  does include interactions  with multi-nucleon initial states such as  2p2h\,\cite{Subedi:2008zz} and any resonant events where additional pions and nucleons are absorbed in the nucleus. 

In the CCQE 2-body kinematic hypothesis, the initial nucleon is assumed to be at rest with a binding energy of 30 MeV in carbon. This allows an estimate of the antineutrino energy $\enuqe$ and momentum transfer squared $Q^2_{QE}$ from the muon kinematics alone\,\cite{Patrick:2018gvi}.

\section{Simulation}
Antineutrino interactions are simulated using the GENIE 2.12.6 event generator\,\cite{Andreopoulos:2015wxa}. This implementation of GENIE uses the relativistic Fermi Gas Model\,\cite{SMITH1972605} with short range correlations included via a Bodek-Ritchie tail\,\cite{Bodek:1980ar}. Multi-nucleon events using the Valencia model\,\cite{Nieves:2004wx},\,\cite{Gran:2013kda},\,\cite{Schwehr:2016pvn} are included.  The default GENIE model is referred to as GENIE 2.12.6 in subsequent figures.  
The simulation used for cross section extraction is  the \minerva Tune-v1 tune described in\,\cite{Carneiro:2019jds}.  This simulation has been tuned to match previous \minerva  data in the neutrino channel and is found to be consistent with our previous LE antineutrino measurement\,\cite{Patrick:2018gvi}. This tune includes:
\\
    {\it i.} Modification of non-resonant pion production  rates based on a combined re-analysis of the ANL\,\cite{Barish:1977qk},\,\cite{Baker:1982ty} and BNL\,\cite{Graczyk:2014dpa} bubble chamber data\,\cite{Rodrigues:2016xjj}. This modification reduces the non-resonant pion production by 57\%. This modification has a negligible effect ($<1$\%) in this analysis and is referred to as ($\pi$ tune) in Table \ref{tab:pzptchi2} but not shown separately in the figures.\\
    {\it ii.} A further empirical enhancement of the Valencia model based on\,\cite{Rodrigues:2015hik} which increases the integrated multi-nucleon (2p2h) event rate by 49\%. This is referred to as (Low Recoil Tune) in the figures.\\
    {\it iii.} Long range correlations modeled by the random phase approximation correction based on\,\cite{Morfin:2012kn} and implemented for \minerva in\,\cite{Gran:2017psn}. This is referred to as RPA in the figures. \\

There is an additional suppression of  pion production at low $Q^{2}$ to decrease an observed tension between data and simulation in previous experiments (\,\cite{Adamson:2007zzb},\,\cite{Paschos:2004md}) using an \textit{ad hoc} fit\,\cite{Stowell:2019zsh}. Addition of this tune makes MINERvA Tune-v2.

The GENIE 2.12.6 model used as a basis for our full detector simulation does not include the $\Delta S = -1 $ hyperon production processes\,\cite{PAIS1971361} $\numubar + n \rightarrow \mu^{+}+ Y$ that contribute 
only in antineutrino scattering.  These processes can contribute up to 6-8\% to the total antineutrino scattering cross section.  However, a generator level study using a more recent version of GENIE (v3.0.6)\,\cite{GENIE:2021zuu}  that does include these processes indicates that almost all hyperon production results in either detectable tracks or significant energy deposition that are vetoed by the multiplicity and recoil selections used to define the CCQE-like signal and data samples.  Any residual hyperon processes are estimated to contribute $\sim 1\%$ to the final CCQE-like data sample.

\begin{figure}[ht]
    \centering
    \includegraphics[width=0.99\linewidth]{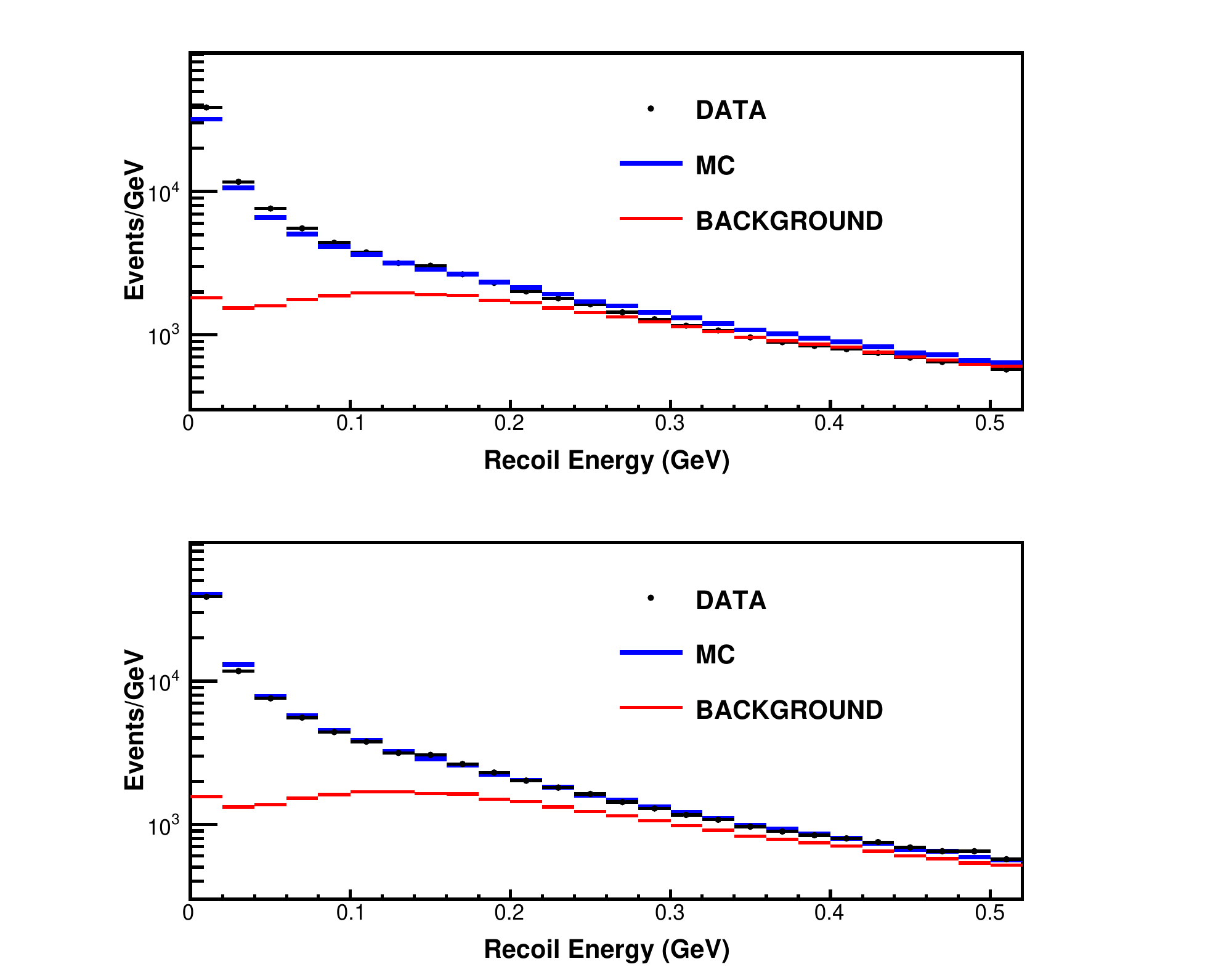}
     \caption{The recoil energy distribution in the bin of $\pperp = 0.25-0.4$ GeV/c,  $\ppar = 1.5-5 $ GeV/c. The top Fig. shows the data compared to default simulation.  The bottom Fig. shows  the  contributions of simulated signal and background  after their relative contributions have been constrained to the  data in the region above 100 MeV.  The $\chi^2$ of this fit is 16 for 19 degrees of freedom. }
    \label{fig:recoilfit}
\end{figure}

A GEANT4\,\cite{AGOSTINELLI2003250} based full detector simulation models the response of the detector\,\cite{Aliaga:2015aqe}. The simulation is tuned to match  test beam data\,\cite{MINERvA:2015yej} and overlain with random detector readouts to reproduce rate dependent backgrounds. 
Background contributions from non-CCQE-like processes are then estimated by fitting the reconstructed recoil distribution between 100 and 500 MeV before recoil selection to simulated signal and background samples in 14  $\pperp$, $\ppar$ bins.
Fig. \ref{fig:recoilfit} illustrates one of the fits for a typical $\pperp,\ppar$ bin.

The signal selection efficiency is estimated to be $\sim 70$\% and the selected sample is 70-80\% pure, with the purity falling at higher $Q^2$.

\section{Cross Section Extraction}  %

A total of 635,592 $\pm$ 1,251 (stat.) $\pm$ 13,850 (syst.) events are selected after  background subtraction compared to 14,839 events in the LE sample\,\cite{Patrick:2018gvi}. The background subtracted sample is unfolded to the true kinematic variables using an iterative Bayesian unfolding  based on the RooUnfold algorithms\,\cite{DAgostini:1994fjx}\cite{Adye:2011gm}. To determine the number of iterations required to unfold the sample into true kinematics variables, various model predictions were used as fake data to unfold using the MINERvA Tune v1. Based on the unfolding studies, 4 and 8 iterations were determined as the optimum number of iterations to unfold the sample into true $(p_{T},p_{||})$ and $(\enuqe,Q^{2}_{QE})$ variables respectively\,\cite{Bashyal:2021tzd}. One dimensional cross sections as a function of $Q^{2}_{QE}$ and $\enuqe$ are projected from the two dimensional $(\enuqe,Q^{2}_{QE})$ cross section. 
The sample is then corrected for geometric efficiency and normalized by the total number of protons on target ($1.12\times 10^{21}$) and the total number of nucleons in the fiducial volume of the detector ($3.23\times10^{30}$ nucleons). The total antineutrino flux integrated from 0 to 120 GeV is used to obtain  the differential cross sections. 

To extract the $\enu$ cross-section presented in section \ref{sec:enu}, the $(\enuqe,Q^{2}_{QE})$ cross section is corrected to a true $({\enu,Q^{2}_{QE}})$ cross section and projected into the $\enu$ phase space.
The neutrino flux is estimated both from simulation and from studies of multiple antineutrino processes and is parameterized as a function of true $\enu$.  The cross section as a function of ${\enu}$ is then the corrected number of events observed as a function of $\enu$, corrected by the flux.   $\enuqe$ is a more robust observable defined solely from muon kinematics but generally offset from the true $\enu$.  Determination of the transformation from the observable $\enuqe$ to the true $\enu$  requires simulation and hence introduces additional model dependencies as illustrated in Fig. \ref{fig:fid error summaryEnu}. 

To illustrate the effect of the $T_p$ and $\theta_\mu$ selections in our signal definition, we also provide the full cross section  as a function of true antineutrino energy  corrected for the $T_p$ and $\theta_\mu$ restrictions. As this additional correction has more model uncertainty, the result has significantly larger uncertainties. %

\begin{figure*}[ht]
    \centering
    \includegraphics[width=0.99\linewidth]{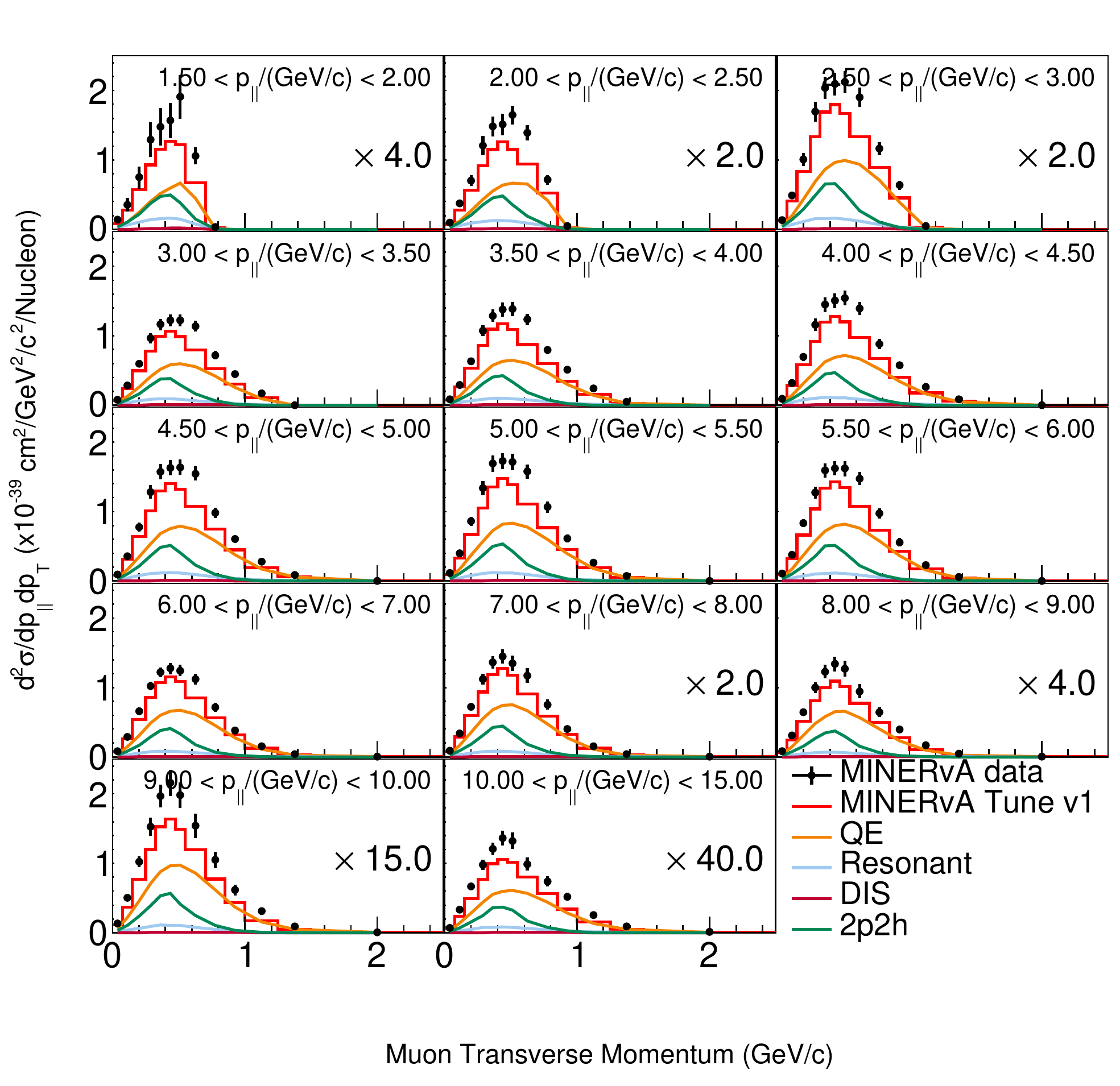}
    \caption{The double differential CCQE-like antineutrino cross section as a function of muon $\pperp$ in  bins of $\ppar$. The black markers are the measured cross section and the solid lines are the model predictions of MINERvA Tune v1 and its individual components. The multipliers on the panels are the scale factors used to scale the histograms.}
    \label{fig:xsection-pt}
\end{figure*}

\begin{figure*}[ht]
    \centering
    \includegraphics[width=0.99\linewidth]{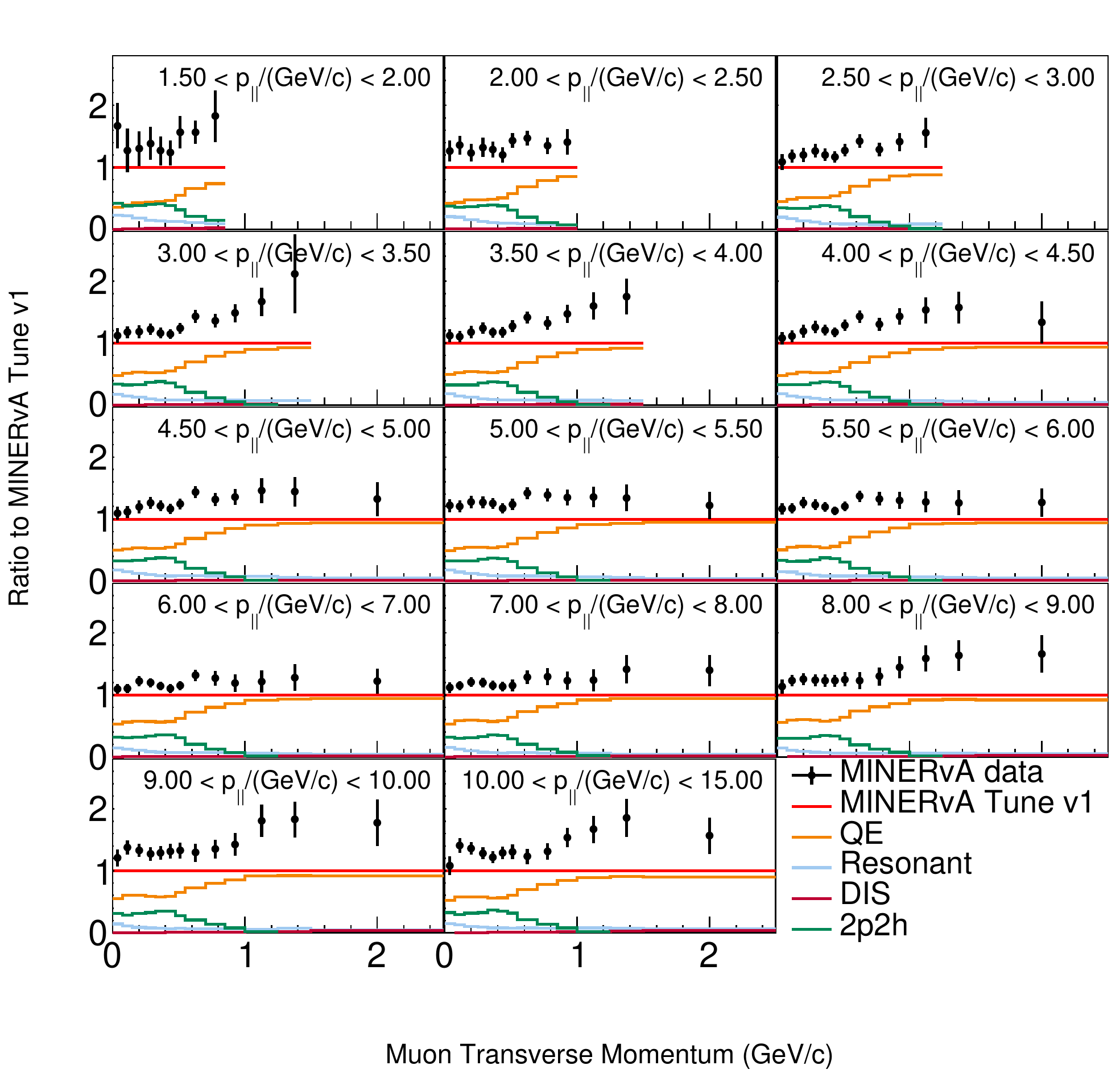}
    \caption{Ratio of the cross section data to MINERvA Tune v1 (points)  and fractional contributions of   components of \minerva Tune v1 to MINERvA Tune v1 (colored lines).  }
    \label{fig:xsection-pt-ratio}
\end{figure*}

\begin{figure*}[ht]
    \centering
    \includegraphics[width=0.99\linewidth]{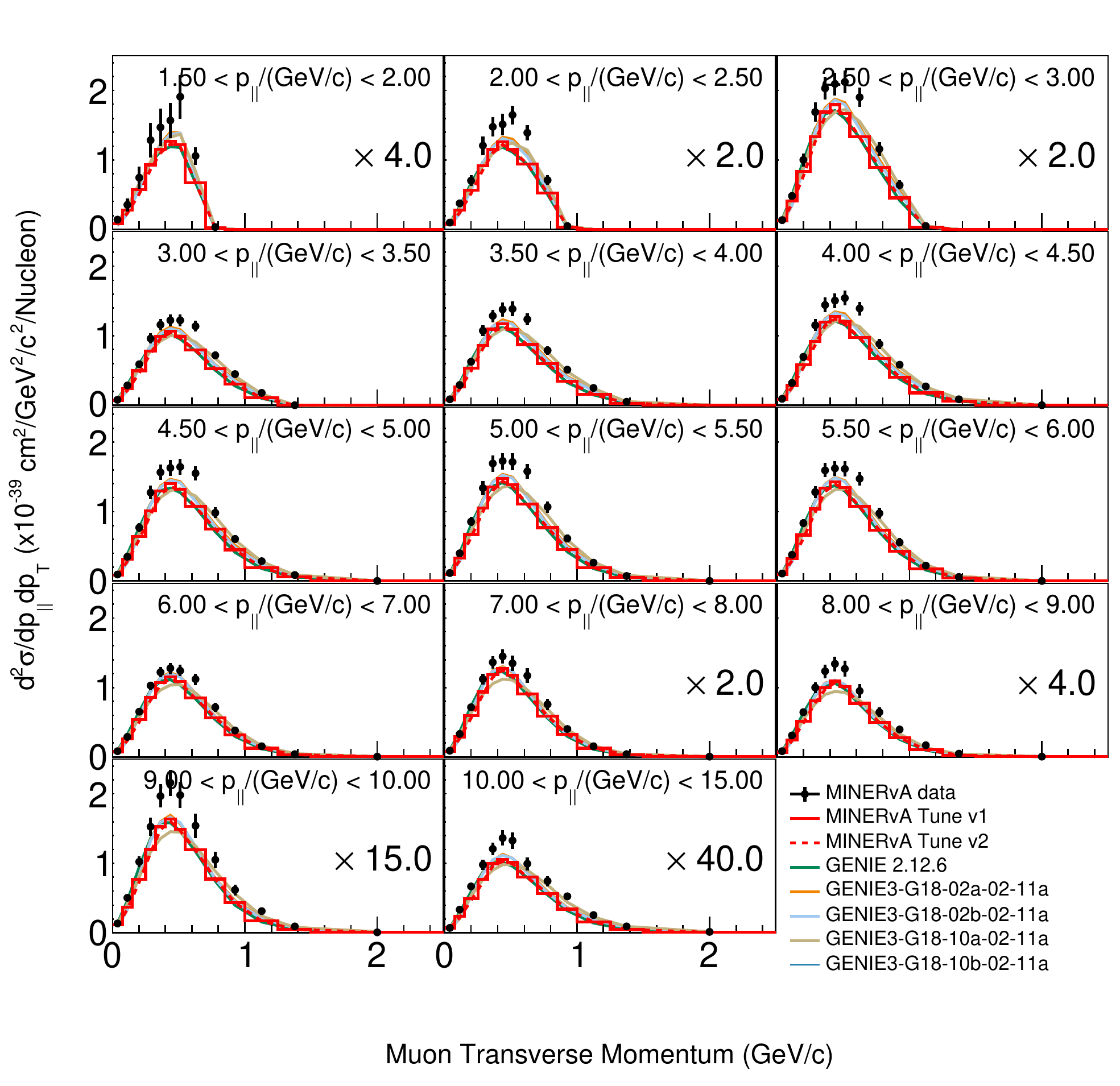}
    \caption{Double differential cross section as a function of muon transverse momentum in  bins of muon longitudinal momentum shown for GENIE 2 and tuned models along with GENIE 3 models. }
    \label{fig:pt-genie3-main}
\end{figure*}

\begin{figure*}[ht]
    \centering
    \includegraphics[width=0.99\linewidth]{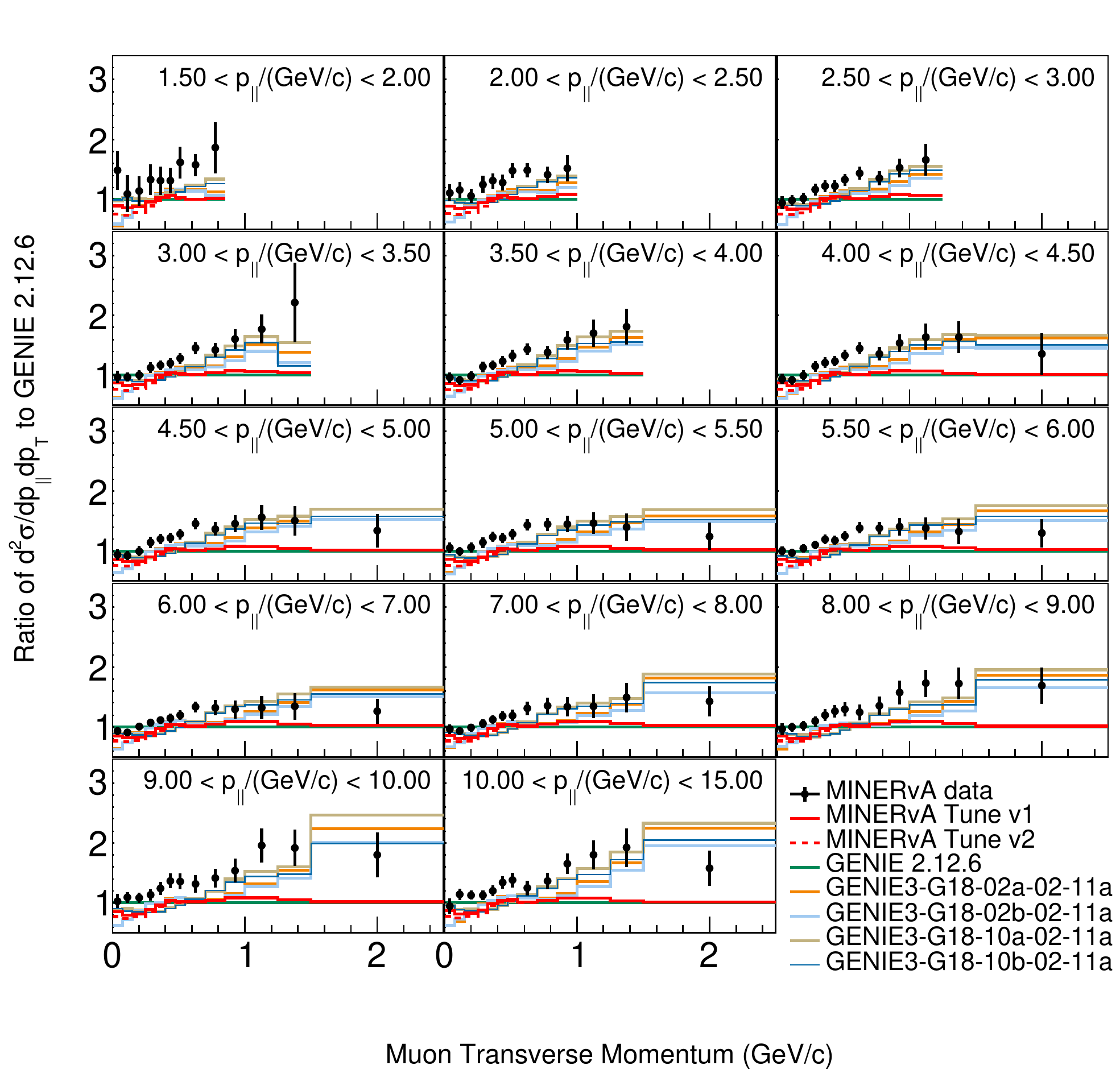}
    \caption{Ratios of double differential cross sections to GENIE 2.12.6 as a function of muon transverse momentum in bins of muon longitudinal momentum shown for GENIE 2 and tuned models along with GENIE 3 (v3.0.6) models. }
    \label{fig:pt-genie3-ratio}
\end{figure*}

\begin{figure*}[ht]
    \centering
    \includegraphics[width=0.9\linewidth]{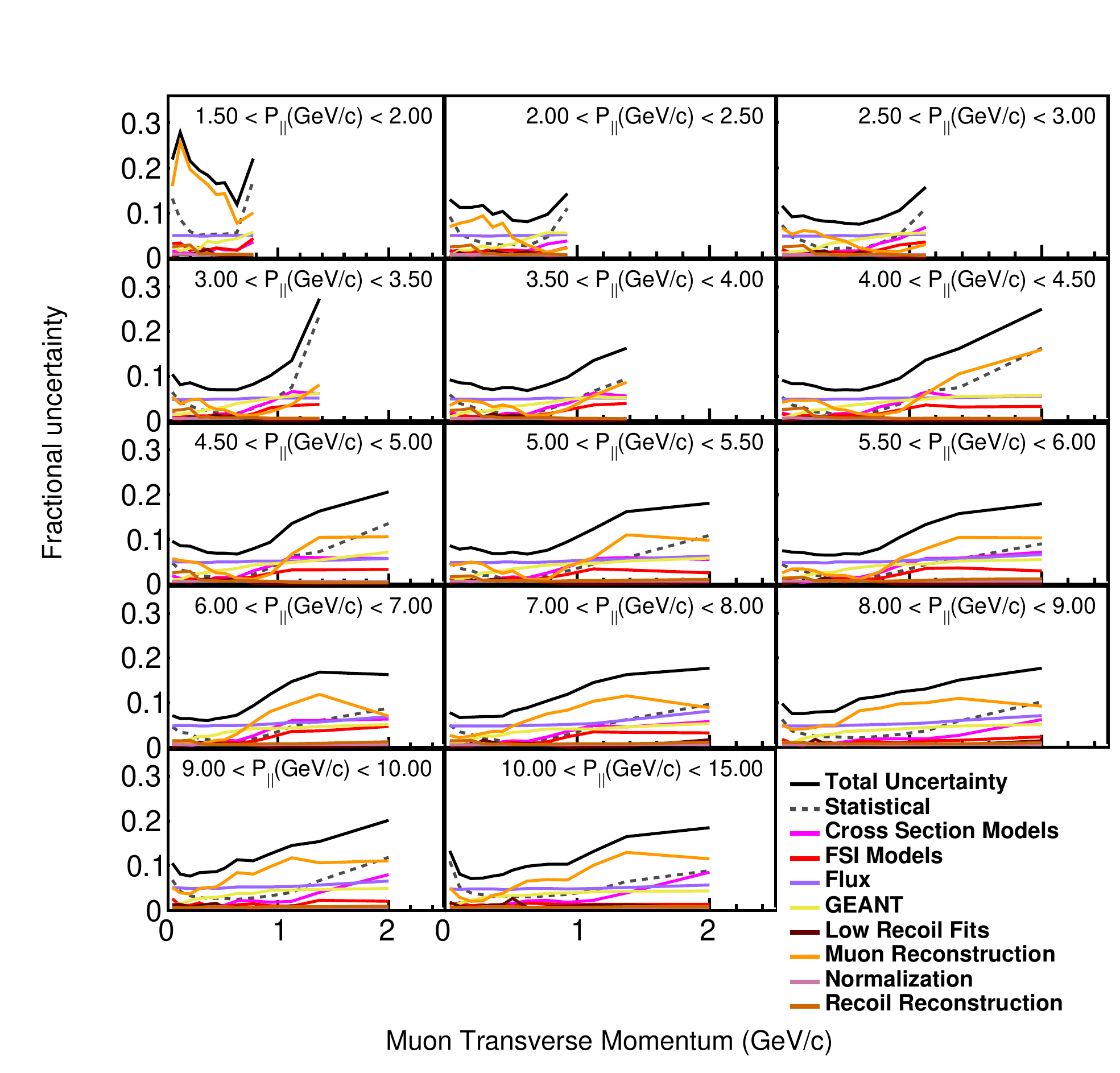}
    \caption{Summary of uncertainties on the $\pperp, \ppar$ cross sections shown in Fig. \ref{fig:xsection-pt}.}
    \label{fig:error-summary-xsec-pt}
\end{figure*}

\begin{figure*}[ht]
    \centering
    \includegraphics[width=0.99\linewidth]{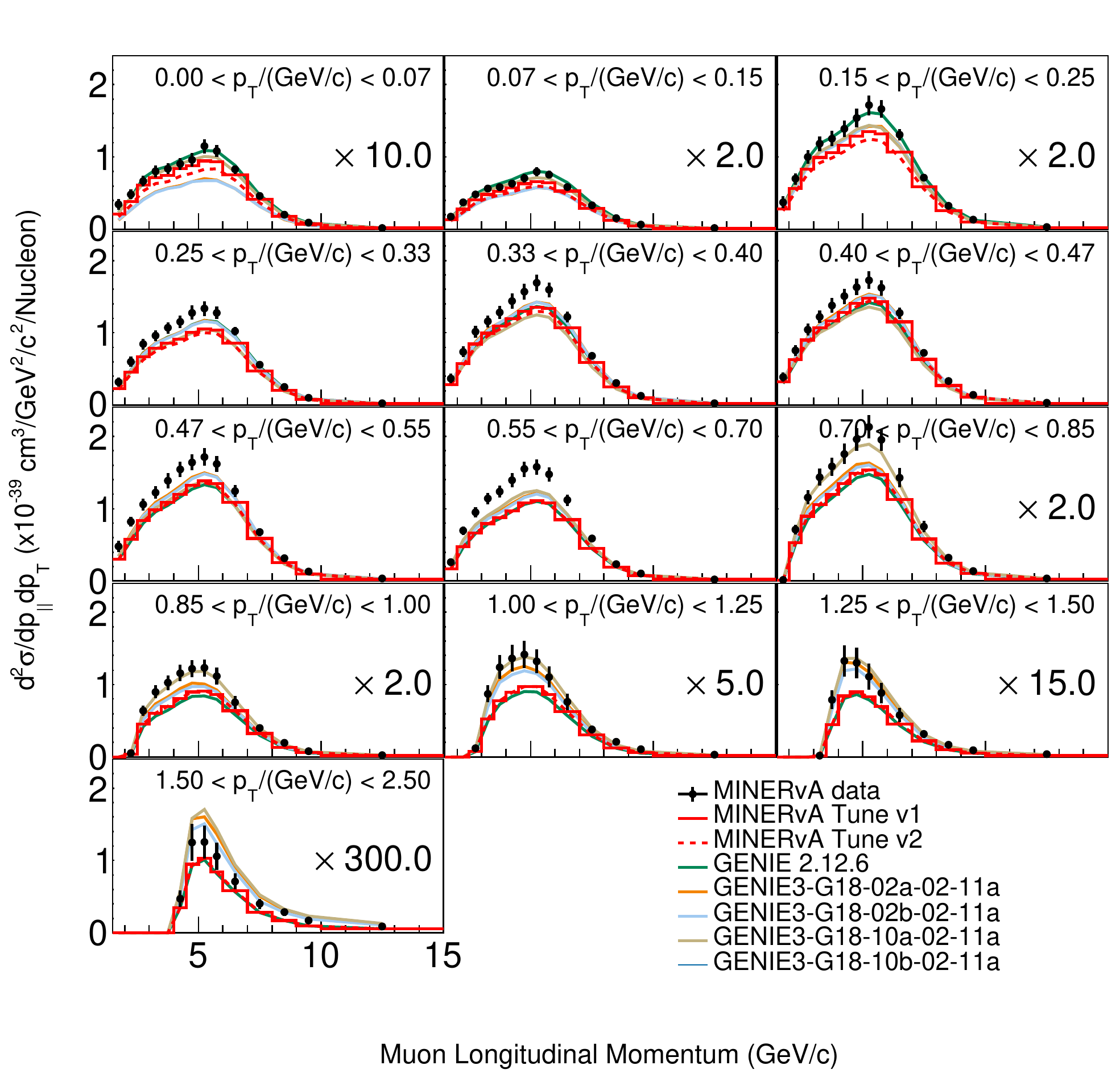}
    \caption{Double differential cross section as a function of muon longitudinal momentum in  bins of muon transverse momentum shown for GENIE 2 and tuned models along with GENIE 3 (v3.0.6) models.}
    \label{fig:pz-genie3}
\end{figure*}

\begin{figure*}[ht]
    \centering
    \includegraphics[width=0.99\linewidth]{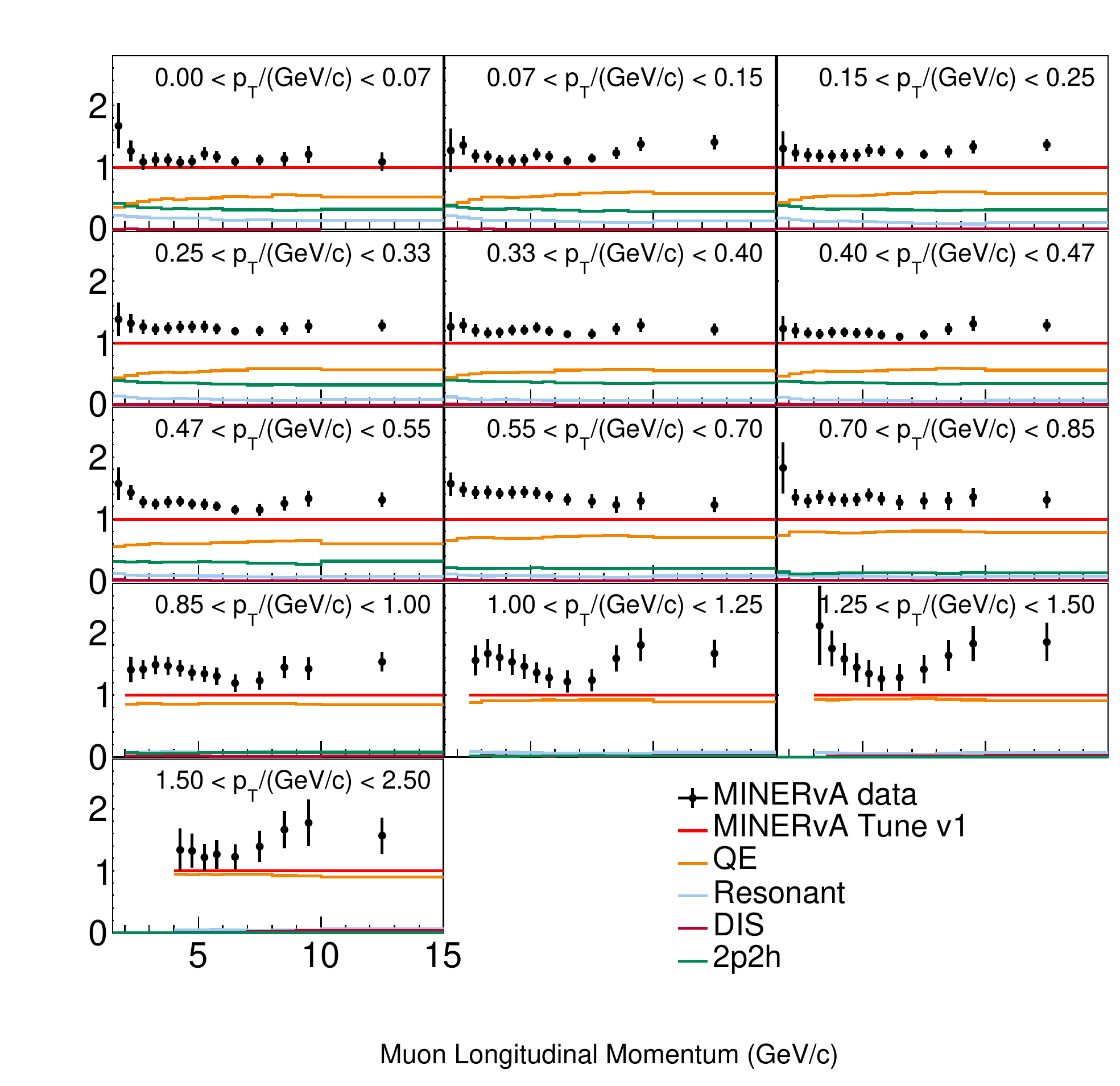}
    \caption{Ratio of cross sections of data and various components of MINERvA Tune v1 to MINERvA Tune v1 in  bins of muon transverse momentum.}
    \label{fig:pz-components}
\end{figure*}

\begin{figure*}[ht]
    \centering
    \includegraphics[width=0.99\linewidth]{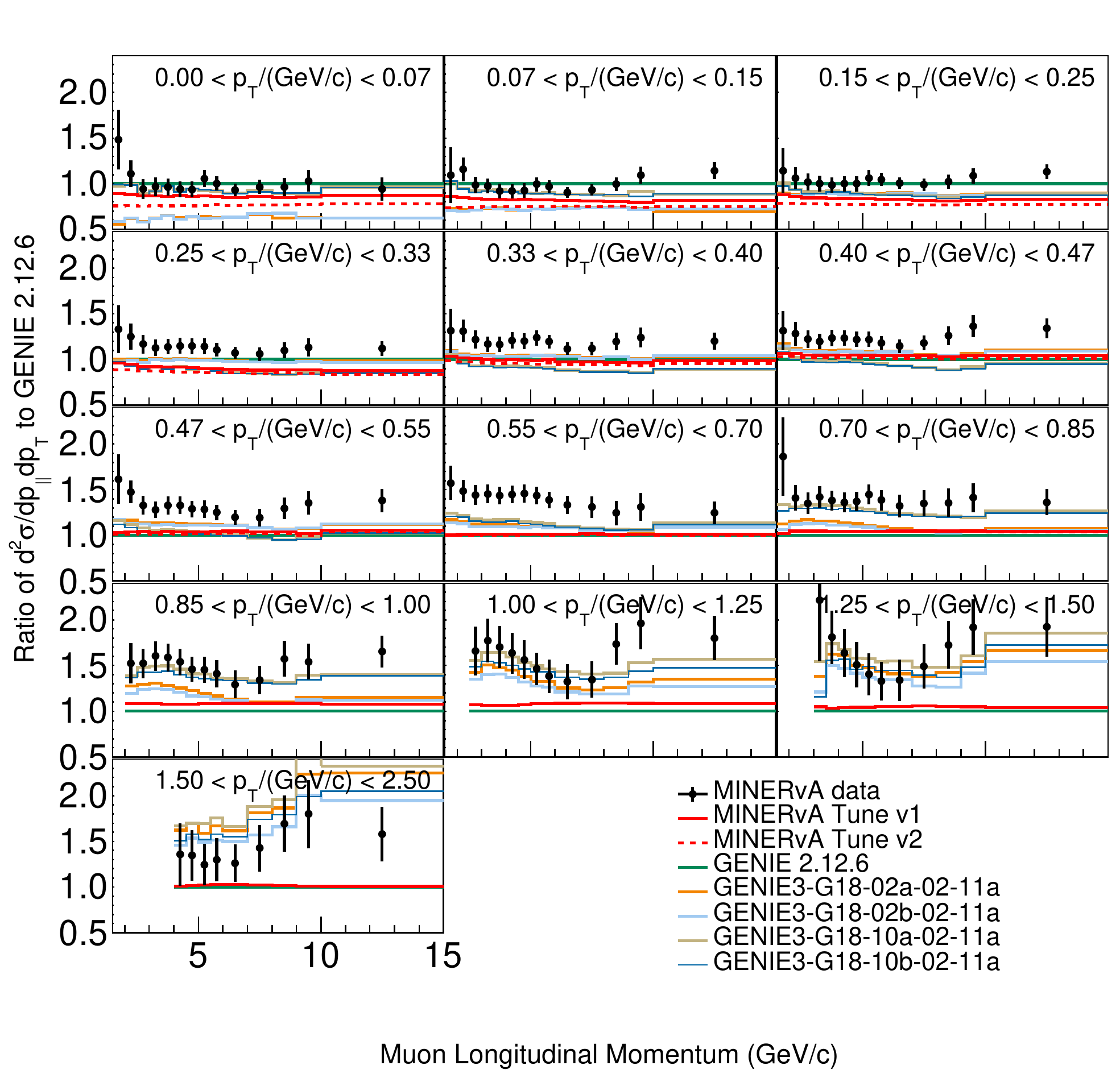}
    \caption{Ratios of double differential cross sections to GENIE 2.12.6 as a function of muon longitudinal momentum in bins of muon transverse momentum shown for GENIE 2 and tuned models along with GENIE 3 (v3.0.6) models.}
    \label{fig:pz-genie3-ratio}
\end{figure*}

\begin{figure*}[h]
    \centering
    \includegraphics[width=0.99\linewidth]{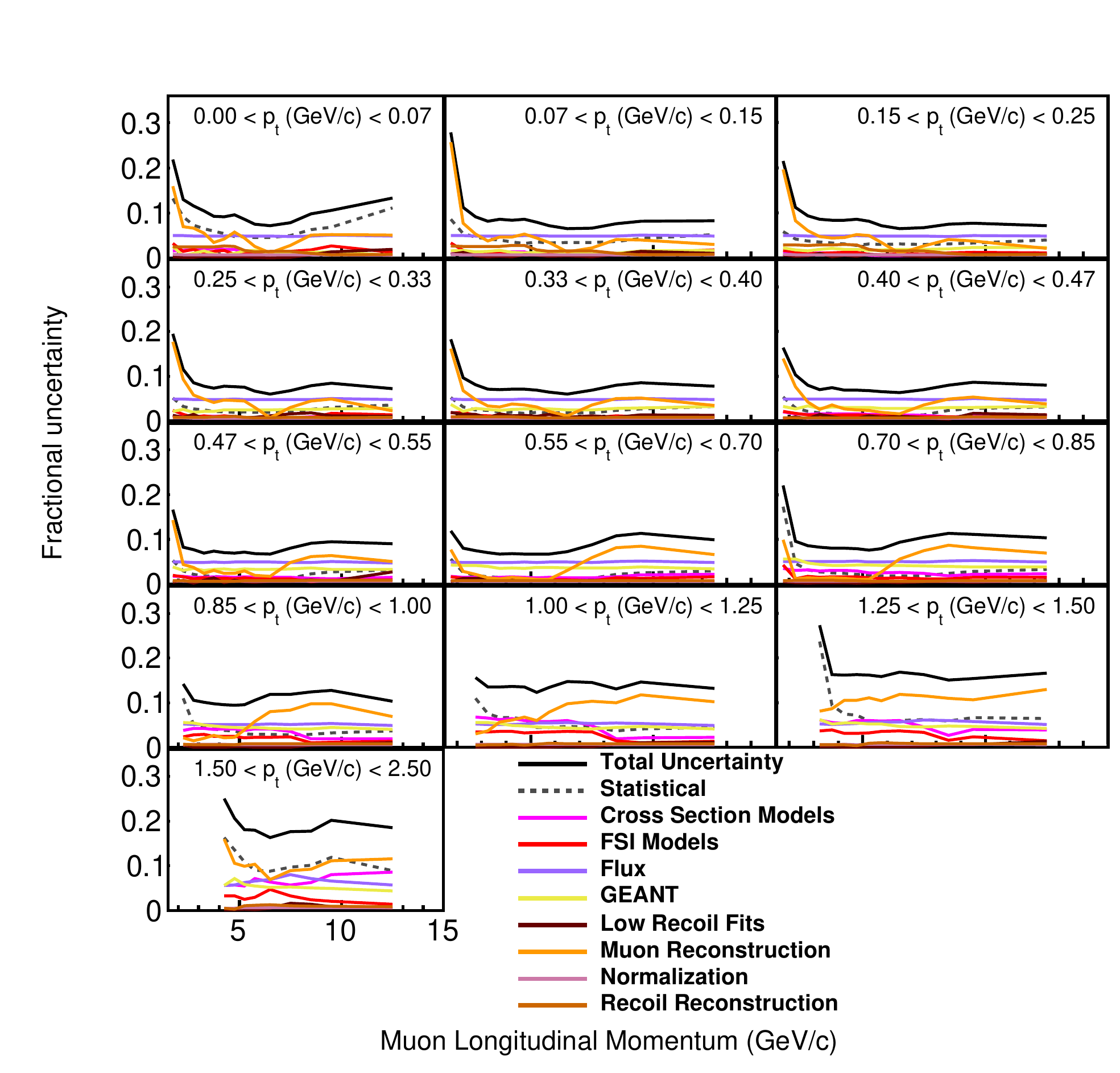}
    \caption{Summary of fractional uncertainties on the cross-section as a function of muon longitudinal momentum in  bins of muon transverse momentum shown in Fig.  \ref{fig:pz-genie3}}
    \label{fig:error summary muon pz}
\end{figure*}

\section{Uncertainties}
Summaries of the uncertainties on the double differential cross section  in Fig. \ref{fig:xsection-pt} are shown in Fig. \ref{fig:error-summary-xsec-pt}. Three major factors influence the uncertainties on this measurement.

First, the incoming neutrino flux has uncertainties in both overall normalization and energy dependence.
 The antineutrino flux is modeled using a GEANT4-based simulation of the target material and focusing systems. Focusing uncertainties are estimated by generating alternate fluxes where the focusing components of the beamline are shifted by $\pm$ 1 $\sigma$ from their nominal positions. Hadron production in the simulation is modified to match  thin target data-sets  following the method of\,\cite{Aliaga:2016oaz} that uses thin target data from the NA49 experiment\,\cite{Stefanek:2014pza} and the Barton data-sets\,\cite{PhysRevD.27.2580}.  %
Recent constraints from 
$\nu+e$\,\cite{MINERvA:2019hhc},  $\overline{\nu}+e$ scattering\,\cite{MINERvA:2022vmb} and inverse muon decay 
\,\cite{MINERvA:2021dhf} measurements in \minerva reduce the overall flux uncertainty from $\sim$8\% to $\sim$5\%. 

Second,  substantial uncertainties related to recoil and  muon reconstruction dominate the overall uncertainty budget and are estimated by varying the muon energy scale and angle, their resolutions and the neutron interaction cross section. 
As described in\,\cite{MINERvA:2021mpk} the muon energy scale is corrected by 3.6\% from its nominal value to resolve a discrepancy between data and the simulation. Correlations between neutrino flux parameters and muon energy scale, which result from that simultaneous fit to the energy scale and flux parameters, are accounted for, when assessing the muon energy uncertainty.
As seen in Fig. \ref{fig:ME nu nu-bar xsection} the resulting uncertainties due to muon reconstruction rise to 9\% at high $Q^2_{QE}$.

Uncertainties in the energy loss  and detection efficiency of final-state hadrons in the \minerva detector also introduce uncertainties in the reconstruction of the recoil energy and are estimated by simulating the effects of differing hadronic interaction cross sections in the detector. The recoil response uncertainty contribution to the cross section is dominated by the model of neutron interactions and ranges from 1\% at low $Q^2_{QE}$ to 4\% at high $Q^2_{QE}$. 

Finally, the cross section extraction depends on proper simulation of neutrino interactions including final state processes, which are estimated by varying input parameters to the event simulations. 
Final state interactions (FSI) effects contribute less than 2-3\% to the final $Q^2_{QE}$ cross-section uncertainty  while GENIE cross-section model parameters (such as the axial mass, random phase approximation and resonance production) contribute from 1\% at low $Q^2_{QE}$ to 6\% at high $Q^2_{QE}$.%
For the cross section as a function of $\enu$, model uncertainties in converting from $\enuqe$ to $\enu$ increase the GENIE and FSI contributions to the uncertainty budget to 4-6\% each.

Uncertainties are propagated through the ``universe" method developed by the MINERvA collaboration and described in Ref.\,\cite{MINERvA:2021ddh}. For each observable,  separate histograms (``universes") of the simulated reconstructed variable are stored for each of the several hundred sources of systematic uncertainty.  For example, the reconstructed muon angle is replicated with $\pm 1 \sigma$ offsets in 2 directions transverse to the beam and several GENIE parameters are similarly varied.  The full analysis chain is applied to each  universe independently. For the flux systematics which depend upon various beamline parameters, 500 ``universes" are simulated  where each universe is drawn from a distribution of focusing parameters that takes into account their uncertainties and their correlations.  These beam parameters and their 1 $\sigma$ values are listed in\,\cite{MINERvA:2021mpk}.
The total systematic uncertainty and covariance for any observable are then estimated by summing the deviations of the modified universes from the central values provided by the simulation in quadrature\,\cite{Aliaga:2016oaz}. Table \ref{tab:Q2_QE cross-section uncertainties} summarizes the significant uncertainties in the measured cross section.

\begin{table*}[ht]
    \centering
    \begin{tabular}{llr}
     Quantity    & Variation ($\pm 1 \sigma$ from CV)  & Effect on cross section (\%)\\
     \hline %
    Angle reconstruction     & $\pm 1$ mr& 0 - 2\\
    MINOS muon energy scale & $\pm 1.0$ \% & 2 - 6\\
    GEANT Neutron & $\pm 10$ \% & 2 - 5\\
    Flux & Focusing/interaction parameters & 5\\
    GENIE Cross Section Models & GENIE cross section parameters  & 1-5 \\
   GENIE 2p2h  & Low recoil fit parameters& 1-3\\
    Final State Interaction Model&  GENIE FSI 
     models  parameters &1-3\\
    \end{tabular}
    \caption{Effect of input uncertainties on the cross section extraction for $d\sigma/dQ^2_{QE}$. Uncertainties which have significant effect on final cross section are listed. The $\pm 1 \sigma$ is the shift of model parameters from their central values (CV).}
    \label{tab:Q2_QE cross-section uncertainties}
\end{table*}

\section{Discussion of results}

\subsection{Differential cross sections in muon kinematics}
The primary result is the  measured double differential cross section in  bins of muon momenta shown in Fig. \ref{fig:xsection-pt}. The relative contributions of various processes to the \minerva  v1 tune are shown in Fig. \ref{fig:xsection-pt-ratio}.
This model predicts that  the CCQE-like cross section is dominated by  pure 1p1h QE and 2p2h processes. The low $\pperp$ region is dominated by QE and 2p2h processes whereas the high $\pperp$ region is dominated by QE processes only. %
Generally, the data lies above the simulation in almost all bins, with the excess growing as $\pperp$ increases. Additional comparisons to models are shown in Fig. \ref{fig:pt-genie3-main}
    and \ref{fig:pt-genie3-ratio} with an uncertainty summary in
    Fig. \ref{fig:error-summary-xsec-pt}  The integrated cross section for  $1.5<\ppar<15$ GeV/c and  $0.0<\pperp<2.5$ GeV/c within our restricted
fiducial region
is $5.28\times \pm 0.02  \pm 0.35 \times 10^{-39}$ cm$^{2}$/nucleon.

Figures \ref{fig:pz-components}, \ref{fig:pz-genie3}, 
    \ref{fig:pz-genie3-ratio}, and 
    \ref{fig:error summary muon pz} show the same data as a function of $\ppar$ in bins of $\pperp$.

    Table \ref{tab:pzptchi2} shows the $\chi^2$ for comparisons of the model variations to the 2-D $\pperp, \ppar$ cross section measurements. The $\chi^2$ is calculated in two ways, once on the values themselves and once on the log of the values. The log method is discussed in more detail in references\,\cite{Patrick:2018gvi,CARLSON20093215} and is more robust in the presence of multiplicative normalization uncertainties.

\subsection{Cross sections in $Q^{2}_{QE}$} 
We also present one dimensional cross sections as a function of $Q^{2}_{QE}$. and $\enu$. Fig. \ref{fig:Q2 model comparison} shows the ratio of data and a suite of cross-section models  to the baseline GENIE 2.12.6 which uses the Valencia 2p2h model as a function of $Q^2_{QE}$. The cross-section models shown are default GENIE 2.12.6, the MINERvA tunes and partial combinations of the components in the MINERvA tunes. 
All of the GENIE 2 models fail to reproduce the high $Q^2_{QE}$ behavior of the cross section. %
The shape of the data appears to favor models that include RPA effects while the $\pi$-tune correction appears to have little effect on the predicted rates. 

\subsection{Comparisons to GENIE 3.0.6}

\begin{figure*}[ht]
    \centering

    \includegraphics[width=0.49\linewidth]{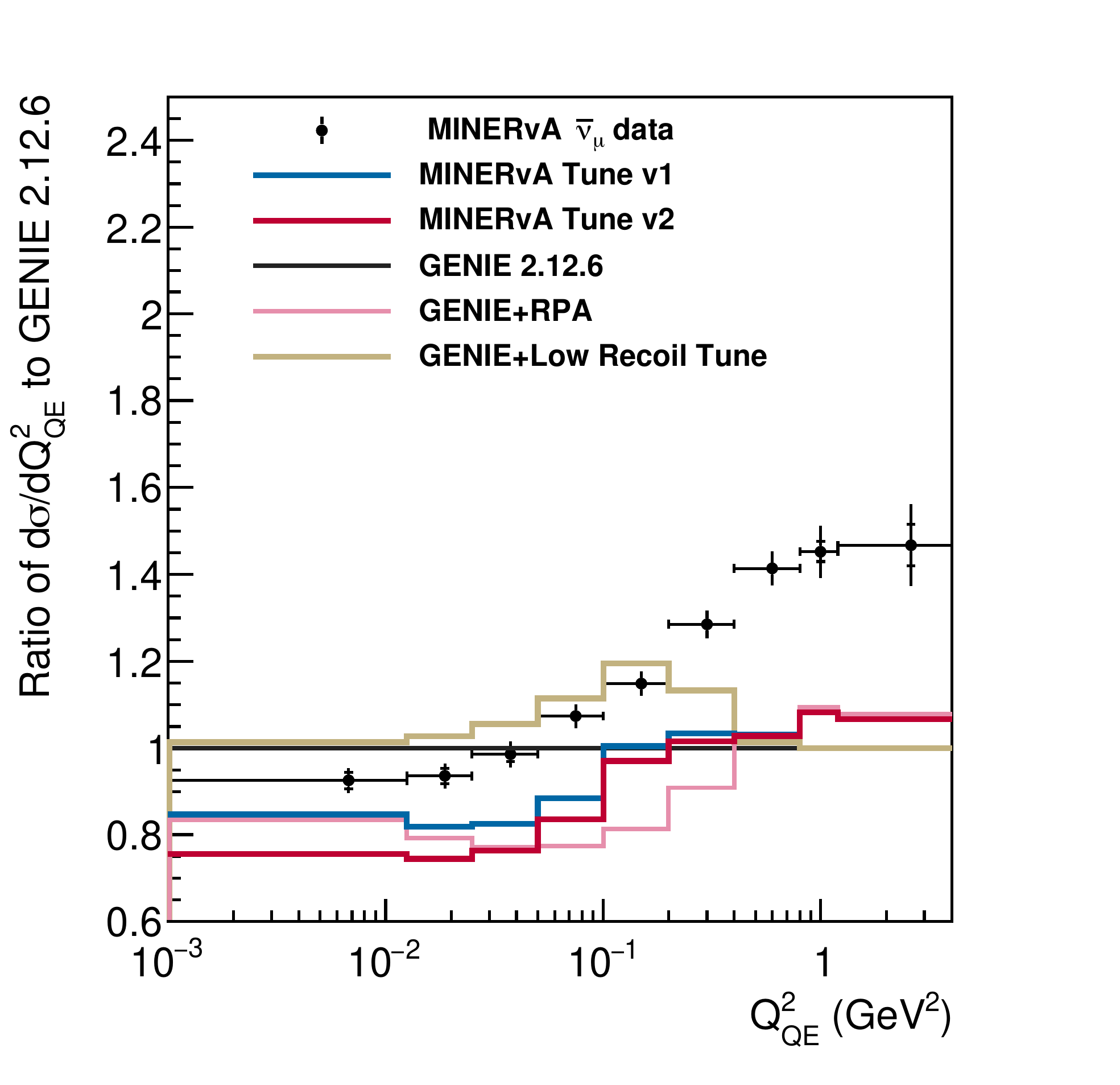}
    \includegraphics[width=0.49\linewidth]{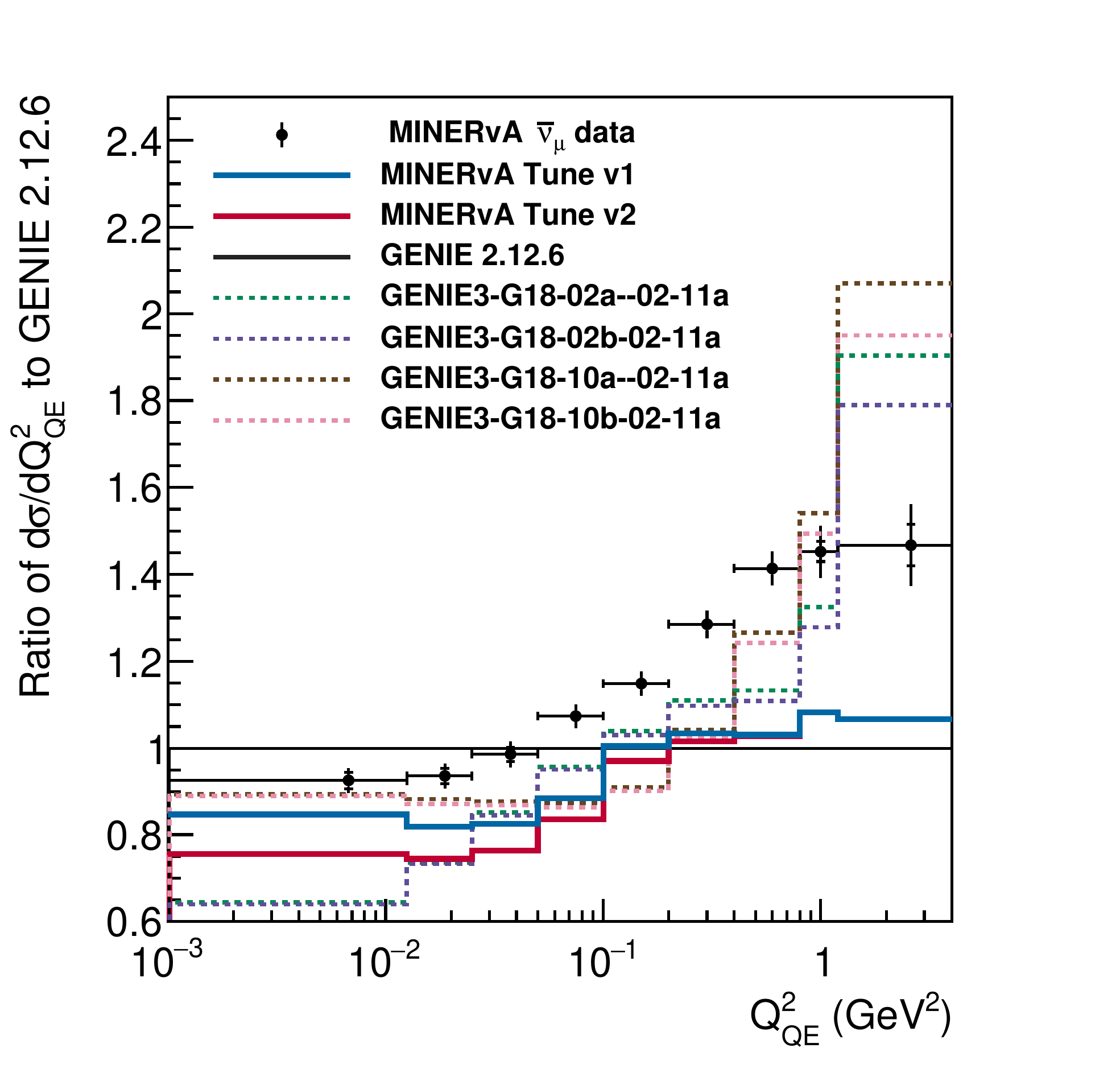}
    \caption{Comparisons of the cross section predicted by various tunes applied on GENIE with respect to the baseline GENIE 2.12.6 (black) as a function of $Q^{2}_{QE}$ (left). \minerva Tune v1 (blue) is the standard simulation tuned to the \minerva low energy data. \minerva Tune v2 (red) is \minerva Tune v1 with  non-resonant pions suppressed in the low $Q^{2}_{QE}$ region\,\cite{Stowell:2019zsh}. The remaining curves show the effect of enabling different corrections to the base model. The plots on the right show comparisons of cross sections predictions for GENIE v3.0.6 (dotted lines) with the \minerva tuned GENIE v2.12.6 predictions. Inner error bars represent statistical uncertainties and  outer error bars represent systematic uncertainties.} 
    \label{fig:Q2 model comparison}
\end{figure*}

Recently available GENIE 3.0.6 models  appear to better reproduce the high $Q^2$ behavior and are shown on the right hand side of Fig. \ref{fig:Q2 model comparison} and as a function of muon kinematics in Fig. \ref{fig:pt-genie3-ratio}. Comparisons are shown with two different cross-section model tunes \,\cite{GENIE:2021zuu}. 
The tune {\tt G18\_02x\_02\_11a} has a 2p2h model similar to that of default GENIE 2. GENIE 3 models {\tt G18\_10a\_02\_11a} and {\tt G18\_10b\_02\_11a} incorporate the default Valencia\,\cite{Nieves:2011pp} model. {\tt G18\_10a\_02\_11a} uses an effective ``hA" intranuclear transport model while {\tt G18\_10a\_02\_11a} incorporates the full intranuclear ``hN" transport model which includes additional processes \,\cite{GENIE:2021zuu}.  
The best agreement at high $Q^2$ among the GENIE 3 models is with model {\tt G18\_10b\_02\_11a} which incorporates the  Valencia model and the ``hN" model for final state interactions. 

\begin{figure*}[ht]
    \centering
    \includegraphics[width=0.49\linewidth]{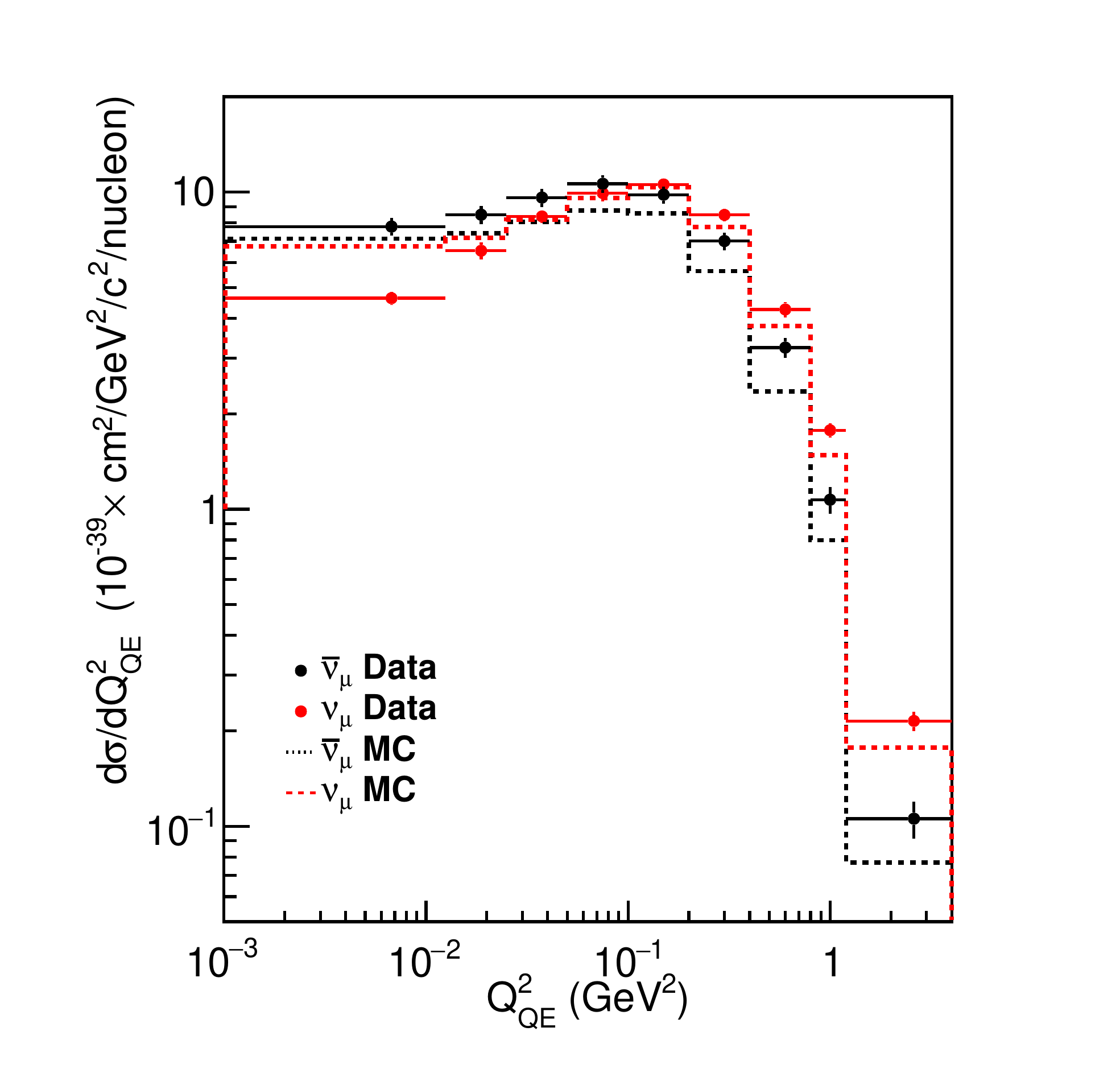}
    \includegraphics[width=0.49\linewidth]{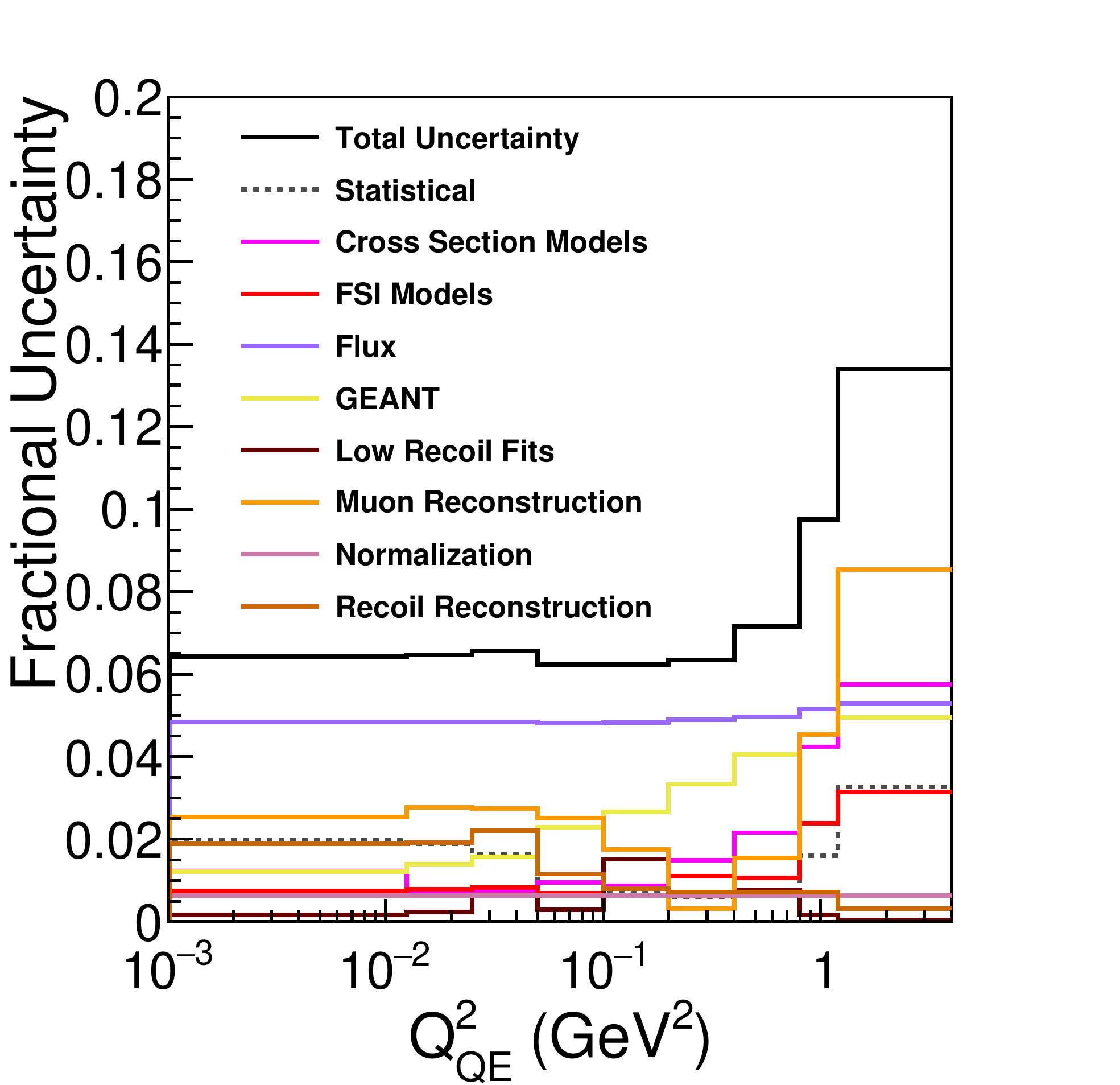}
    \caption{Left, measured (data points) and \minerva Tune v1 prediction (dotted lines) of CCQE-like $d\sigma$/$dQ^{2}_{QE}$ for neutrinos (reference \cite{Carneiro:2019jds}, red) and antineutrinos (this measurement, black) extracted with the \numi neutrino and antineutrinos energies $\sim$ 6 GeV.  Right, summary of fractional uncertainties on the differential antineutrino cross-section as a function of $Q^2_{QE}$ hypothesis.}
    \label{fig:ME nu nu-bar xsection}
\end{figure*}

Fig. \ref{fig:ME nu nu-bar xsection} shows comparisons of data to simulation for the corresponding neutrino data\,\cite{Carneiro:2019jds} sample and this sample as a function of $Q^{2}_{QE}$. Both $\nu_{\mu}$ and $\bar{\nu_{\mu}}$ cross sections are extracted from the same fiducial region of the \minerva detector. The $\nu_{\mu}$ cross section includes CCQE-like events with any number of protons, whereas  the $\bar{\nu}_{\mu}$  requires no protons above $T_p$ of 120 MeV. Only antineutrino events with $p_{||}$ less than 15 GeV/c for $\bar\nu$ are selected to reduce the larger wrong sign neutrino contamination in the antineutrino beams. As the beam  energy is peaked well below this threshold, this restriction has negligible effect except for a $\sim$ 1\% reduction in the highest $Q^{2}_{QE}$ bin. 
Signal selection cuts for the $\numu$ events are given in \,\cite{Carneiro:2019jds}.

\subsection{Cross section as a function of  $E_{\overline{\nu}}$}\label{sec:enu}
\begin{figure*}[ht]
    \centering
      \includegraphics[width=0.49\linewidth]{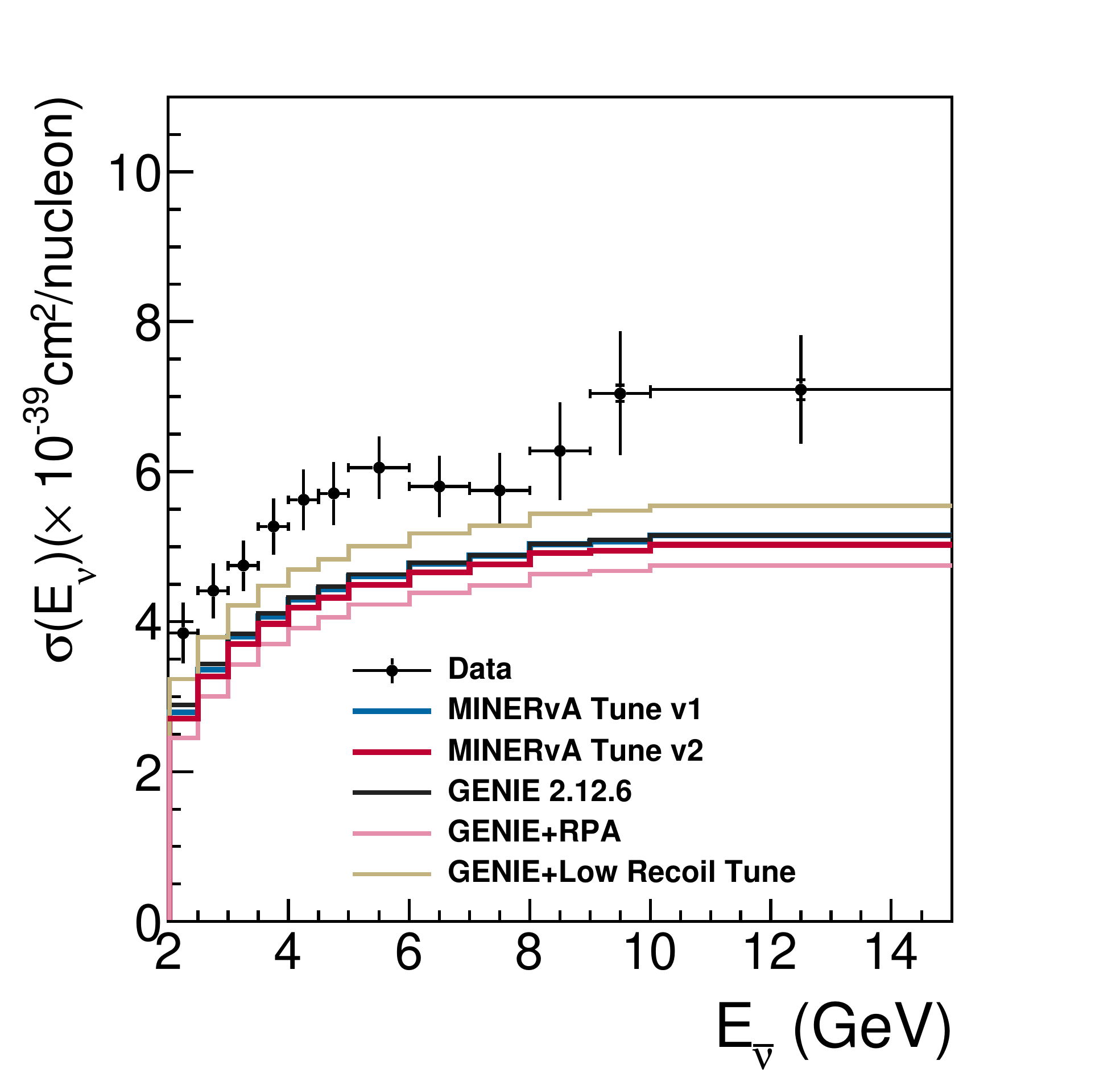}
      \includegraphics[width=0.49\linewidth]{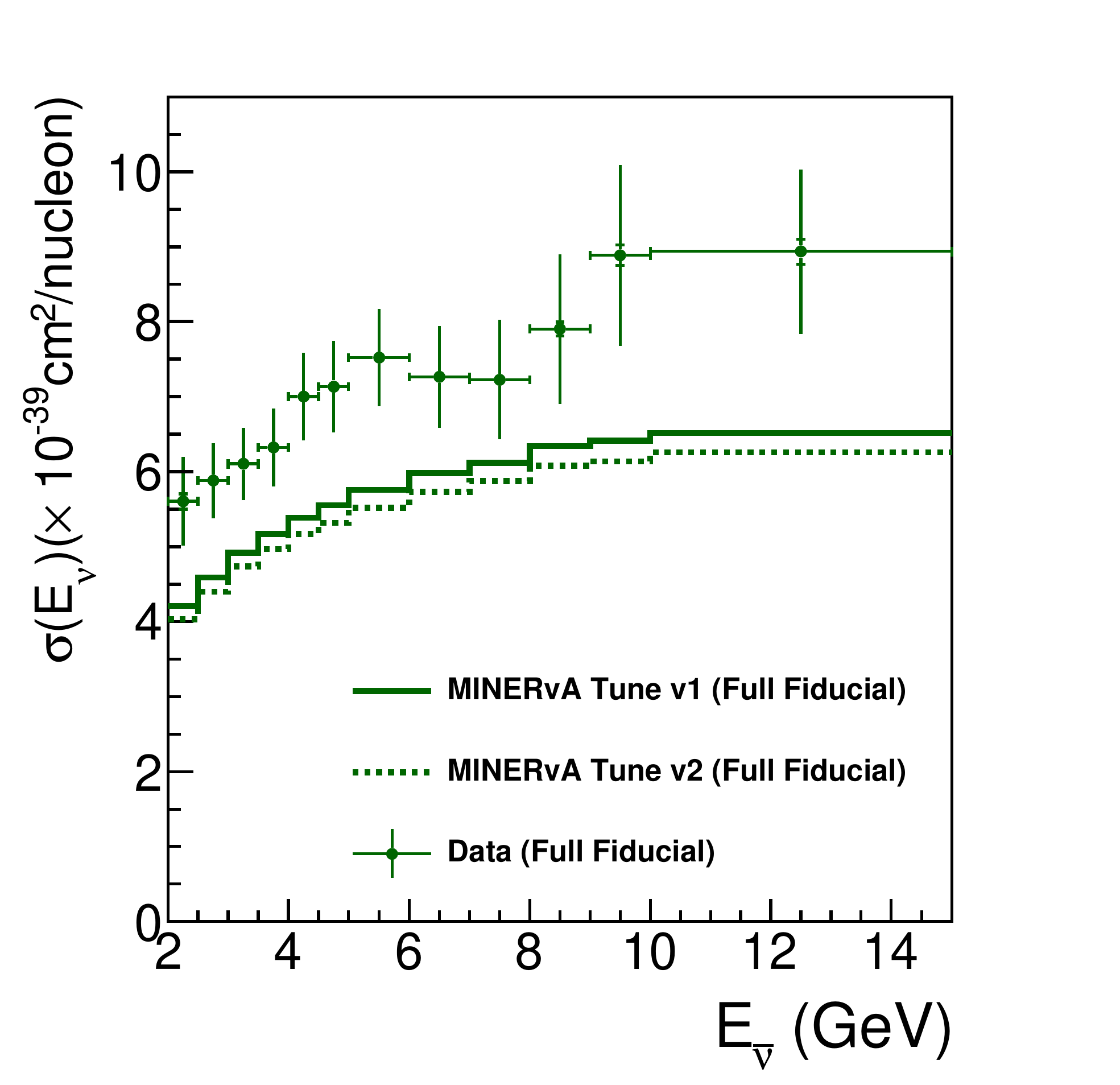}
    \caption{CCQE-like cross section as a function of  true $\bar{\nu}$ energy. The black data points on the left pane use our fiducial final state definition with $\theta_{\mu}< 20^{\degrees}$ and $T_p<120$ MeV. The green  points  on the right pane are data and the  \minerva tunes v1 and v2 model predictions (straight and dotted lines) for the total cross section with the $\theta_\mu$ and $T_p$ cuts removed.}
    \label{fig:True Q2 and Enu xsection}
\end{figure*}

In Fig. \ref{fig:True Q2 and Enu xsection} we show the cross section as a function of true neutrino energy with and without corrections for the $T_p$ and $\theta_\mu$ selections. 
The left panel on Fig. \ref{fig:True Q2 and Enu xsection} compares the data as a function of true neutrino energy with various models. Here the $T_p$ and $\theta_\mu$ selections are applied as in the previous results. In the right panel, the green data points and lines show the cross section and \minerva models corrected (by 25-40\%) to all $\theta_{\mu}$ and $T_p$ for easier comparison to inclusive models and other experiments.  

\begin{figure*}[ht]
    \centering
    \includegraphics[width=0.49\linewidth]{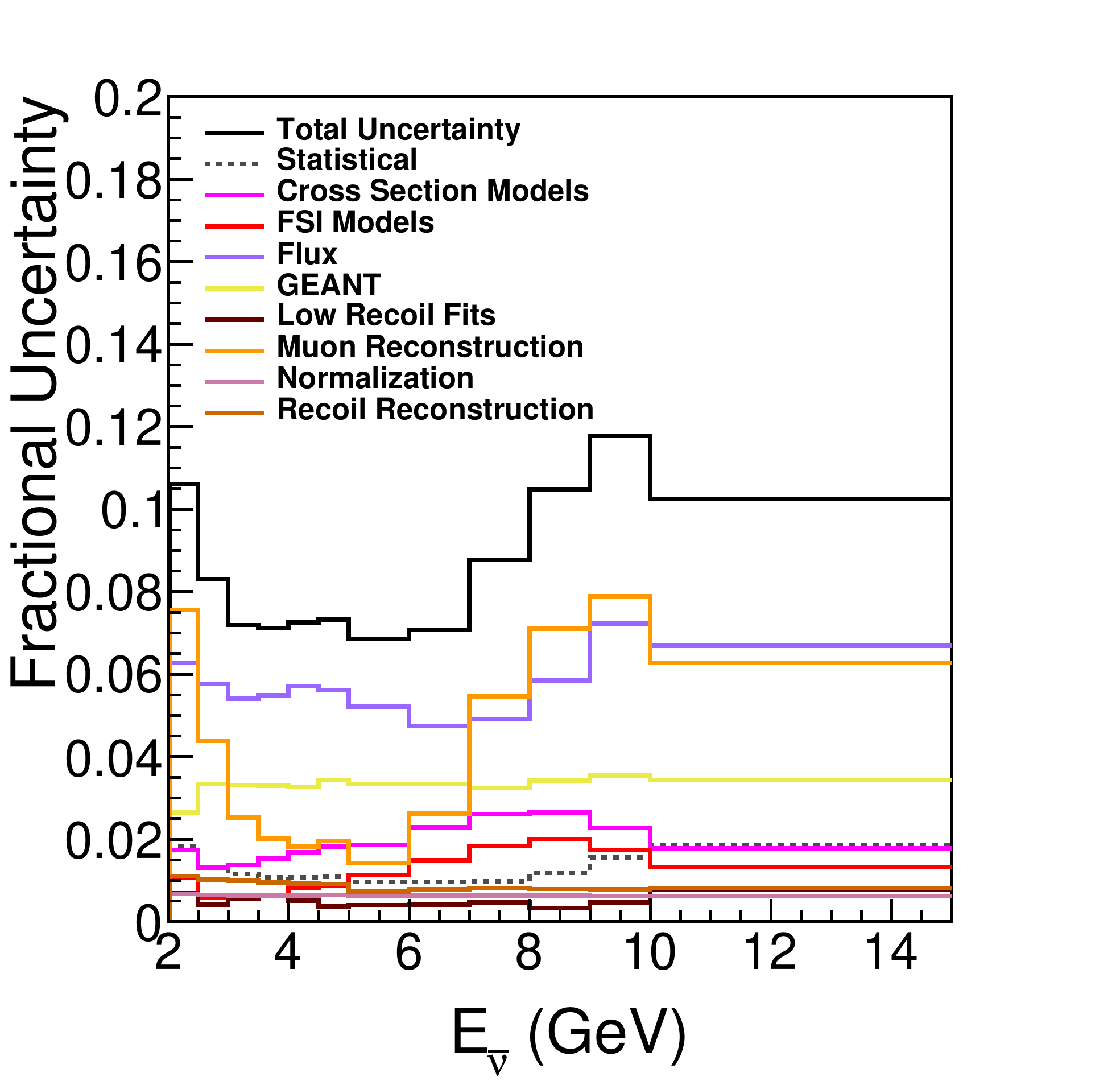}
    \includegraphics[width=0.49\linewidth]{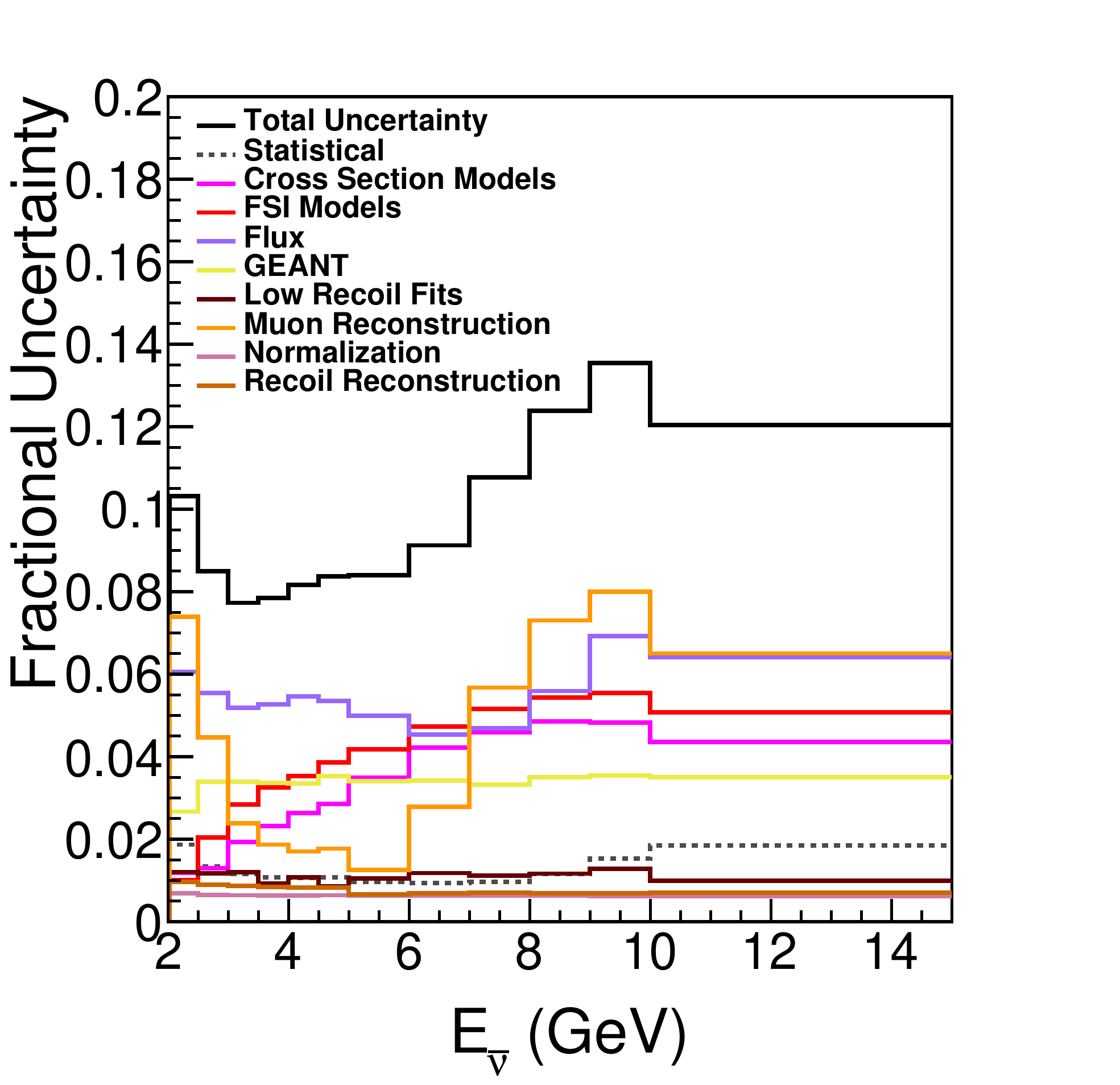}
    \caption{Error summary on the measured fiducial restricted (left)  and full fiducial (right) cross-section as a function of true neutrino energy.}
    
    \label{fig:fid error summaryEnu}

\end{figure*}

The theoretical uncertainties on the $\enu$ results  are larger than those expressed directly in terms of muon kinematics. %
First, due to the conversion from $\enuqe$ to $\enu$, which is necessary to normalize by the antineutrino flux $\Phi(\enu)$, but is affected by uncertainties in nuclear models and second, because the correction for the $T_p$ and $\theta_\mu$ selections necessary to go from the restricted cross section in the left panel to the full cross section in the right panel introduces additional theoretical uncertainties.  Fig. \ref{fig:fid error summaryEnu}  shows the corresponding  uncertainties for the restricted phase space (left) and the full phase space (right). Note that the uncertainties due to the cross section model and FSI increase substantially when one removes the fiducial restrictions.  %

\begin{table}[ht]
\begin{tabular}{lrrrr}
Model &	 $\chi^2$ - linear &	 $\chi^2$ - log  \\ 
\hline%
{\bf GENIE 2.12.6 Tunes} \\
MINERvA Tune v1 &	      362.6 &	      580.4 	 \\
MINERvA Tune v2 &	      364.4 &	      601.4 	 \\
GENIE  w/o 2p2h &	      226.5 &	      473.2 	 \\
GENIE (Default) &	      346.4 &	      550.6 	 \\
GENIE+$\pi$tune &	      354.3 &	      568.5 	 \\
GENIE+RPA &	      230.0 &	      406.7 	 \\
GENIE+RPA+$\pi$tune &	      231.7 &	      414.6 	 \\
GENIE+Low Recoil Tune &	      755.4 &	     1059.4 	 \\
GENIE+Low Recoil Tune+RPA &	      361.2 &	      570.0 	 \\
GENIE+Low Recoil Tune+$\pi$tune &	      760.6 &	     1081.8 	 \\
\\
{\bf GENIE 3.0.6 Tunes} \\
GENIE 3.0.6 G18\_02a\_02\_11a &	      602.9 &	      865.0 	 \\
GENIE 3.0.6 G18\_02b\_02\_11a &	      586.9 &	      878.3 	 \\
GENIE 3.0.6 G18\_10a\_02\_11a &	      353.1 &	      447.5 	 \\
GENIE 3.0.6 G18\_10b\_02\_11a &      312.8 &     421.7   \\
\end{tabular}

\caption{$p_\parallel - p_\perp$ $\chi^2$  between data and model variants derived from GENIE. The number of degrees of freedom is 171. Both the $\chi^2$ between the values and between the logs of the values are listed.} \label{tab:pzptchi2}
\end{table}

\subsection{Summary}  
We have presented the antineutrino CCQE-like cross section measured with the \numi $<\enu >\sim 6 $ GeV beam in the \minerva detector. The measurement extends our  past measurements in the antineutrino channel made in the lower neutrino energy beam and complements the neutrino cross section measurement at higher energy.
We have extended the measurement of $Q^{2}_{QE}$  to 4 GeV$^{2}$. These measurements indicate that none of the variations of GENIE v2 based on the \minerva low energy data are sufficient to  describe the \minerva medium energy data. However, GENIE v3 models appear to better describe the data especially in the high $Q^{2}_{QE}$ region which is dominated by the QE process. This high statistics double differential cross section   provides  useful model inputs  for future neutrino oscillation experiments.  Tables of values are available in the supplemental materials\,\cite{supplement}.  %

\begin{acknowledgments}
This document was prepared by members of the MINERvA Collaboration using the resources of the Fermi National Accelerator Laboratory (Fermilab), a U.S. Department of Energy, Office of Science, HEP User Facility. Fermilab is managed by Fermi Research Alliance, LLC (FRA), acting under Contract No. DE-AC02-07CH11359.
These resources included support for the MINERvA construction project, and support
for construction also
was granted by the United States National Science Foundation under
Award No. PHY-0619727 and by the University of Rochester. Support for
participating scientists was provided by NSF and DOE (USA); by CAPES
and CNPq (Brazil); by CoNaCyT (Mexico); by Proyecto Basal FB 0821, CONICYT PIA ACT1413, and Fondecyt 3170845 and 11130133 (Chile); 
by CONCYTEC (Consejo Nacional de Ciencia, Tecnolog\'ia e Innovaci\'on Tecnol\'ogica), DGI-PUCP (Direcci\'on de Gesti\'on de la Investigaci\'on  - Pontificia Universidad Cat\'olica del Peru), and VRI-UNI (Vice-Rectorate for Research of National University of Engineering) (Peru); NCN Opus Grant No. 2016/21/B/ST2/01092 (Poland); by Science and Technology Facilities Council (UK).  We thank the MINOS Collaboration for use of its near detector data. Finally, we thank the staff of
Fermilab for support of the beam line, the detector, and computing infrastructure.

\end{acknowledgments}
\clearpage
\ 
\bibliographystyle{apsrev4-2}
\bibliography{main}
\input{supplement.tex}

\end{document}

%% file: supplement.tex
\section*{Supplemental Material}

\subsection*{More information on the $Q^2_{QE}$ distributions}

\begin{figure*}[ht]
    \centering
    \includegraphics[width=0.99\linewidth]{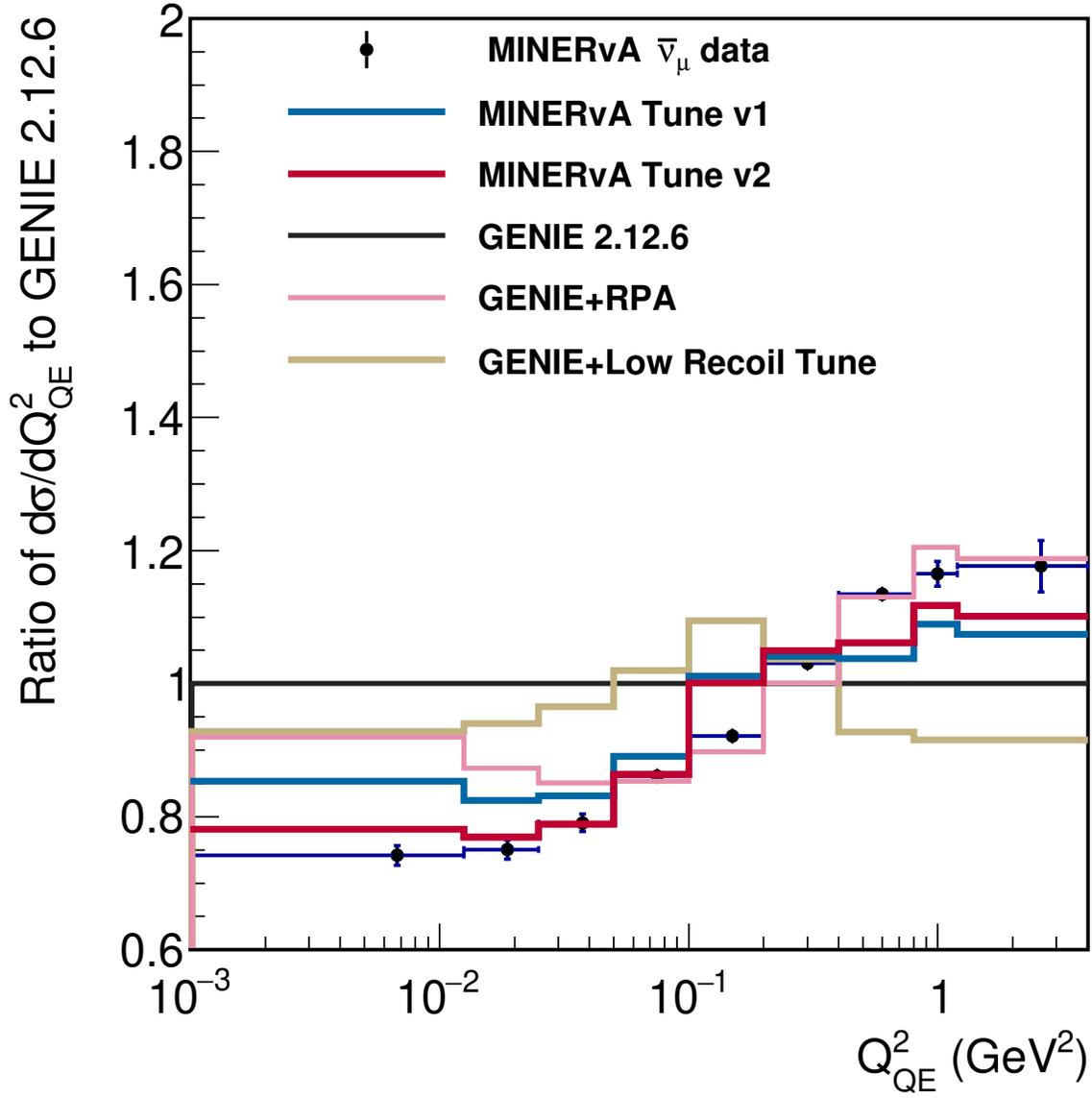}
    \caption{Comparisons of the cross section shapes predicted by various tunes applied on GENIE with respect to baseline GENIE 2.12.6 (black) as a function of $Q^{2}_{QE}$. The cross sections are area normalized before taking the ratio for the shape comparison.}
    \label{fig:q2 ratio area normalized}
\end{figure*}

\begin{figure*}[ht]
    \centering
    \includegraphics[width=0.99\linewidth]{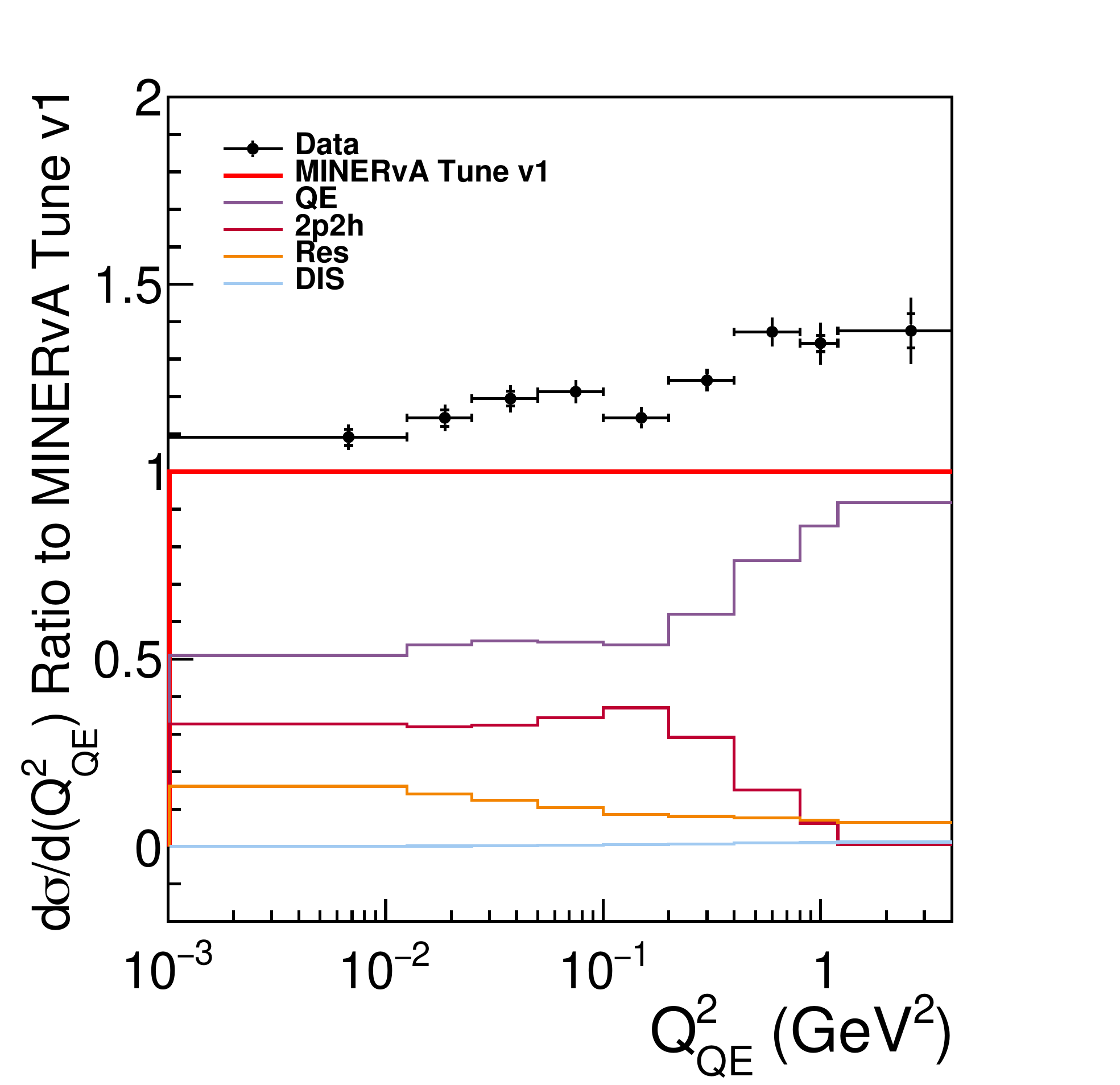}
    \caption{Ratio of measured and various components of MINERvA Tune v1 to MINERvA Tune v1}
    \label{fig:q2 mc components}
\end{figure*}

Fig. \ref{fig:q2 ratio area normalized} shows the ratio of the data to various GENIE 2 models.  Fig. \ref{fig:q2 mc components} shows the contribution of different model components to the GENIE 2 prediction.

\subsection*{Effects of removing the $T_p$ selection}

\begin{figure*}[ht]
    \centering
    \includegraphics[width=0.99\linewidth]{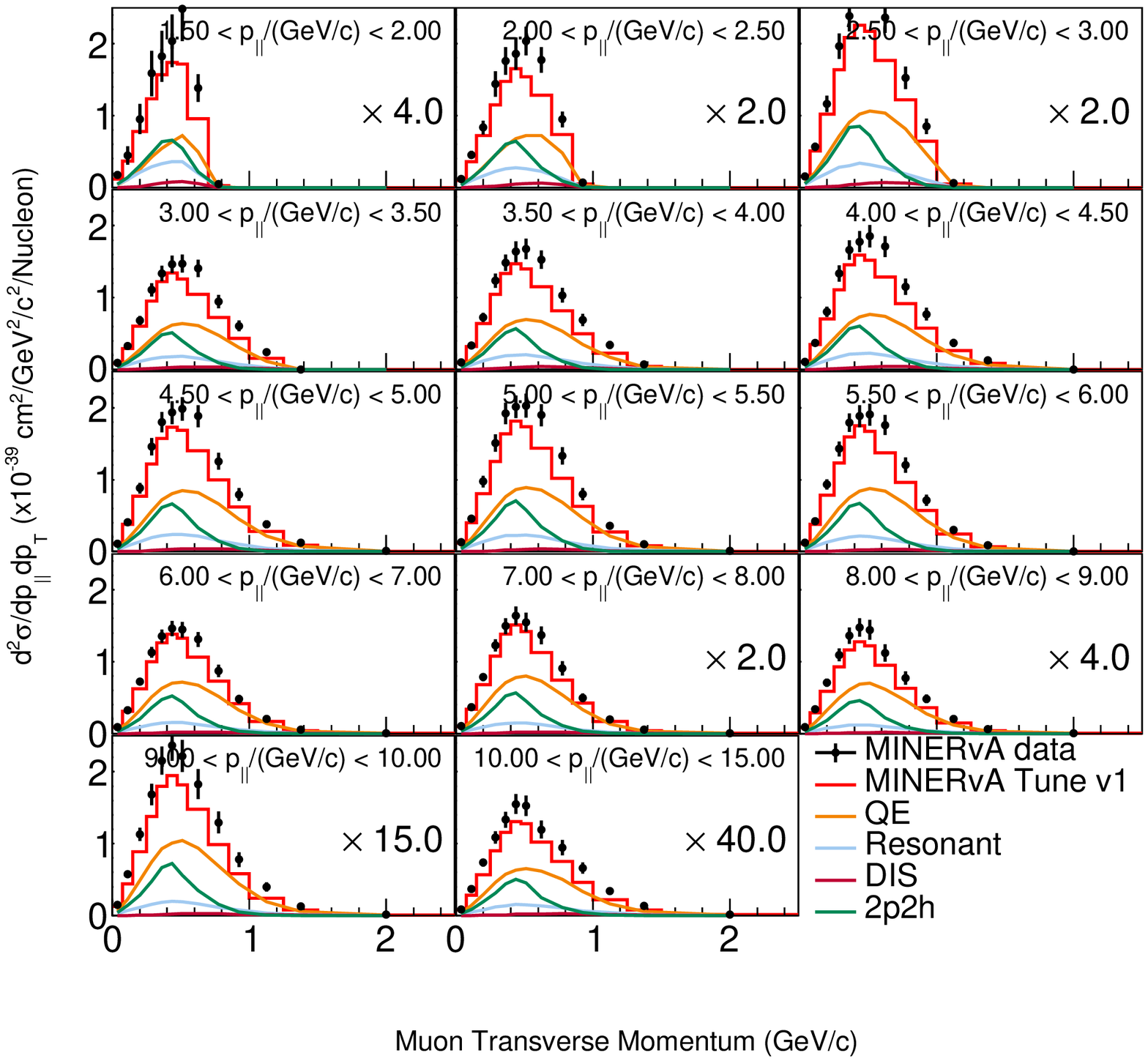}
    \caption{Measured and model prediction of double differential cross section as a function of muon transverse momentum in  bins of muon longitudinal momentum. The requirement to exclude proton kinetic energy greater than 120 MeV is removed from the definition of CCQELike events.}
    \label{fig:pt-noprotonKE-xsection}
\end{figure*}

\begin{figure*}[ht]
    \centering
    \includegraphics[width=0.99\linewidth]{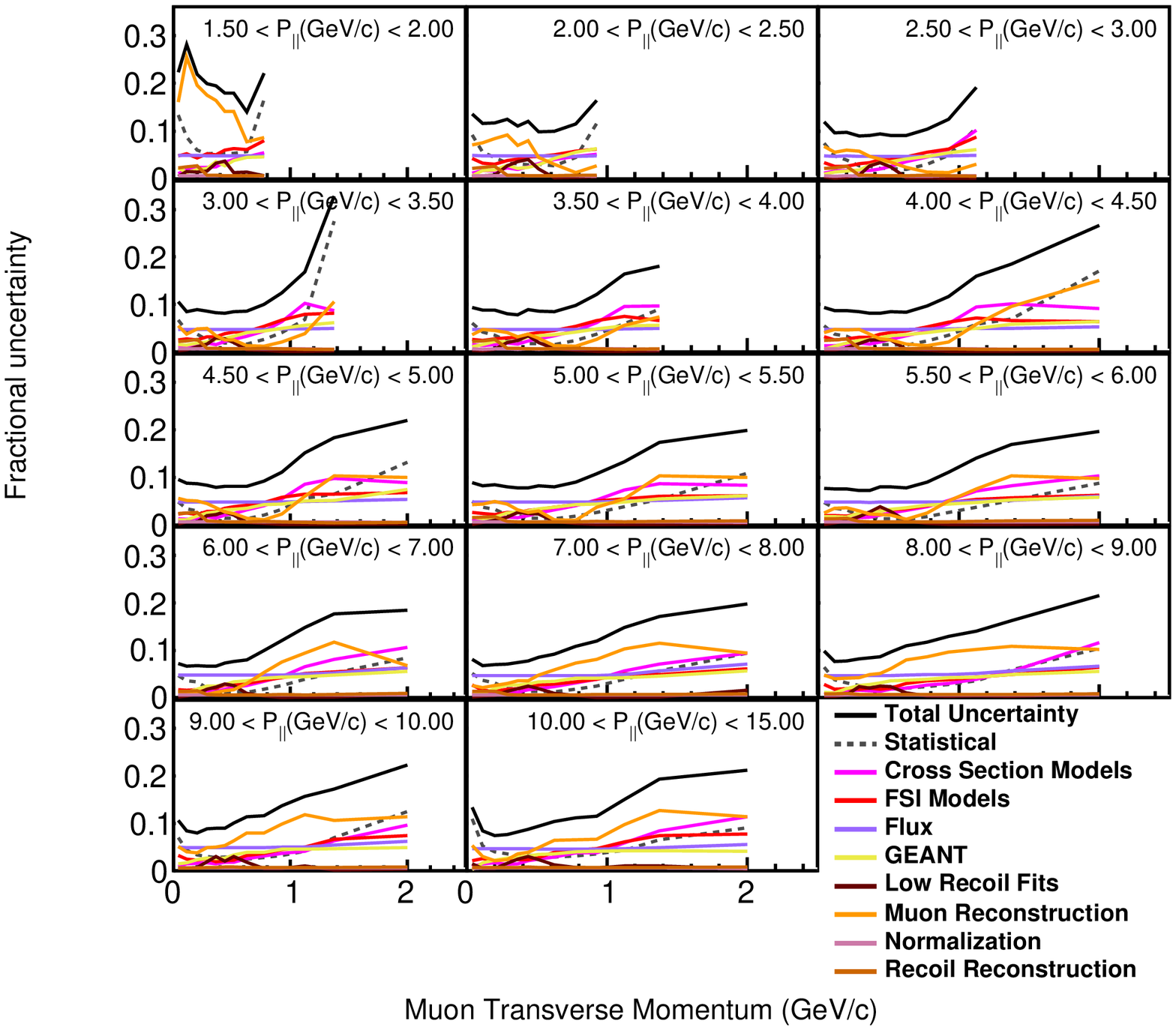}
    \caption{Summary of uncertainties for the measured cross section shown in Fig. \ref{fig:pt-noprotonKE-xsection}.}
    \label{fig:pt-noprotonKE-xsection-errorSummary}
\end{figure*}

\begin{figure*}[ht]
    \centering
    \includegraphics[width=0.99\linewidth]{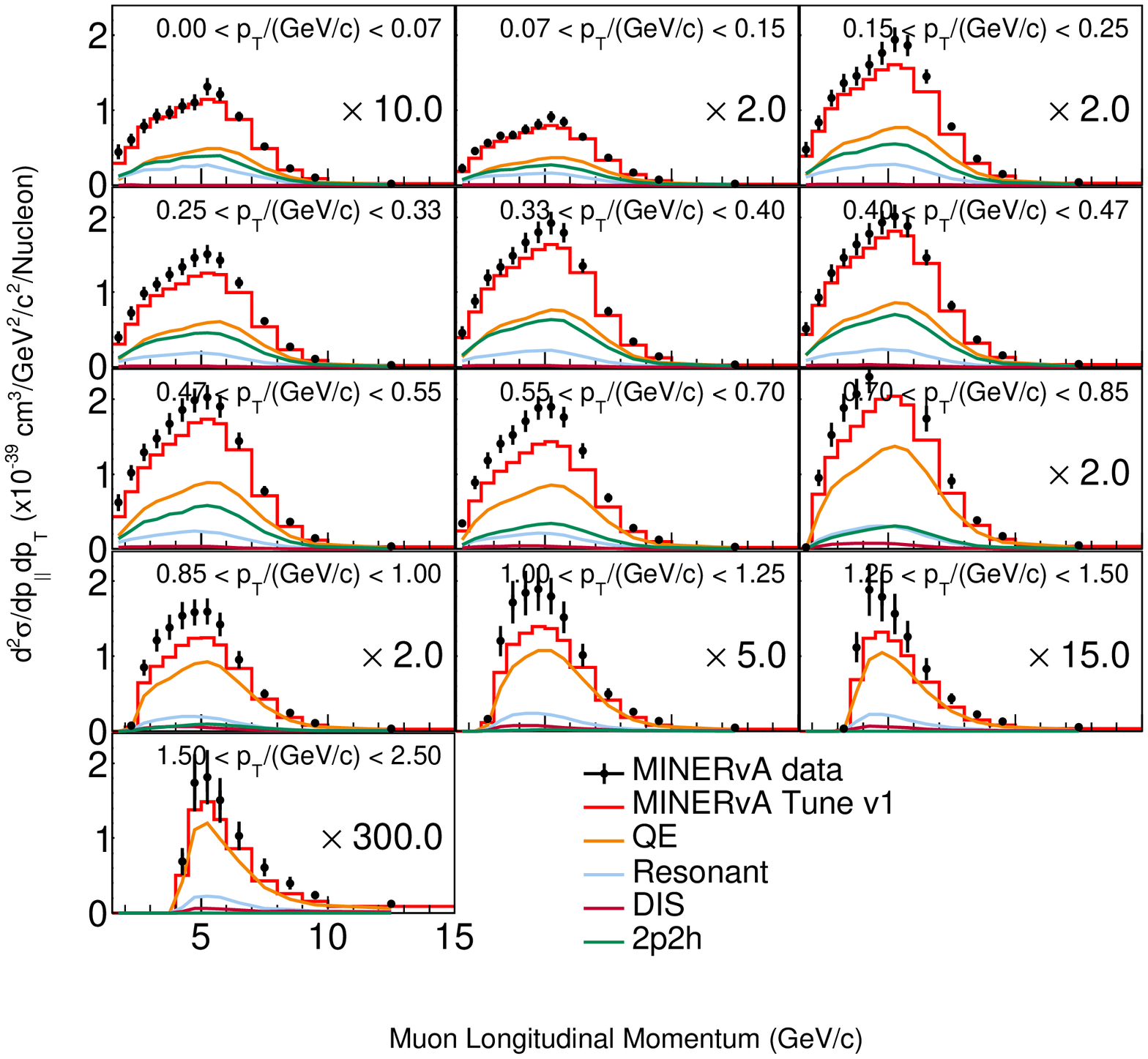}
    \caption{Measured and model prediction of double differential cross section as a function of muon longitudinal momentum in  bins of muon transverse momentum. The requirement to exclude proton kinetic energy greater than 120 MeV is removed from the definition of CCQELike events.}
    \label{fig:pz-noprotonKE-xsection}
\end{figure*}

\begin{figure*}[ht]
    \centering
    \includegraphics[width=0.99\linewidth]{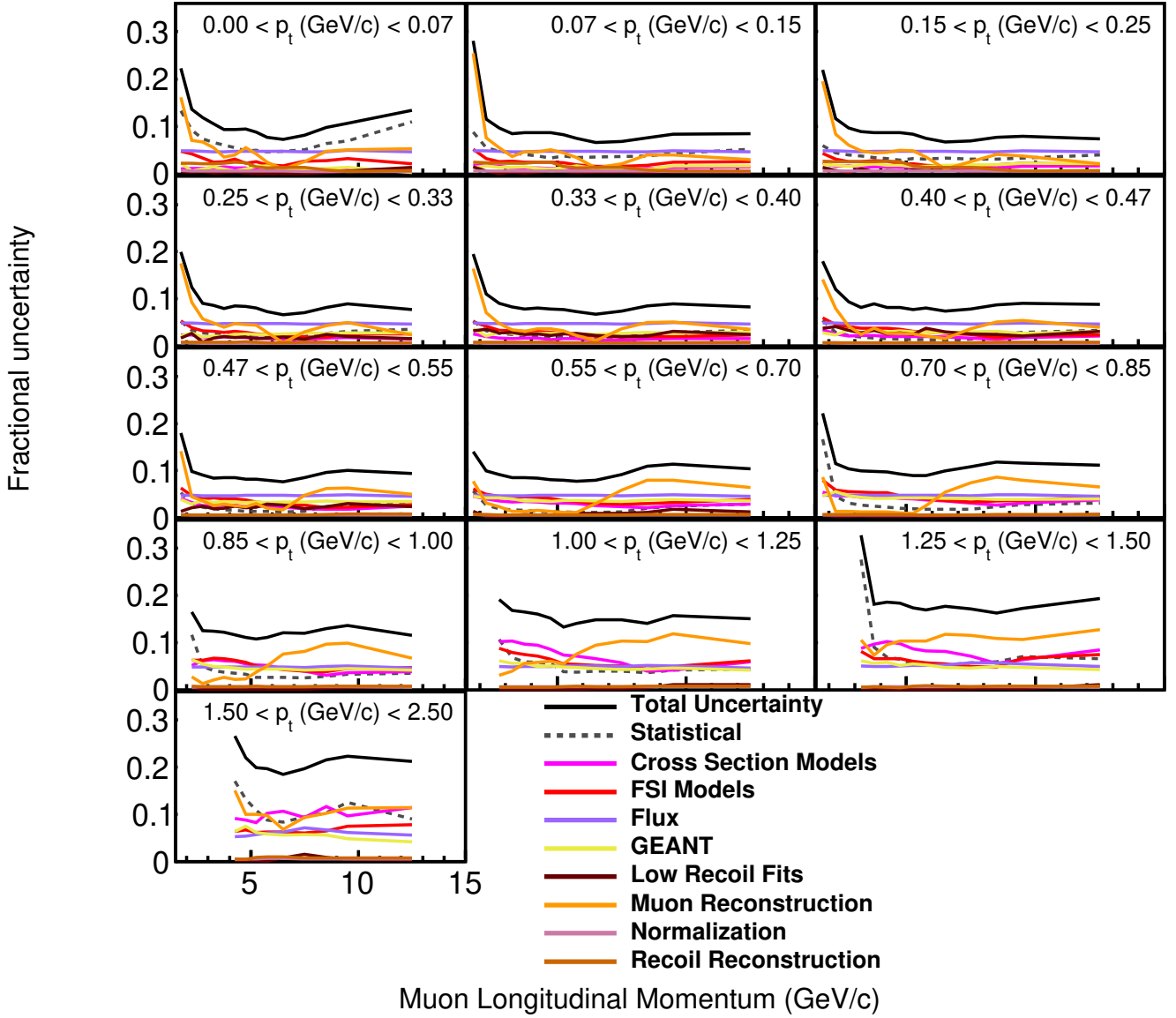}
    \caption{Summary of estimated uncertainties for the measured cross section shown in Fig. \ref{fig:pz-noprotonKE-xsection}.}
    \label{fig:pz-noprotonKE-xsection-errorSummary}
\end{figure*}

Figures \ref{fig:pt-noprotonKE-xsection}, \ref{fig:pt-noprotonKE-xsection-errorSummary}, \ref{fig:pz-noprotonKE-xsection} and \ref{fig:pz-noprotonKE-xsection-errorSummary} are the cross section measurement without the $T_p$ selection in the signal definition.

\begin{figure*}[ht]
    \centering
    \includegraphics[width=0.99\linewidth]{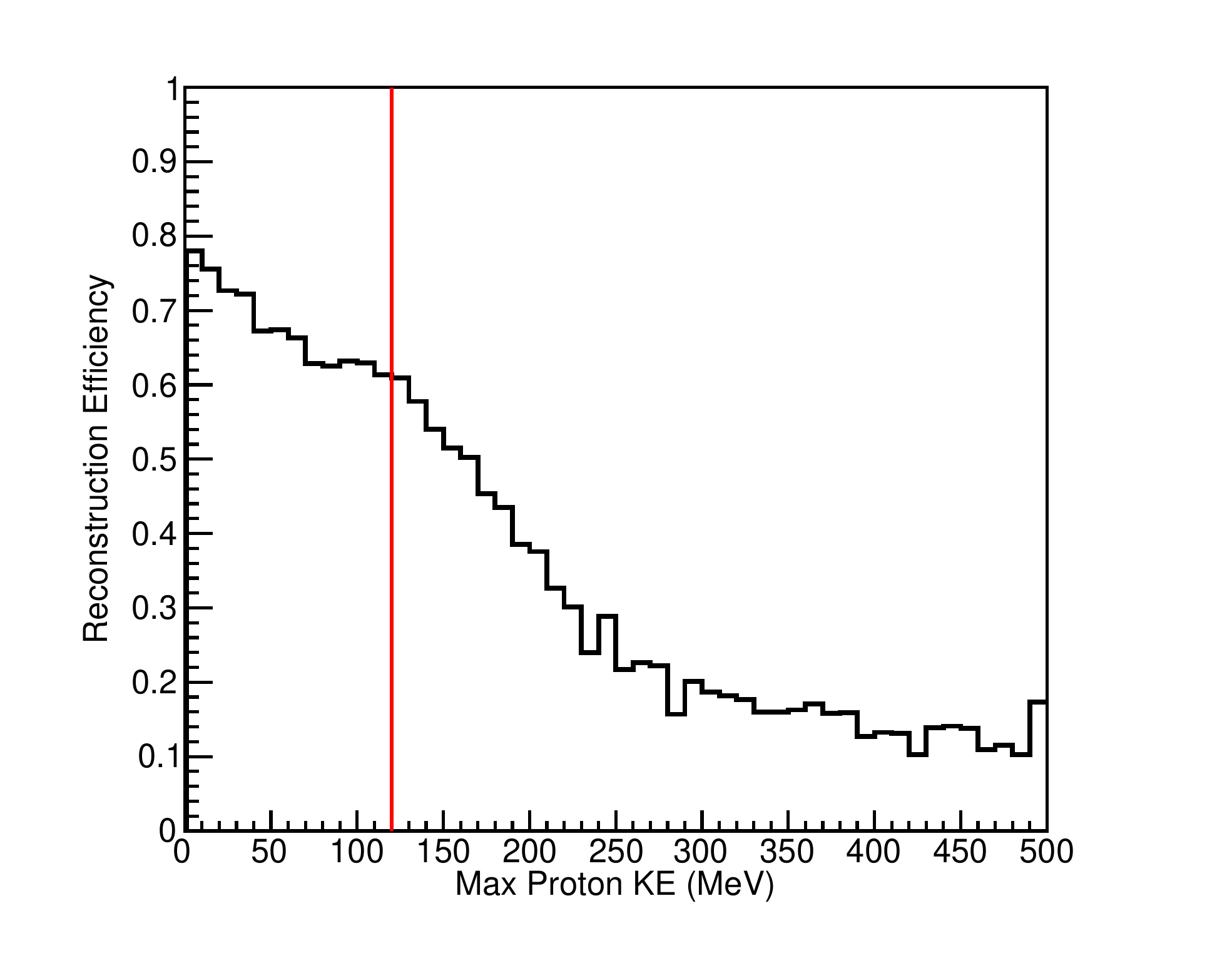}
    \caption{Efficiency of signal reconstruction as a function of maximum proton kinetic energy in a given neutrino event, where the signal definition has no cut on the maximum proton kinetic energy.  The drop in efficiency comes from the analysis cuts, and the signal definition of 120MeV maximum kinetic energy, shown on the plot, is used because the efficiency is large for events with protons below that energy.}
    \label{fig:reconstruction_efficiency}
\end{figure*}
\subsection{Data Tables} 
\begin{sidewaystable}
\caption{Double differential cross-section as a function of muon kinematics. The horizontal rows are muon transverse momentum ($p_{T}$ GeV/c) and vertical columns are muon longitudinal momentum ($p_{||}$ GeV/c). The cross-section is in units of $10^{-44}/ \nu/$cm$^{2}$$/nucleon$.}
\begin{tabular}{r||r| r| r| r| r| r| r| r| r| r| r| r| r| r}
$  p_{T} $(GeV/c)&$ 0.000-$&$ 0.075-$&$ 0.150-$&$ 0.250-$&$ 0.325-$&$ 0.400-$&$ 0.475-$&$ 0.550-$&$ 0.700-$&$ 0.850-$&$ 1.000-$&$ 1.250-$&$ 1.500-$\\
&$ 0.075$&$ 0.150$&$ 0.250$&$ 0.325$&$ 0.400$&$ 0.475$&$ 0.550$&$ 0.700$&$ 0.850$&$ 1.000$&$ 1.250$&$ 1.500$&$ 2.500$\\
$  p_{||} $(GeV/c)\quad&   &   &   &  &   &  &  &   &   &   &   &   &  \\
\hline
$1.5-2.0$ & 130.52 & 328.25 & 931.06 & 1205.31 & 1376.74 & 1470.16 & 1787.53 & 1981.04 & 60.40 & 0.00  & 0.0 0 & 0.00 & 0.00 \\
$2.0-2.5$ & 181.76 & 700.1 & 1746.15 & 2258.88 & 2769.3 & 2838.26 & 3085.13 & 5213.17 & 2665.23 & 190.63 & 0.0 & 0.0 & 0.0 \\
$2.5-3.0$ & 249.81 & 905.52 & 2501.51 & 3170.07 & 3803.29 & 3924.45 & 3978.03 & 7128.81 & 4336.56 & 2402.79 & 299.11 & 0.0 & 0.0 \\
$3.0-3.5$ & 300.01 & 1059.25 & 2956.31 & 3606.02 & 4347.17 & 4584.21 & 4586.75 & 8533.7 & 5386.23 & 3374.5 & 2184.13 & 14.84 & 0.0 \\
$3.5-4.0$ & 314.09 & 1091.56 & 3136.79 & 4012.71 & 4815.41 & 5161.33 & 5197.75 & 9263.07 & 5933.24 & 3853.42 & 3104.04 & 655.95 & 0.0 \\
$4.0-4.5$ & 338.78 & 1187.85 & 3473.6 & 4324.5 & 5409.12 & 5647.05 & 5782.02 & 10435.52 & 6590.36 & 4349.37 & 3407.67 & 1105.51 & 77.9 \\
$4.5-5.0$ & 360.82 & 1326.55 & 3852.56 & 4788.27 & 5881.98 & 6114.71 & 6151.66 & 11608.18 & 7344.08 & 4562.96 & 3535.42 & 1082.03 & 208.35 \\
$5.0-5.5$ & 431.44 & 1492.6 & 4286.34 & 5017.57 & 6342.21 & 6479.59 & 6431.56 & 11825.67 & 7995.34 & 4607.4 & 3296.76 & 925.33 & 209.2 \\
$5.5-6.0$ & 406.59 & 1417.4 & 4163.27 & 4792.03 & 5986.46 & 6085.06 & 6067.18 & 11054.51 & 7313.74 & 4191.56 & 2761.59 & 734.19 & 176.62 \\
$6.0-7.0$ & 620.77 & 2183.3 & 6538.27 & 7683.95 & 9159.91 & 9543.37 & 9323.31 & 16764.79 & 10724.18 & 5670.15 & 3792.87 & 967.23 & 237.31 \\
$7.0-8.0$ & 345.42 & 1252.58 & 3585.33 & 4203.27 & 5109.34 & 5427.12 & 5059.43 & 8780.23 & 5658.18 & 2990.91 & 1890.95 & 525.36 & 132.9 \\
$8.0-9.0$ & 152.4 & 574.53 & 1614.67 & 1873.48 & 2306.48 & 2514.01 & 2377.27 & 3549.41 & 2413.97 & 1487.38 & 1031.91 & 279.26 & 94.32 \\
$9.0-10.0$ & 67.65 & 252.53 & 681.71 & 764.64 & 984.67 & 1079.07 & 992.08 & 1540.63 & 1049.36 & 615.74 & 519.38 & 156.69 & 56.88 \\
$10.0-15.0$ & 69.01 & 315.18 & 833.15 & 918.73 & 1132.69 & 1279.41 & 1241.48 & 1857.1 & 1392.57 & 979.05 & 799.72 & 294.45 & 142.76 \\
\end{tabular}
\end{sidewaystable}
\begin{sidewaystable}
\caption{Statistical uncertainties on the double differential cross-section as a function of muon kinematics. The horizontal rows are muon transverse momentum ($p_{T}$ GeV/c) and vertical columns are muon longitudinal momentum ($p_{||}$ GeV/c). The cross-section is in units of $\mathrm{10^{-44}/ \nu/cm^{2}/nucleon}$.}
\begin{tabular}{r||r| r| r| r| r| r| r| r| r| r| r| r| r| r}
$  p_{T} $(GeV/c)&$ 0.000-$&$ 0.075-$&$ 0.150-$&$ 0.250-$&$ 0.325-$&$ 0.400-$&$ 0.475-$&$ 0.550-$&$ 0.700-$&$ 0.850-$&$ 1.000-$&$ 1.250-$&$ 1.500-$\\
&$ 0.075$&$ 0.150$&$ 0.250$&$ 0.325$&$ 0.400$&$ 0.475$&$ 0.550$&$ 0.700$&$ 0.850$&$ 1.000$&$ 1.250$&$ 1.500$&$ 2.500$\\
$  p_{||} $(GeV/c)\quad&   &   &   &  &   &  &  &   &   &   &   &   &  \\
\hline
$1.5-2.0$ & 17.16 & 28.21 & 53.87 & 61.32 & 71.51 & 78.05 & 93.6 & 111.24 & 10.46 & 0.0 & 0.0 & 0.0 & 0.0 \\
$2.0-2.5$ & 16.53 & 37.38 & 73.67 & 73.13 & 86.42 & 81.37 & 91.26 & 139.17 & 122.39 & 20.83 & 0.0 & 0.0 & 0.0 \\
$2.5-3.0$ & 17.95 & 41.62 & 92.88 & 84.54 & 95.91 & 83.18 & 86.7 & 138.84 & 139.01 & 123.43 & 32.76 & 0.0 & 0.0 \\
$3.0-3.5$ & 19.36 & 44.27 & 103.43 & 88.07 & 101.1 & 84.61 & 85.82 & 141.58 & 150.76 & 147.27 & 165.2 & 3.52 & 0.0 \\
$3.5-4.0$ & 18.52 & 43.16 & 102.66 & 88.15 & 100.55 & 86.61 & 88.57 & 141.4 & 152.98 & 156.87 & 207.22 & 62.12 & 0.0 \\
$4.0-4.5$ & 18.03 & 42.13 & 106.03 & 86.46 & 103.42 & 85.75 & 87.97 & 143.63 & 158.1 & 165.77 & 226.75 & 81.86 & 12.7 \\
$4.5-5.0$ & 16.94 & 42.75 & 105.65 & 89.61 & 102.49 & 83.39 & 84.83 & 145.96 & 161.79 & 165.48 & 223.45 & 78.16 & 28.22 \\
$5.0-5.5$ & 20.49 & 52.52 & 131.15 & 100.62 & 121.1 & 86.02 & 87.38 & 148.86 & 157.95 & 136.35 & 147.16 & 54.68 & 22.78 \\
$5.5-6.0$ & 18.09 & 46.45 & 121.76 & 92.15 & 107.52 & 77.61 & 78.85 & 131.55 & 136.34 & 120.88 & 121.91 & 41.61 & 15.9 \\
$6.0-7.0$ & 28.16 & 73.91 & 202.85 & 150.39 & 169.6 & 116.16 & 112.67 & 196.85 & 203.63 & 163.9 & 178.55 & 57.97 & 20.84 \\
$7.0-8.0$ & 17.12 & 42.91 & 105.96 & 80.31 & 92.74 & 75.15 & 68.13 & 113.46 & 108.06 & 83.23 & 85.38 & 32.89 & 12.81 \\
$8.0-9.0$ & 9.53 & 21.05 & 49.18 & 48.33 & 54.45 & 58.3 & 54.13 & 76.02 & 58.64 & 43.18 & 37.77 & 16.54 & 9.6 \\
$9.0-10.0$ & 4.61 & 10.67 & 22.24 & 22.85 & 26.9 & 29.81 & 27.31 & 39.26 & 30.23 & 20.24 & 21.27 & 10.47 & 6.78 \\
$10.0-15.0$ & 7.63 & 16.32 & 32.77 & 32.44 & 36.28 & 40.66 & 39.49 & 55.97 & 46.29 & 35.87 & 36.0 & 18.96 & 12.65 \\
\end{tabular}
\end{sidewaystable}
\begin{sidewaystable}
\caption{Systematic uncertainties on the double differential cross-section as a function of muon kinematics. The horizontal rows are muon transverse momentum ($p_{T}$ GeV/c) and vertical columns are muon longitudinal momentum ($p_{||}$ GeV/c). The cross-section is in units of $\mathrm{10^{-44}/ \nu/cm^{2}/nucleon}$.}
\begin{tabular}{r||r| r| r| r| r| r| r| r| r| r| r| r| r| r }
$  p_{T} $(GeV/c)&$ 0.000-$&$ 0.075-$&$ 0.150-$&$ 0.250-$&$ 0.325-$&$ 0.400-$&$ 0.475-$&$ 0.550-$&$ 0.700-$&$ 0.850-$&$ 1.000-$&$ 1.250-$&$ 1.500-$\\
&$ 0.075$&$ 0.150$&$ 0.250$&$ 0.325$&$ 0.400$&$ 0.475$&$ 0.550$&$ 0.700$&$ 0.850$&$ 1.000$&$ 1.250$&$ 1.500$&$ 2.500$\\
$  p_{||} $(GeV/c)\quad&   &   &   &  &   &  &  &   &   &   &   &   &  \\
\hline
$1.5-2.0$ & 22.66 & 87.05 & 192.45 & 226.27 & 241.72 & 228.04 & 283.42 & 206.78 & 8.29 & 0.0 & 0.0 & 0.0 & 0.0 \\
$2.0-2.5$ & 16.64 & 69.18 & 180.23 & 250.91 & 251.78 & 281.38 & 239.65 & 393.15 & 224.7 & 17.33 & 0.0 & 0.0 & 0.0 \\
$2.5-3.0$ & 22.4 & 71.38 & 213.48 & 255.59 & 293.42 & 300.74 & 294.13 & 513.73 & 348.16 & 221.37 & 33.44 & 0.0 & 0.0 \\
$3.0-3.5$ & 24.39 & 73.86 & 229.25 & 267.81 & 289.85 & 309.99 & 308.6 & 585.23 & 416.92 & 307.67 & 245.49 & 2.0 & 0.0 \\
$3.5-4.0$ & 22.08 & 82.18 & 239.64 & 280.89 & 322.47 & 374.52 & 374.28 & 608.73 & 453.3 & 341.69 & 365.7 & 86.85 & 0.0 \\
$4.0-4.5$ & 25.03 & 89.77 & 269.05 & 323.82 & 373.35 & 381.92 & 395.5 & 699.93 & 504.6 & 377.92 & 406.45 & 158.69 & 14.77 \\
$4.5-5.0$ & 29.96 & 105.45 & 310.43 & 355.91 & 404.43 & 413.05 & 419.01 & 771.24 & 562.0 & 394.84 & 422.6 & 158.21 & 32.34 \\
$5.0-5.5$ & 30.83 & 104.95 & 321.51 & 363.16 & 417.52 & 433.32 & 455.68 & 782.04 & 585.58 & 415.63 & 378.09 & 139.43 & 30.22 \\
$5.5-6.0$ & 24.53 & 90.22 & 269.15 & 304.35 & 370.93 & 388.2 & 403.66 & 729.1 & 564.05 & 424.3 & 348.01 & 108.18 & 27.57 \\
$6.0-7.0$ & 34.19 & 119.23 & 368.77 & 439.93 & 521.93 & 598.53 & 615.1 & 1195.49 & 983.04 & 652.1 & 528.92 & 151.85 & 32.52 \\
$7.0-8.0$ & 20.76 & 70.69 & 216.63 & 275.85 & 339.63 & 374.75 & 402.67 & 766.94 & 581.92 & 343.61 & 260.24 & 78.66 & 19.68 \\
$8.0-9.0$ & 11.32 & 37.7 & 110.46 & 138.84 & 177.17 & 193.76 & 211.05 & 375.78 & 268.33 & 178.94 & 128.86 & 38.47 & 13.65 \\
$9.0-10.0$ & 5.45 & 17.41 & 47.18 & 60.65 & 79.63 & 88.96 & 89.66 & 169.66 & 113.28 & 75.5 & 72.55 & 21.73 & 9.22 \\
$10.0-15.0$ & 5.11 & 19.97 & 50.0 & 58.43 & 79.6 & 93.97 & 104.75 & 174.5 & 136.59 & 94.74 & 99.19 & 44.98 & 23.14 \\
\end{tabular}
\end{sidewaystable}
\begin{sidewaystable}
\caption{Total uncertainties on the double differential cross-section as a function of muon kinematics. The horizontal rows are muon transverse momentum ($p_{T}$ GeV/c) and vertical columns are muon longitudinal momentum ($p_{||}$ GeV/c). The cross-section is in units of $\mathrm{10^{-44}/ \nu/cm^{2}/nucleon}$.}
\begin{tabular}{r||r| r| r| r| r| r| r| r| r| r| r| r| r| r}
$  p_{T} $(GeV/c)&$ 0.000-$&$ 0.075-$&$ 0.150-$&$ 0.250-$&$ 0.325-$&$ 0.400-$&$ 0.475-$&$ 0.550-$&$ 0.700-$&$ 0.850-$&$ 1.000-$&$ 1.250-$&$ 1.500-$\\
&$ 0.075$&$ 0.150$&$ 0.250$&$ 0.325$&$ 0.400$&$ 0.475$&$ 0.550$&$ 0.700$&$ 0.850$&$ 1.000$&$ 1.250$&$ 1.500$&$ 2.500$\\
$  p_{||} $(GeV/c)\quad&   &   &   &  &   &  &  &   &   &   &   &   &  \\
\hline
$1.5-2.0$ & 28.42 & 91.5 & 199.85 & 234.43 & 252.07 & 241.03 & 298.48 & 234.81 & 13.35 & 0.0 & 0.0 & 0.0 & 0.0 \\
$2.0-2.5$ & 23.45 & 78.63 & 194.7 & 261.35 & 266.2 & 292.91 & 256.44 & 417.06 & 255.87 & 27.1 & 0.0 & 0.0 & 0.0 \\
$2.5-3.0$ & 28.7 & 82.63 & 232.81 & 269.21 & 308.7 & 312.03 & 306.64 & 532.16 & 374.89 & 253.45 & 46.82 & 0.0 & 0.0 \\
$3.0-3.5$ & 31.14 & 86.11 & 251.51 & 281.92 & 306.98 & 321.33 & 320.31 & 602.11 & 443.34 & 341.1 & 295.9 & 4.05 & 0.0 \\
$3.5-4.0$ & 28.82 & 92.82 & 260.71 & 294.4 & 337.79 & 384.4 & 384.61 & 624.94 & 478.41 & 375.98 & 420.33 & 106.78 & 0.0 \\
$4.0-4.5$ & 30.85 & 99.17 & 289.19 & 335.16 & 387.41 & 391.43 & 405.17 & 714.51 & 528.78 & 412.67 & 465.42 & 178.56 & 19.48 \\
$4.5-5.0$ & 34.42 & 113.79 & 327.91 & 367.02 & 417.21 & 421.38 & 427.51 & 784.93 & 584.82 & 428.11 & 478.04 & 176.47 & 42.93 \\
$5.0-5.5$ & 37.02 & 117.36 & 347.23 & 376.84 & 434.72 & 441.77 & 463.98 & 796.09 & 606.51 & 437.42 & 405.72 & 149.77 & 37.85 \\
$5.5-6.0$ & 30.48 & 101.47 & 295.41 & 318.0 & 386.2 & 395.88 & 411.29 & 740.87 & 580.3 & 441.18 & 368.74 & 115.91 & 31.83 \\
$6.0-7.0$ & 44.29 & 140.28 & 420.88 & 464.93 & 548.79 & 609.7 & 625.34 & 1211.59 & 1003.91 & 672.38 & 558.24 & 162.54 & 38.63 \\
$7.0-8.0$ & 26.91 & 82.7 & 241.16 & 287.3 & 352.06 & 382.21 & 408.39 & 775.29 & 591.87 & 353.55 & 273.89 & 85.26 & 23.48 \\
$8.0-9.0$ & 14.8 & 43.17 & 120.91 & 147.01 & 185.34 & 202.34 & 217.88 & 383.39 & 274.66 & 184.07 & 134.28 & 41.87 & 16.69 \\
$9.0-10.0$ & 7.14 & 20.42 & 52.16 & 64.81 & 84.06 & 93.82 & 93.73 & 174.14 & 117.24 & 78.16 & 75.61 & 24.12 & 11.45 \\
$10.0-15.0$ & 9.18 & 25.79 & 59.79 & 66.83 & 87.48 & 102.39 & 111.95 & 183.26 & 144.22 & 101.31 & 105.52 & 48.82 & 26.37 \\
\end{tabular}
\end{sidewaystable}

\begin{sidewaystable}
\caption{Double differential cross-section as a function of muon kinematics. The horizontal rows are muon transverse momentum ($p_{T}$ GeV/c) and vertical columns are muon longitudinal momentum ($p_{||}$ GeV/c). The cross-section is in units of $10^{-42}/ \nu/$cm$^{2}$$/nucleon$. The signal definition does not require to reject protons with KE more than 120 MeV. However samples with tracks other than muon are not included.}
\begin{tabular}{r||r| r| r| r| r| r| r| r| r| r| r| r| r| r }
$  p_{T} $(GeV/c)&$ 0.000-$&$ 0.075-$&$ 0.150-$&$ 0.250-$&$ 0.325-$&$ 0.400-$&$ 0.475-$&$ 0.550-$&$ 0.700-$&$ 0.850-$&$ 1.000-$&$ 1.250-$&$ 1.500-$\\
&$ 0.075$&$ 0.150$&$ 0.250$&$ 0.325$&$ 0.400$&$ 0.475$&$ 0.550$&$ 0.700$&$ 0.850$&$ 1.000$&$ 1.250$&$ 1.500$&$ 2.500$\\
$  p_{||} $(GeV/c)\quad&   &   &   &  &   &  &  &   &   &   &   &   &  \\
\hline
$1.5-2.0$ & 166.29 & 418.54 & 1190.05 & 1488.48 & 1710.1 & 1910.63 & 2327.59 & 2589.58 & 86.76 & 0.0 & 0.0 & 0.0 & 0.0 \\
$2.0-2.5$ & 225.83 & 852.37 & 2084.53 & 2699.15 & 3300.76 & 3487.33 & 3820.64 & 6647.15 & 3554.64 & 260.1 & 0.0 & 0.0 & 0.0 \\
$2.5-3.0$ & 295.92 & 1060.32 & 2905.65 & 3682.87 & 4474.72 & 4699.33 & 4844.54 & 8872.24 & 5718.4 & 3183.76 & 405.21 & 0.0 & 0.0 \\
$3.0-3.5$ & 346.12 & 1235.41 & 3408.75 & 4152.72 & 5009.85 & 5482.83 & 5520.94 & 10546.03 & 7072.53 & 4532.03 & 3005.37 & 24.01 & 0.0 \\
$3.5-4.0$ & 363.64 & 1252.29 & 3638.86 & 4638.9 & 5567.09 & 6147.13 & 6278.17 & 11439.59 & 7753.18 & 5184.9 & 4295.67 & 931.69 & 0.0 \\
$4.0-4.5$ & 396.52 & 1394.63 & 4015.65 & 5014.12 & 6234.08 & 6672.38 & 6959.41 & 12818.57 & 8635.72 & 5779.74 & 4606.71 & 1572.92 & 114.22 \\
$4.5-5.0$ & 414.59 & 1522.28 & 4402.87 & 5472.67 & 6748.84 & 7243.69 & 7439.12 & 14124.04 & 9414.77 & 5940.73 & 4736.84 & 1494.86 & 289.4 \\
$5.0-5.5$ & 491.93 & 1708.46 & 4859.97 & 5660.21 & 7204.4 & 7539.68 & 7606.91 & 14229.72 & 9982.82 & 5986.6 & 4501.34 & 1302.38 & 302.26 \\
$5.5-6.0$ & 455.66 & 1584.0 & 4665.08 & 5357.43 & 6722.33 & 7056.0 & 7140.12 & 13202.25 & 9012.72 & 5343.32 & 3792.03 & 1048.89 & 251.08 \\
$6.0-7.0$ & 687.38 & 2418.22 & 7251.58 & 8451.94 & 10145.09 & 10962.72 & 10830.23 & 19651.93 & 13056.39 & 7151.21 & 5058.43 & 1377.36 & 343.84 \\
$7.0-8.0$ & 385.34 & 1376.87 & 3915.29 & 4596.07 & 5601.54 & 6149.41 & 5810.14 & 10247.66 & 6802.8 & 3731.86 & 2473.03 & 718.86 & 202.97 \\
$8.0-9.0$ & 165.21 & 628.13 & 1767.62 & 2049.71 & 2550.42 & 2764.52 & 2703.18 & 4192.76 & 2886.83 & 1810.16 & 1288.69 & 375.34 & 132.69 \\
$9.0-10.0$ & 73.16 & 286.16 & 752.11 & 841.72 & 1077.49 & 1182.92 & 1109.62 & 1827.45 & 1295.17 & 777.42 & 663.84 & 213.82 & 79.36 \\
$10.0-15.0$ & 77.55 & 348.23 & 918.38 & 1018.84 & 1249.22 & 1454.25 & 1429.35 & 2241.71 & 1772.94 & 1237.22 & 1058.46 & 415.25 & 208.16 \\
\end{tabular}
\end{sidewaystable}

\begin{sidewaystable}
\caption{Statistical uncertainty on double differential cross-section as a function of muon kinematics. The horizontal rows are muon transverse momentum ($p_{T}$ GeV/c) and vertical columns are muon longitudinal momentum ($p_{||}$ GeV/c). The cross-section is in units of $10^{-44}/ \nu/$cm$^{2}$$/nucleon$. The signal definition does not require to reject protons with KE more than 120 MeV. However samples with tracks other than muon are not included.}
\begin{tabular}{r||r| r| r| r| r| r| r| r| r| r| r| r| r| r }
$  p_{T} $(GeV/c)&$ 0.000-$&$ 0.075-$&$ 0.150-$&$ 0.250-$&$ 0.325-$&$ 0.400-$&$ 0.475-$&$ 0.550-$&$ 0.700-$&$ 0.850-$&$ 1.000-$&$ 1.250-$&$ 1.500-$\\
&$ 0.075$&$ 0.150$&$ 0.250$&$ 0.325$&$ 0.400$&$ 0.475$&$ 0.550$&$ 0.700$&$ 0.850$&$ 1.000$&$ 1.250$&$ 1.500$&$ 2.500$\\
$  p_{||} $(GeV/c)\quad&   &   &   &  &   &  &  &   &   &   &   &   &  \\
\hline
$1.5-2.0$ & 22.18 & 36.74 & 71.18 & 77.29 & 91.12 & 102.16 & 123.83 & 146.69 & 14.46 & 0.0 & 0.0 & 0.0 & 0.0 \\
$2.0-2.5$ & 20.86 & 47.41 & 92.99 & 89.5 & 105.37 & 101.18 & 114.63 & 178.15 & 160.1 & 30.12 & 0.0 & 0.0 & 0.0 \\
$2.5-3.0$ & 21.84 & 50.86 & 115.37 & 100.69 & 115.81 & 100.95 & 107.18 & 173.11 & 175.22 & 154.67 & 43.3 & 0.0 & 0.0 \\
$3.0-3.5$ & 22.97 & 54.36 & 127.91 & 104.2 & 119.19 & 102.74 & 104.27 & 174.99 & 186.7 & 182.61 & 203.57 & 6.61 & 0.0 \\
$3.5-4.0$ & 21.98 & 52.25 & 127.86 & 104.74 & 119.19 & 104.12 & 108.57 & 174.62 & 187.82 & 192.47 & 254.22 & 84.15 & 0.0 \\
$4.0-4.5$ & 21.8 & 52.42 & 131.62 & 103.19 & 122.18 & 102.11 & 106.89 & 176.35 & 193.4 & 199.6 & 268.48 & 108.75 & 19.51 \\
$4.5-5.0$ & 20.1 & 51.84 & 129.88 & 105.37 & 121.0 & 100.36 & 103.68 & 177.54 & 193.88 & 194.32 & 261.49 & 96.39 & 38.28 \\
$5.0-5.5$ & 24.3 & 63.97 & 160.19 & 116.98 & 141.96 & 101.1 & 103.47 & 177.46 & 185.01 & 158.68 & 170.55 & 69.87 & 32.82 \\
$5.5-6.0$ & 21.23 & 55.27 & 147.09 & 106.47 & 124.93 & 91.56 & 93.65 & 156.17 & 158.1 & 136.73 & 140.31 & 54.26 & 21.99 \\
$6.0-7.0$ & 32.51 & 87.29 & 243.12 & 170.43 & 194.07 & 133.65 & 130.66 & 227.29 & 230.42 & 182.41 & 197.18 & 72.49 & 28.81 \\
$7.0-8.0$ & 19.76 & 50.07 & 124.52 & 90.61 & 105.16 & 86.77 & 78.07 & 130.68 & 122.72 & 94.2 & 95.89 & 40.75 & 19.13 \\
$8.0-9.0$ & 10.6 & 23.7 & 55.37 & 54.03 & 61.97 & 65.43 & 62.51 & 85.22 & 66.77 & 50.39 & 45.7 & 22.12 & 13.74 \\
$9.0-10.0$ & 5.1 & 12.51 & 25.05 & 25.72 & 30.37 & 34.13 & 31.08 & 45.23 & 36.07 & 25.36 & 27.16 & 14.94 & 9.94 \\
$10.0-15.0$ & 8.53 & 18.29 & 36.53 & 36.12 & 40.57 & 47.35 & 45.96 & 64.9 & 56.1 & 43.62 & 46.05 & 27.14 & 18.84 \\
\end{tabular}
\end{sidewaystable}

\begin{sidewaystable}
\caption{Total estimated uncertainty on double differential cross-section as a function of muon kinematics. The horizontal rows are muon transverse momentum ($p_{T}$ GeV/c) and vertical columns are muon longitudinal momentum ($p_{||}$ GeV/c). The cross-section is in units of $10^{-44}/ \nu/$cm$^{2}$$/nucleon$. The signal definition does not require to reject protons with KE more than 120 MeV. However samples with tracks other than muon are not included.}
\begin{tabular}{r||r| r| r| r| r| r| r| r| r| r| r| r| r| r }
$  p_{T} $(GeV/c)&$ 0.000-$&$ 0.075-$&$ 0.150-$&$ 0.250-$&$ 0.325-$&$ 0.400-$&$ 0.475-$&$ 0.550-$&$ 0.700-$&$ 0.850-$&$ 1.000-$&$ 1.250-$&$ 1.500-$\\
&$ 0.075$&$ 0.150$&$ 0.250$&$ 0.325$&$ 0.400$&$ 0.475$&$ 0.550$&$ 0.700$&$ 0.850$&$ 1.000$&$ 1.250$&$ 1.500$&$ 2.500$\\
$  p_{||} $(GeV/c)\quad&   &   &   &  &   &  &  &   &   &   &   &   &  \\
\hline
$1.5-2.0$ & 37.08 & 117.59 & 260.74 & 297.09 & 332.5 & 342.22 & 418.69 & 362.06 & 19.22 & 0.0 & 0.0 & 0.0 & 0.0 \\
$2.0-2.5$ & 30.68 & 98.93 & 244.32 & 335.17 & 363.89 & 418.88 & 377.45 & 662.36 & 408.95 & 42.81 & 0.0 & 0.0 & 0.0 \\
$2.5-3.0$ & 35.27 & 102.35 & 282.6 & 330.83 & 405.31 & 440.3 & 438.91 & 809.55 & 594.26 & 396.75 & 77.53 & 0.0 & 0.0 \\
$3.0-3.5$ & 36.58 & 104.49 & 304.63 & 354.92 & 411.75 & 447.65 & 465.22 & 899.35 & 703.33 & 560.94 & 506.08 & 7.87 & 0.0 \\
$3.5-4.0$ & 33.93 & 109.78 & 318.1 & 370.15 & 434.7 & 550.11 & 535.1 & 967.6 & 760.52 & 629.59 & 705.46 & 168.66 & 0.0 \\
$4.0-4.5$ & 37.0 & 120.87 & 347.95 & 422.49 & 504.25 & 541.63 & 591.52 & 1076.59 & 839.16 & 673.18 & 737.06 & 291.61 & 30.44 \\
$4.5-5.0$ & 39.5 & 133.04 & 381.26 & 455.2 & 529.49 & 589.65 & 605.79 & 1147.14 & 869.04 & 658.09 & 719.09 & 274.78 & 63.63 \\
$5.0-5.5$ & 43.5 & 141.16 & 409.49 & 455.41 & 558.94 & 583.8 & 622.64 & 1132.92 & 892.75 & 644.67 & 599.38 & 226.57 & 60.16 \\
$5.5-6.0$ & 35.19 & 119.79 & 351.3 & 388.8 & 488.41 & 566.91 & 569.72 & 1028.47 & 810.06 & 592.6 & 532.6 & 177.17 & 49.44 \\
$6.0-7.0$ & 49.9 & 161.04 & 491.15 & 560.29 & 680.44 & 805.0 & 826.97 & 1562.91 & 1296.85 & 860.42 & 749.42 & 243.08 & 63.57 \\
$7.0-8.0$ & 31.26 & 93.94 & 272.24 & 325.23 & 415.42 & 480.81 & 495.44 & 944.93 & 738.69 & 445.12 & 367.47 & 122.97 & 40.05 \\
$8.0-9.0$ & 16.29 & 48.21 & 136.81 & 167.21 & 215.59 & 240.2 & 259.19 & 458.12 & 340.23 & 234.64 & 180.86 & 60.96 & 28.65 \\
$9.0-10.0$ & 7.81 & 24.09 & 59.77 & 74.7 & 96.53 & 106.69 & 111.96 & 207.92 & 150.25 & 106.2 & 104.26 & 36.88 & 17.7 \\
$10.0-15.0$ & 10.38 & 29.4 & 68.17 & 77.94 & 102.76 & 127.62 & 134.24 & 233.51 & 198.11 & 142.19 & 159.33 & 80.35 & 44.23 \\
\end{tabular}
\end{sidewaystable}

\begin{sidewaystable}
\caption{Estimated Systematic uncertainty for the double differential cross-section as a function of muon kinematics. The horizontal rows are muon transverse momentum ($p_{T}$ GeV/c) and vertical columns are muon longitudinal momentum ($p_{||}$ GeV/c). The cross-section is in units of $10^{-44}/ \nu/$cm$^{2}$$/nucleon$. The signal definition does not require to reject protons with KE more than 120 MeV. However samples with tracks other than muon are not included.}
\begin{tabular}{r||r| r| r| r| r| r| r| r| r| r| r| r| r| r }
$  p_{T} $(GeV/c)&$ 0.000-$&$ 0.075-$&$ 0.150-$&$ 0.250-$&$ 0.325-$&$ 0.400-$&$ 0.475-$&$ 0.550-$&$ 0.700-$&$ 0.850-$&$ 1.000-$&$ 1.250-$&$ 1.500-$\\
&$ 0.075$&$ 0.150$&$ 0.250$&$ 0.325$&$ 0.400$&$ 0.475$&$ 0.550$&$ 0.700$&$ 0.850$&$ 1.000$&$ 1.250$&$ 1.500$&$ 2.500$\\
$  p_{||} $(GeV/c)\quad&   &   &   &  &   &  &  &   &   &   &   &   &  \\
\hline
$1.5-2.0$ & 29.72 & 111.7 & 250.84 & 286.86 & 319.77 & 326.62 & 399.96 & 331.01 & 12.67 & 0.0 & 0.0 & 0.0 & 0.0 \\
$2.0-2.5$ & 22.5 & 86.83 & 225.93 & 323.0 & 348.3 & 406.47 & 359.62 & 637.96 & 376.3 & 30.42 & 0.0 & 0.0 & 0.0 \\
$2.5-3.0$ & 27.69 & 88.81 & 257.97 & 315.14 & 388.41 & 428.57 & 425.62 & 790.83 & 567.84 & 365.36 & 64.32 & 0.0 & 0.0 \\
$3.0-3.5$ & 28.47 & 89.24 & 276.48 & 339.28 & 394.12 & 435.7 & 453.39 & 882.16 & 678.09 & 530.38 & 463.33 & 4.27 & 0.0 \\
$3.5-4.0$ & 25.85 & 96.54 & 291.28 & 355.02 & 418.05 & 540.17 & 523.98 & 951.72 & 736.96 & 599.45 & 658.07 & 146.17 & 0.0 \\
$4.0-4.5$ & 29.9 & 108.91 & 322.1 & 409.7 & 489.22 & 531.92 & 581.78 & 1062.04 & 816.57 & 642.91 & 686.42 & 270.58 & 23.36 \\
$4.5-5.0$ & 34.0 & 122.53 & 358.46 & 442.83 & 515.48 & 581.05 & 596.85 & 1133.31 & 847.13 & 628.74 & 669.86 & 257.32 & 50.83 \\
$5.0-5.5$ & 36.08 & 125.83 & 376.85 & 440.13 & 540.61 & 574.98 & 613.98 & 1118.93 & 873.37 & 624.84 & 574.6 & 215.52 & 50.42 \\
$5.5-6.0$ & 28.06 & 106.28 & 319.03 & 373.94 & 472.16 & 559.47 & 561.97 & 1016.55 & 794.48 & 576.61 & 513.78 & 168.66 & 44.28 \\
$6.0-7.0$ & 37.86 & 135.33 & 426.76 & 533.74 & 652.18 & 793.82 & 816.59 & 1546.29 & 1276.21 & 840.86 & 723.02 & 232.02 & 56.66 \\
$7.0-8.0$ & 24.23 & 79.48 & 242.09 & 312.35 & 401.89 & 472.91 & 489.25 & 935.85 & 728.42 & 435.04 & 354.73 & 116.03 & 35.19 \\
$8.0-9.0$ & 12.37 & 41.98 & 125.1 & 158.24 & 206.5 & 231.12 & 251.54 & 450.12 & 333.62 & 229.17 & 174.99 & 56.8 & 25.14 \\
$9.0-10.0$ & 5.91 & 20.58 & 54.27 & 70.13 & 91.62 & 101.08 & 107.56 & 202.94 & 145.86 & 103.12 & 100.66 & 33.71 & 14.64 \\
$10.0-15.0$ & 5.91 & 23.03 & 57.55 & 69.07 & 94.41 & 118.51 & 126.13 & 224.31 & 190.0 & 135.34 & 152.53 & 75.63 & 40.02 \\
\end{tabular}
\end{sidewaystable}

%% file: main.bbl
\begin{thebibliography}{49}%
\makeatletter
\providecommand \@ifxundefined [1]{%
 \@ifx{#1\undefined}
}%
\providecommand \@ifnum [1]{%
 \ifnum #1\expandafter \@firstoftwo
 \else \expandafter \@secondoftwo
 \fi
}%
\providecommand \@ifx [1]{%
 \ifx #1\expandafter \@firstoftwo
 \else \expandafter \@secondoftwo
 \fi
}%
\providecommand \natexlab [1]{#1}%
\providecommand \enquote  [1]{``#1''}%
\providecommand \bibnamefont  [1]{#1}%
\providecommand \bibfnamefont [1]{#1}%
\providecommand \citenamefont [1]{#1}%
\providecommand \href@noop [0]{\@secondoftwo}%
\providecommand \href [0]{\begingroup \@sanitize@url \@href}%
\providecommand \@href[1]{\@@startlink{#1}\@@href}%
\providecommand \@@href[1]{\endgroup#1\@@endlink}%
\providecommand \@sanitize@url [0]{\catcode `\\12\catcode `\$12\catcode
  `\&12\catcode `\#12\catcode `\^12\catcode `\_12\catcode `\%12\relax}%
\providecommand \@@startlink[1]{}%
\providecommand \@@endlink[0]{}%
\providecommand \url  [0]{\begingroup\@sanitize@url \@url }%
\providecommand \@url [1]{\endgroup\@href {#1}{\urlprefix }}%
\providecommand \urlprefix  [0]{URL }%
\providecommand \Eprint [0]{\href }%
\providecommand \doibase [0]{https://doi.org/}%
\providecommand \selectlanguage [0]{\@gobble}%
\providecommand \bibinfo  [0]{\@secondoftwo}%
\providecommand \bibfield  [0]{\@secondoftwo}%
\providecommand \translation [1]{[#1]}%
\providecommand \BibitemOpen [0]{}%
\providecommand \bibitemStop [0]{}%
\providecommand \bibitemNoStop [0]{.\EOS\space}%
\providecommand \EOS [0]{\spacefactor3000\relax}%
\providecommand \BibitemShut  [1]{\csname bibitem#1\endcsname}%
\let\auto@bib@innerbib\@empty
\bibitem [{\citenamefont {Acero}\ \emph {et~al.}(2022)\citenamefont {Acero}
  \emph {et~al.}}]{NOvA:2021nfi}%
  \BibitemOpen
  \bibfield  {author} {\bibinfo {author} {\bibfnamefont {M.~A.}\ \bibnamefont
  {Acero}} \emph {et~al.} (\bibinfo {collaboration} {NOvA}),\ }\href
  {https://doi.org/10.1103/PhysRevD.106.032004} {\bibfield  {journal} {\bibinfo
   {journal} {Phys. Rev. D}\ }\textbf {\bibinfo {volume} {106}},\ \bibinfo
  {pages} {032004} (\bibinfo {year} {2022})},\ \Eprint
  {https://arxiv.org/abs/2108.08219} {arXiv:2108.08219 [hep-ex]} \BibitemShut
  {NoStop}%
\bibitem [{\citenamefont {Abe}\ \emph {et~al.}(2020{\natexlab{a}})\citenamefont
  {Abe} \emph {et~al.}}]{Abe:2019vii}%
  \BibitemOpen
  \bibfield  {author} {\bibinfo {author} {\bibfnamefont {K.}~\bibnamefont
  {Abe}} \emph {et~al.} (\bibinfo {collaboration} {T2K}),\ }\href
  {https://doi.org/10.1038/s41586-020-2177-0} {\bibfield  {journal} {\bibinfo
  {journal} {Nature}\ }\textbf {\bibinfo {volume} {580}},\ \bibinfo {pages}
  {339} (\bibinfo {year} {2020}{\natexlab{a}})},\ \bibinfo {note} {[Erratum:
  Nature 583, E16 (2020)]},\ \Eprint {https://arxiv.org/abs/1910.03887}
  {arXiv:1910.03887 [hep-ex]} \BibitemShut {NoStop}%
\bibitem [{\citenamefont {Workman}\ \emph {et~al.}(2022)\citenamefont {Workman}
  \emph {et~al.}}]{Workman:2022ynf}%
  \BibitemOpen
  \bibfield  {author} {\bibinfo {author} {\bibfnamefont {R.~L.}\ \bibnamefont
  {Workman}} \emph {et~al.} (\bibinfo {collaboration} {Particle Data Group}),\
  }\href {https://doi.org/10.1093/ptep/ptac097} {\bibfield  {journal} {\bibinfo
   {journal} {PTEP}\ }\textbf {\bibinfo {volume} {2022}},\ \bibinfo {pages}
  {083C01} (\bibinfo {year} {2022})}\BibitemShut {NoStop}%
\bibitem [{\citenamefont {Abi}\ \emph {et~al.}(2020)\citenamefont {Abi} \emph
  {et~al.}}]{DUNE:2020jqi}%
  \BibitemOpen
  \bibfield  {author} {\bibinfo {author} {\bibfnamefont {B.}~\bibnamefont
  {Abi}} \emph {et~al.} (\bibinfo {collaboration} {DUNE}),\ }\href
  {https://doi.org/10.1140/epjc/s10052-020-08456-z} {\bibfield  {journal}
  {\bibinfo  {journal} {Eur. Phys. J. C}\ }\textbf {\bibinfo {volume} {80}},\
  \bibinfo {pages} {978} (\bibinfo {year} {2020})},\ \Eprint
  {https://arxiv.org/abs/2006.16043} {arXiv:2006.16043 [hep-ex]} \BibitemShut
  {NoStop}%
\bibitem [{\citenamefont {Yokoyama}(2017)}]{Yokoyama:2017mnt}%
  \BibitemOpen
  \bibfield  {author} {\bibinfo {author} {\bibfnamefont {M.}~\bibnamefont
  {Yokoyama}} (\bibinfo {collaboration} {Hyper-Kamiokande Proto}),\ }in\
  \href@noop {} {\emph {\bibinfo {booktitle} {{Prospects in Neutrino
  Physics}}}}\ (\bibinfo {year} {2017})\ \Eprint
  {https://arxiv.org/abs/1705.00306} {arXiv:1705.00306 [hep-ex]} \BibitemShut
  {NoStop}%
\bibitem [{\citenamefont {Formaggio}\ and\ \citenamefont
  {Zeller}(2012)}]{RevModPhys.84.1307}%
  \BibitemOpen
  \bibfield  {author} {\bibinfo {author} {\bibfnamefont {J.~A.}\ \bibnamefont
  {Formaggio}}\ and\ \bibinfo {author} {\bibfnamefont {G.~P.}\ \bibnamefont
  {Zeller}},\ }\href {https://doi.org/10.1103/RevModPhys.84.1307} {\bibfield
  {journal} {\bibinfo  {journal} {Rev. Mod. Phys.}\ }\textbf {\bibinfo {volume}
  {84}},\ \bibinfo {pages} {1307} (\bibinfo {year} {2012})}\BibitemShut
  {NoStop}%
\bibitem [{\citenamefont {Abe}\ \emph {et~al.}(2020{\natexlab{b}})\citenamefont
  {Abe} \emph {et~al.}}]{T2K:2020jav}%
  \BibitemOpen
  \bibfield  {author} {\bibinfo {author} {\bibfnamefont {K.}~\bibnamefont
  {Abe}} \emph {et~al.} (\bibinfo {collaboration} {T2K}),\ }\href
  {https://doi.org/10.1103/PhysRevD.101.112004} {\bibfield  {journal} {\bibinfo
   {journal} {Phys. Rev. D}\ }\textbf {\bibinfo {volume} {101}},\ \bibinfo
  {pages} {112004} (\bibinfo {year} {2020}{\natexlab{b}})},\ \Eprint
  {https://arxiv.org/abs/2004.05434} {arXiv:2004.05434 [hep-ex]} \BibitemShut
  {NoStop}%
\bibitem [{\citenamefont {Aguilar-Arevalo}\ \emph {et~al.}(2013)\citenamefont
  {Aguilar-Arevalo} \emph {et~al.}}]{MiniBooNE:2013qnd}%
  \BibitemOpen
  \bibfield  {author} {\bibinfo {author} {\bibfnamefont {A.~A.}\ \bibnamefont
  {Aguilar-Arevalo}} \emph {et~al.} (\bibinfo {collaboration} {MiniBooNE}),\
  }\href {https://doi.org/10.1103/PhysRevD.88.032001} {\bibfield  {journal}
  {\bibinfo  {journal} {Phys. Rev. D}\ }\textbf {\bibinfo {volume} {88}},\
  \bibinfo {pages} {032001} (\bibinfo {year} {2013})},\ \Eprint
  {https://arxiv.org/abs/1301.7067} {arXiv:1301.7067 [hep-ex]} \BibitemShut
  {NoStop}%
\bibitem [{\citenamefont {Papadopoulou}(2022)}]{Papadopoulou:2022aoo}%
  \BibitemOpen
  \bibfield  {author} {\bibinfo {author} {\bibfnamefont {A.}~\bibnamefont
  {Papadopoulou}},\ }\href {https://doi.org/10.22323/1.402.0189} {\bibfield
  {journal} {\bibinfo  {journal} {PoS}\ }\textbf {\bibinfo {volume}
  {NuFact2021}},\ \bibinfo {pages} {189} (\bibinfo {year} {2022})}\BibitemShut
  {NoStop}%
\bibitem [{\citenamefont {Patrick}\ \emph {et~al.}(2018)\citenamefont {Patrick}
  \emph {et~al.}}]{Patrick:2018gvi}%
  \BibitemOpen
  \bibfield  {author} {\bibinfo {author} {\bibfnamefont {C.}~\bibnamefont
  {Patrick}} \emph {et~al.} (\bibinfo {collaboration} {MINERvA}),\ }\href
  {https://doi.org/10.1103/PhysRevD.97.052002} {\bibfield  {journal} {\bibinfo
  {journal} {Phys. Rev. D}\ }\textbf {\bibinfo {volume} {97}},\ \bibinfo
  {pages} {052002} (\bibinfo {year} {2018})},\ \Eprint
  {https://arxiv.org/abs/1801.01197} {arXiv:1801.01197 [hep-ex]} \BibitemShut
  {NoStop}%
\bibitem [{\citenamefont {Carneiro}\ \emph {et~al.}(2020)\citenamefont
  {Carneiro} \emph {et~al.}}]{Carneiro:2019jds}%
  \BibitemOpen
  \bibfield  {author} {\bibinfo {author} {\bibfnamefont {M.}~\bibnamefont
  {Carneiro}} \emph {et~al.} (\bibinfo {collaboration} {MINERvA}),\ }\href
  {https://doi.org/10.1103/PhysRevLett.124.121801} {\bibfield  {journal}
  {\bibinfo  {journal} {Phys. Rev. Lett.}\ }\textbf {\bibinfo {volume} {124}},\
  \bibinfo {pages} {121801} (\bibinfo {year} {2020})},\ \Eprint
  {https://arxiv.org/abs/1912.09890} {arXiv:1912.09890 [hep-ex]} \BibitemShut
  {NoStop}%
\bibitem [{\citenamefont {Adamson}\ \emph {et~al.}(2016)\citenamefont {Adamson}
  \emph {et~al.}}]{Adamson:2015dkw}%
  \BibitemOpen
  \bibfield  {author} {\bibinfo {author} {\bibfnamefont {P.}~\bibnamefont
  {Adamson}} \emph {et~al.},\ }\href
  {https://doi.org/10.1016/j.nima.2015.08.063} {\bibfield  {journal} {\bibinfo
  {journal} {Nucl. Instrum. Meth. A}\ }\textbf {\bibinfo {volume} {806}},\
  \bibinfo {pages} {279} (\bibinfo {year} {2016})},\ \Eprint
  {https://arxiv.org/abs/1507.06690} {arXiv:1507.06690} \BibitemShut {NoStop}%
\bibitem [{\citenamefont {Wojcicki}(1999)}]{WOJCICKI1999182}%
  \BibitemOpen
  \bibfield  {author} {\bibinfo {author} {\bibfnamefont {S.~G.}\ \bibnamefont
  {Wojcicki}},\ }\href
  {https://doi.org/https://doi.org/10.1016/S0920-5632(99)00416-8} {\bibfield
  {journal} {\bibinfo  {journal} {Nuclear Physics B - Proceedings Supplements}\
  }\textbf {\bibinfo {volume} {77}},\ \bibinfo {pages} {182} (\bibinfo {year}
  {1999})}\BibitemShut {NoStop}%
\bibitem [{\citenamefont {Subedi}\ \emph {et~al.}(2008)\citenamefont {Subedi}
  \emph {et~al.}}]{Subedi:2008zz}%
  \BibitemOpen
  \bibfield  {author} {\bibinfo {author} {\bibfnamefont {R.}~\bibnamefont
  {Subedi}} \emph {et~al.},\ }\href {https://doi.org/10.1126/science.1156675}
  {\bibfield  {journal} {\bibinfo  {journal} {Science}\ }\textbf {\bibinfo
  {volume} {320}},\ \bibinfo {pages} {1476} (\bibinfo {year} {2008})},\ \Eprint
  {https://arxiv.org/abs/0908.1514} {arXiv:0908.1514 [nucl-ex]} \BibitemShut
  {NoStop}%
\bibitem [{\citenamefont {Andreopoulos}\ \emph {et~al.}(2015)\citenamefont
  {Andreopoulos} \emph {et~al.}}]{Andreopoulos:2015wxa}%
  \BibitemOpen
  \bibfield  {author} {\bibinfo {author} {\bibfnamefont {C.}~\bibnamefont
  {Andreopoulos}} \emph {et~al.},\ }\href@noop {} {\bibinfo {title} {{The GENIE
  Neutrino Monte Carlo Generator: Physics and User Manual}}} (\bibinfo {year}
  {2015}),\ \Eprint {https://arxiv.org/abs/1510.05494} {arXiv:1510.05494
  [hep-ph]} \BibitemShut {NoStop}%
\bibitem [{\citenamefont {and E.J.~Moniz}(1972)}]{SMITH1972605}%
  \BibitemOpen
  \bibfield  {author} {\bibinfo {author} {\bibfnamefont {R.~S.}\ \bibnamefont
  {and E.J.~Moniz}},\ }\href {https://doi.org/10.1016/0550-3213(72)90040-5}
  {\bibfield  {journal} {\bibinfo  {journal} {Nucl. Phys.}\ }\textbf {\bibinfo
  {volume} {B43}},\ \bibinfo {pages} {605 } (\bibinfo {year}
  {1972})}\BibitemShut {NoStop}%
\bibitem [{\citenamefont {Bodek}\ and\ \citenamefont
  {Ritchie}(1981)}]{Bodek:1980ar}%
  \BibitemOpen
  \bibfield  {author} {\bibinfo {author} {\bibfnamefont {A.}~\bibnamefont
  {Bodek}}\ and\ \bibinfo {author} {\bibfnamefont {J.}~\bibnamefont
  {Ritchie}},\ }\href {https://doi.org/10.1103/PhysRevD.23.1070} {\bibfield
  {journal} {\bibinfo  {journal} {Phys. Rev. D}\ }\textbf {\bibinfo {volume}
  {23}},\ \bibinfo {pages} {1070} (\bibinfo {year} {1981})}\BibitemShut
  {NoStop}%
\bibitem [{\citenamefont {Nieves}\ \emph {et~al.}(2004)\citenamefont {Nieves},
  \citenamefont {Amaro},\ and\ \citenamefont {Valverde}}]{Nieves:2004wx}%
  \BibitemOpen
  \bibfield  {author} {\bibinfo {author} {\bibfnamefont {J.}~\bibnamefont
  {Nieves}}, \bibinfo {author} {\bibfnamefont {J.~E.}\ \bibnamefont {Amaro}},\
  and\ \bibinfo {author} {\bibfnamefont {M.}~\bibnamefont {Valverde}},\ }\href
  {https://doi.org/10.1103/PhysRevC.70.055503} {\bibfield  {journal} {\bibinfo
  {journal} {Phys. Rev. C}\ }\textbf {\bibinfo {volume} {70}},\ \bibinfo
  {pages} {055503} (\bibinfo {year} {2004})},\ \bibinfo {note} {[Erratum: Phys.
  Rev.C 72,019902(2005)]},\ \Eprint {https://arxiv.org/abs/nucl-th/0408005}
  {arXiv:nucl-th/0408005} \BibitemShut {NoStop}%
\bibitem [{\citenamefont {Gran}\ \emph {et~al.}(2013)\citenamefont {Gran},
  \citenamefont {Nieves}, \citenamefont {Sanchez},\ and\ \citenamefont
  {Vicente~Vacas}}]{Gran:2013kda}%
  \BibitemOpen
  \bibfield  {author} {\bibinfo {author} {\bibfnamefont {R.}~\bibnamefont
  {Gran}}, \bibinfo {author} {\bibfnamefont {J.}~\bibnamefont {Nieves}},
  \bibinfo {author} {\bibfnamefont {F.}~\bibnamefont {Sanchez}},\ and\ \bibinfo
  {author} {\bibfnamefont {M.}~\bibnamefont {Vicente~Vacas}},\ }\href
  {https://doi.org/10.1103/PhysRevD.88.113007} {\bibfield  {journal} {\bibinfo
  {journal} {Phys. Rev. D}\ }\textbf {\bibinfo {volume} {88}},\ \bibinfo
  {pages} {113007} (\bibinfo {year} {2013})},\ \Eprint
  {https://arxiv.org/abs/1307.8105} {arXiv:1307.8105 [hep-ph]} \BibitemShut
  {NoStop}%
\bibitem [{\citenamefont {Schwehr}\ \emph {et~al.}(2016)\citenamefont
  {Schwehr}, \citenamefont {Cherdack},\ and\ \citenamefont
  {Gran}}]{Schwehr:2016pvn}%
  \BibitemOpen
  \bibfield  {author} {\bibinfo {author} {\bibfnamefont {J.}~\bibnamefont
  {Schwehr}}, \bibinfo {author} {\bibfnamefont {D.}~\bibnamefont {Cherdack}},\
  and\ \bibinfo {author} {\bibfnamefont {R.}~\bibnamefont {Gran}},\ }\href@noop
  {} {\bibinfo {title} {{GENIE implementation of IFIC Valencia model for
  QE-like 2p2h neutrino-nucleus cross section}}} (\bibinfo {year} {2016}),\
  \Eprint {https://arxiv.org/abs/1601.02038} {arXiv:1601.02038 [hep-ph]}
  \BibitemShut {NoStop}%
\bibitem [{\citenamefont {Barish}\ \emph {et~al.}(1977)\citenamefont {Barish}
  \emph {et~al.}}]{Barish:1977qk}%
  \BibitemOpen
  \bibfield  {author} {\bibinfo {author} {\bibfnamefont {S.}~\bibnamefont
  {Barish}} \emph {et~al.},\ }\href {https://doi.org/10.1103/PhysRevD.16.3103}
  {\bibfield  {journal} {\bibinfo  {journal} {Phys. Rev. D}\ }\textbf {\bibinfo
  {volume} {16}},\ \bibinfo {pages} {3103} (\bibinfo {year}
  {1977})}\BibitemShut {NoStop}%
\bibitem [{\citenamefont {Baker}\ \emph {et~al.}(1982)\citenamefont {Baker},
  \citenamefont {Connolly}, \citenamefont {Kahn}, \citenamefont {Murtagh},
  \citenamefont {Palmer}, \citenamefont {Samios},\ and\ \citenamefont
  {Tanaka}}]{Baker:1982ty}%
  \BibitemOpen
  \bibfield  {author} {\bibinfo {author} {\bibfnamefont {N.}~\bibnamefont
  {Baker}}, \bibinfo {author} {\bibfnamefont {P.}~\bibnamefont {Connolly}},
  \bibinfo {author} {\bibfnamefont {S.}~\bibnamefont {Kahn}}, \bibinfo {author}
  {\bibfnamefont {M.}~\bibnamefont {Murtagh}}, \bibinfo {author} {\bibfnamefont
  {R.}~\bibnamefont {Palmer}}, \bibinfo {author} {\bibfnamefont
  {N.}~\bibnamefont {Samios}},\ and\ \bibinfo {author} {\bibfnamefont
  {M.}~\bibnamefont {Tanaka}},\ }\href
  {https://doi.org/10.1103/PhysRevD.25.617} {\bibfield  {journal} {\bibinfo
  {journal} {Phys. Rev. D}\ }\textbf {\bibinfo {volume} {25}},\ \bibinfo
  {pages} {617} (\bibinfo {year} {1982})}\BibitemShut {NoStop}%
\bibitem [{\citenamefont {Graczyk}\ \emph {et~al.}(2014)\citenamefont
  {Graczyk}, \citenamefont {\.Zmuda},\ and\ \citenamefont
  {Sobczyk}}]{Graczyk:2014dpa}%
  \BibitemOpen
  \bibfield  {author} {\bibinfo {author} {\bibfnamefont {K.~M.}\ \bibnamefont
  {Graczyk}}, \bibinfo {author} {\bibfnamefont {J.}~\bibnamefont {\.Zmuda}},\
  and\ \bibinfo {author} {\bibfnamefont {J.~T.}\ \bibnamefont {Sobczyk}},\
  }\href {https://doi.org/10.1103/PhysRevD.90.093001} {\bibfield  {journal}
  {\bibinfo  {journal} {Phys. Rev. D}\ }\textbf {\bibinfo {volume} {90}},\
  \bibinfo {pages} {093001} (\bibinfo {year} {2014})},\ \Eprint
  {https://arxiv.org/abs/1407.5445} {arXiv:1407.5445 [hep-ph]} \BibitemShut
  {NoStop}%
\bibitem [{\citenamefont {Rodrigues}\ \emph
  {et~al.}(2016{\natexlab{a}})\citenamefont {Rodrigues}, \citenamefont
  {Wilkinson},\ and\ \citenamefont {McFarland}}]{Rodrigues:2016xjj}%
  \BibitemOpen
  \bibfield  {author} {\bibinfo {author} {\bibfnamefont {P.}~\bibnamefont
  {Rodrigues}}, \bibinfo {author} {\bibfnamefont {C.}~\bibnamefont
  {Wilkinson}},\ and\ \bibinfo {author} {\bibfnamefont {K.}~\bibnamefont
  {McFarland}},\ }\href {https://doi.org/10.1140/epjc/s10052-016-4314-3}
  {\bibfield  {journal} {\bibinfo  {journal} {Eur. Phys. J. C}\ }\textbf
  {\bibinfo {volume} {76}},\ \bibinfo {pages} {474} (\bibinfo {year}
  {2016}{\natexlab{a}})},\ \Eprint {https://arxiv.org/abs/1601.01888}
  {arXiv:1601.01888 [hep-ex]} \BibitemShut {NoStop}%
\bibitem [{\citenamefont {Rodrigues}\ \emph
  {et~al.}(2016{\natexlab{b}})\citenamefont {Rodrigues} \emph
  {et~al.}}]{Rodrigues:2015hik}%
  \BibitemOpen
  \bibfield  {author} {\bibinfo {author} {\bibfnamefont {P.}~\bibnamefont
  {Rodrigues}} \emph {et~al.} (\bibinfo {collaboration} {MINERvA}),\ }\href
  {https://doi.org/10.1103/PhysRevLett.116.071802} {\bibfield  {journal}
  {\bibinfo  {journal} {Phys. Rev. Lett.}\ }\textbf {\bibinfo {volume} {116}},\
  \bibinfo {pages} {071802} (\bibinfo {year} {2016}{\natexlab{b}})},\ \bibinfo
  {note} {[Addendum: Phys.Rev.Lett. 121, 209902 (2018)]},\ \Eprint
  {https://arxiv.org/abs/1511.05944} {arXiv:1511.05944 [hep-ex]} \BibitemShut
  {NoStop}%
\bibitem [{\citenamefont {Morfin}\ \emph {et~al.}(2012)\citenamefont {Morfin},
  \citenamefont {Nieves},\ and\ \citenamefont {Sobczyk}}]{Morfin:2012kn}%
  \BibitemOpen
  \bibfield  {author} {\bibinfo {author} {\bibfnamefont {J.~G.}\ \bibnamefont
  {Morfin}}, \bibinfo {author} {\bibfnamefont {J.}~\bibnamefont {Nieves}},\
  and\ \bibinfo {author} {\bibfnamefont {J.~T.}\ \bibnamefont {Sobczyk}},\
  }\href {https://doi.org/10.1155/2012/934597} {\bibfield  {journal} {\bibinfo
  {journal} {Adv. High Energy Phys.}\ }\textbf {\bibinfo {volume} {2012}},\
  \bibinfo {pages} {934597} (\bibinfo {year} {2012})},\ \Eprint
  {https://arxiv.org/abs/1209.6586} {arXiv:1209.6586 [hep-ex]} \BibitemShut
  {NoStop}%
\bibitem [{\citenamefont {Gran}(2017)}]{Gran:2017psn}%
  \BibitemOpen
  \bibfield  {author} {\bibinfo {author} {\bibfnamefont {R.}~\bibnamefont
  {Gran}},\ }\href@noop {} {\bibinfo {title} {{Model Uncertainties for Valencia
  RPA Effect for MINERvA}}} (\bibinfo {year} {2017}),\ \Eprint
  {https://arxiv.org/abs/1705.02932} {arXiv:1705.02932} \BibitemShut {NoStop}%
\bibitem [{\citenamefont {Adamson}\ \emph {et~al.}(2007)\citenamefont {Adamson}
  \emph {et~al.}}]{Adamson:2007zzb}%
  \BibitemOpen
  \bibfield  {author} {\bibinfo {author} {\bibfnamefont {P.}~\bibnamefont
  {Adamson}} \emph {et~al.} (\bibinfo {collaboration} {MINOS}),\ }\href
  {https://doi.org/10.1103/PhysRevD.76.072005} {\bibfield  {journal} {\bibinfo
  {journal} {Phys. Rev. D}\ }\textbf {\bibinfo {volume} {76}},\ \bibinfo
  {pages} {072005} (\bibinfo {year} {2007})},\ \Eprint
  {https://arxiv.org/abs/0706.0437} {arXiv:0706.0437 [hep-ex]} \BibitemShut
  {NoStop}%
\bibitem [{\citenamefont {Paschos}\ \emph {et~al.}(2005)\citenamefont
  {Paschos}, \citenamefont {Sakuda}, \citenamefont {Schienbein},\ and\
  \citenamefont {Yu}}]{Paschos:2004md}%
  \BibitemOpen
  \bibfield  {author} {\bibinfo {author} {\bibfnamefont {E.}~\bibnamefont
  {Paschos}}, \bibinfo {author} {\bibfnamefont {M.}~\bibnamefont {Sakuda}},
  \bibinfo {author} {\bibfnamefont {I.}~\bibnamefont {Schienbein}},\ and\
  \bibinfo {author} {\bibfnamefont {J.}~\bibnamefont {Yu}},\ }\href
  {https://doi.org/10.1016/j.nuclphysbps.2004.11.205} {\bibfield  {journal}
  {\bibinfo  {journal} {Nucl. Phys. B Proc. Suppl.}\ }\textbf {\bibinfo
  {volume} {139}},\ \bibinfo {pages} {125} (\bibinfo {year} {2005})},\ \Eprint
  {https://arxiv.org/abs/hep-ph/0408185} {arXiv:hep-ph/0408185} \BibitemShut
  {NoStop}%
\bibitem [{\citenamefont {Stowell}\ \emph {et~al.}(2019)\citenamefont {Stowell}
  \emph {et~al.}}]{Stowell:2019zsh}%
  \BibitemOpen
  \bibfield  {author} {\bibinfo {author} {\bibfnamefont {P.}~\bibnamefont
  {Stowell}} \emph {et~al.} (\bibinfo {collaboration} {MINERvA}),\ }\href
  {https://doi.org/10.1103/PhysRevD.100.072005} {\bibfield  {journal} {\bibinfo
   {journal} {Phys. Rev. D}\ }\textbf {\bibinfo {volume} {100}},\ \bibinfo
  {pages} {072005} (\bibinfo {year} {2019})},\ \Eprint
  {https://arxiv.org/abs/1903.01558} {arXiv:1903.01558 [hep-ex]} \BibitemShut
  {NoStop}%
\bibitem [{\citenamefont {Pais}(1971)}]{PAIS1971361}%
  \BibitemOpen
  \bibfield  {author} {\bibinfo {author} {\bibfnamefont {A.}~\bibnamefont
  {Pais}},\ }\href
  {https://doi.org/https://doi.org/10.1016/0003-4916(71)90018-2} {\bibfield
  {journal} {\bibinfo  {journal} {Annals of Physics}\ }\textbf {\bibinfo
  {volume} {63}},\ \bibinfo {pages} {361} (\bibinfo {year} {1971})}\BibitemShut
  {NoStop}%
\bibitem [{\citenamefont {Tena-Vidal}\ \emph {et~al.}(2021)\citenamefont
  {Tena-Vidal} \emph {et~al.}}]{GENIE:2021zuu}%
  \BibitemOpen
  \bibfield  {author} {\bibinfo {author} {\bibfnamefont {J.}~\bibnamefont
  {Tena-Vidal}} \emph {et~al.} (\bibinfo {collaboration} {GENIE}),\ }\href
  {https://doi.org/10.1103/PhysRevD.104.072009} {\bibfield  {journal} {\bibinfo
   {journal} {Phys. Rev. D}\ }\textbf {\bibinfo {volume} {104}},\ \bibinfo
  {pages} {072009} (\bibinfo {year} {2021})},\ \Eprint
  {https://arxiv.org/abs/2104.09179} {arXiv:2104.09179 [hep-ph]} \BibitemShut
  {NoStop}%
\bibitem [{\citenamefont {{\it et al.}}(2003)}]{AGOSTINELLI2003250}%
  \BibitemOpen
  \bibfield  {author} {\bibinfo {author} {\bibfnamefont {S.~A.}\ \bibnamefont
  {{\it et al.}}},\ }\href
  {https://doi.org/https://doi.org/10.1016/S0168-9002(03)01368-8} {\bibfield
  {journal} {\bibinfo  {journal} {Nuclear Instruments and Methods in Physics
  Research Section A: Accelerators, Spectrometers, Detectors and Associated
  Equipment}\ }\textbf {\bibinfo {volume} {506}},\ \bibinfo {pages} {250 }
  (\bibinfo {year} {2003})}\BibitemShut {NoStop}%
\bibitem [{\citenamefont {Aliaga}\ \emph
  {et~al.}(2015{\natexlab{a}})\citenamefont {Aliaga} \emph
  {et~al.}}]{Aliaga:2015aqe}%
  \BibitemOpen
  \bibfield  {author} {\bibinfo {author} {\bibfnamefont {L.}~\bibnamefont
  {Aliaga}} \emph {et~al.} (\bibinfo {collaboration} {MINERvA Collaboration}),\
  }\href {https://doi.org/10.1016/j.nima.2015.04.003} {\bibfield  {journal}
  {\bibinfo  {journal} {Nucl.~Instrum.~Meth.~A}\ }\textbf {\bibinfo {volume}
  {789}},\ \bibinfo {pages} {28} (\bibinfo {year} {2015}{\natexlab{a}})},\
  \Eprint {https://arxiv.org/abs/1501.06431} {arXiv:1501.06431
  [physics.ins-det]} \BibitemShut {NoStop}%
\bibitem [{\citenamefont {Aliaga}\ \emph
  {et~al.}(2015{\natexlab{b}})\citenamefont {Aliaga} \emph
  {et~al.}}]{MINERvA:2015yej}%
  \BibitemOpen
  \bibfield  {author} {\bibinfo {author} {\bibfnamefont {L.}~\bibnamefont
  {Aliaga}} \emph {et~al.} (\bibinfo {collaboration} {MINERvA}),\ }\href
  {https://doi.org/10.1016/j.nima.2015.04.003} {\bibfield  {journal} {\bibinfo
  {journal} {Nucl. Instrum. Meth. A}\ }\textbf {\bibinfo {volume} {789}},\
  \bibinfo {pages} {28} (\bibinfo {year} {2015}{\natexlab{b}})},\ \Eprint
  {https://arxiv.org/abs/1501.06431} {arXiv:1501.06431 [physics.ins-det]}
  \BibitemShut {NoStop}%
\bibitem [{\citenamefont {D’Agostini}(1995)}]{DAgostini:1994fjx}%
  \BibitemOpen
  \bibfield  {author} {\bibinfo {author} {\bibfnamefont {G.}~\bibnamefont
  {D’Agostini}},\ }\href {https://doi.org/10.1016/0168-9002(95)00274-X}
  {\bibfield  {journal} {\bibinfo  {journal} {Nucl. Instrum. Meth. A}\ }\textbf
  {\bibinfo {volume} {362}},\ \bibinfo {pages} {487} (\bibinfo {year}
  {1995})}\BibitemShut {NoStop}%
\bibitem [{\citenamefont {Adye}(2011)}]{Adye:2011gm}%
  \BibitemOpen
  \bibfield  {author} {\bibinfo {author} {\bibfnamefont {T.}~\bibnamefont
  {Adye}},\ }in\ \href {https://doi.org/10.5170/CERN-2011-006.313} {\emph
  {\bibinfo {booktitle} {{PHYSTAT 2011}}}}\ (\bibinfo  {publisher} {CERN},\
  \bibinfo {address} {Geneva},\ \bibinfo {year} {2011})\ pp.\ \bibinfo {pages}
  {313--318},\ \Eprint {https://arxiv.org/abs/1105.1160} {arXiv:1105.1160
  [physics.data-an]} \BibitemShut {NoStop}%
\bibitem [{\citenamefont {Bashyal}(2021)}]{Bashyal:2021tzd}%
  \BibitemOpen
  \bibfield  {author} {\bibinfo {author} {\bibfnamefont {A.}~\bibnamefont
  {Bashyal}},\ }\emph {\bibinfo {title} {{DUNE and MINERvA Flux Studies and a
  Measurement of the Charged-Current Quasielastic Antineutrino Scattering Cross
  Section with \ensuremath{<}$E_\nu$ \ensuremath{>} \textasciitilde{} 6 GeV on
  a CH Target}}},\ \href {https://doi.org/10.2172/1779472} {Ph.D. thesis},\
  \bibinfo  {school} {Oregon State U.} (\bibinfo {year} {2021})\BibitemShut
  {NoStop}%
\bibitem [{\citenamefont {Aliaga}\ \emph {et~al.}(2016)\citenamefont {Aliaga}
  \emph {et~al.}}]{Aliaga:2016oaz}%
  \BibitemOpen
  \bibfield  {author} {\bibinfo {author} {\bibfnamefont {L.}~\bibnamefont
  {Aliaga}} \emph {et~al.} (\bibinfo {collaboration} {MINERvA Collaboration}),\
  }\href {https://doi.org/10.1103/PhysRevD.94.092005} {\bibfield  {journal}
  {\bibinfo  {journal} {Phys. Rev. D}\ }\textbf {\bibinfo {volume} {94}},\
  \bibinfo {pages} {092005} (\bibinfo {year} {2016})},\ \bibinfo {note}
  {[Addendum: Phys. Rev. D 95,no.3,039903(2017)]},\ \Eprint
  {https://arxiv.org/abs/1607.00704} {arXiv:1607.00704} \BibitemShut {NoStop}%
\bibitem [{\citenamefont {Stefanek}(2016)}]{Stefanek:2014pza}%
  \BibitemOpen
  \bibfield  {author} {\bibinfo {author} {\bibfnamefont {G.}~\bibnamefont
  {Stefanek}} (\bibinfo {collaboration} {NA49, NA61/SHINE}),\ }\href
  {https://doi.org/10.1016/j.nuclphysbps.2015.10.001} {\bibfield  {journal}
  {\bibinfo  {journal} {Nucl. Part. Phys. Proc.}\ }\textbf {\bibinfo {volume}
  {273-275}},\ \bibinfo {pages} {2596} (\bibinfo {year} {2016})},\ \Eprint
  {https://arxiv.org/abs/1411.2396} {arXiv:1411.2396 [nucl-ex]} \BibitemShut
  {NoStop}%
\bibitem [{\citenamefont {Barton}\ \emph {et~al.}(1983)\citenamefont {Barton}
  \emph {et~al.}}]{PhysRevD.27.2580}%
  \BibitemOpen
  \bibfield  {author} {\bibinfo {author} {\bibfnamefont {D.~S.}\ \bibnamefont
  {Barton}} \emph {et~al.},\ }\href {https://doi.org/10.1103/PhysRevD.27.2580}
  {\bibfield  {journal} {\bibinfo  {journal} {Phys. Rev. D}\ }\textbf {\bibinfo
  {volume} {27}},\ \bibinfo {pages} {2580} (\bibinfo {year}
  {1983})}\BibitemShut {NoStop}%
\bibitem [{\citenamefont {Valencia}\ \emph {et~al.}(2019)\citenamefont
  {Valencia} \emph {et~al.}}]{MINERvA:2019hhc}%
  \BibitemOpen
  \bibfield  {author} {\bibinfo {author} {\bibfnamefont {E.}~\bibnamefont
  {Valencia}} \emph {et~al.} (\bibinfo {collaboration} {MINERvA}),\ }\href
  {https://doi.org/10.1103/PhysRevD.100.092001} {\bibfield  {journal} {\bibinfo
   {journal} {Phys. Rev. D}\ }\textbf {\bibinfo {volume} {100}},\ \bibinfo
  {pages} {092001} (\bibinfo {year} {2019})},\ \Eprint
  {https://arxiv.org/abs/1906.00111} {arXiv:1906.00111 [hep-ex]} \BibitemShut
  {NoStop}%
\bibitem [{\citenamefont {Zazueta}\ \emph {et~al.}(2023)\citenamefont {Zazueta}
  \emph {et~al.}}]{MINERvA:2022vmb}%
  \BibitemOpen
  \bibfield  {author} {\bibinfo {author} {\bibfnamefont {L.}~\bibnamefont
  {Zazueta}} \emph {et~al.} (\bibinfo {collaboration} {MINERvA}),\ }\href
  {https://doi.org/10.1103/PhysRevD.107.012001} {\bibfield  {journal} {\bibinfo
   {journal} {Phys. Rev. D}\ }\textbf {\bibinfo {volume} {107}},\ \bibinfo
  {pages} {012001} (\bibinfo {year} {2023})},\ \Eprint
  {https://arxiv.org/abs/2209.05540} {arXiv:2209.05540 [hep-ex]} \BibitemShut
  {NoStop}%
\bibitem [{\citenamefont {Ruterbories}\ \emph {et~al.}(2021)\citenamefont
  {Ruterbories} \emph {et~al.}}]{MINERvA:2021dhf}%
  \BibitemOpen
  \bibfield  {author} {\bibinfo {author} {\bibfnamefont {D.}~\bibnamefont
  {Ruterbories}} \emph {et~al.} (\bibinfo {collaboration} {MINERvA}),\ }\href
  {https://doi.org/10.1103/PhysRevD.104.092010} {\bibfield  {journal} {\bibinfo
   {journal} {Phys. Rev. D}\ }\textbf {\bibinfo {volume} {104}},\ \bibinfo
  {pages} {092010} (\bibinfo {year} {2021})},\ \Eprint
  {https://arxiv.org/abs/2107.01059} {arXiv:2107.01059 [hep-ex]} \BibitemShut
  {NoStop}%
\bibitem [{\citenamefont {Bashyal}\ \emph {et~al.}(2021)\citenamefont {Bashyal}
  \emph {et~al.}}]{MINERvA:2021mpk}%
  \BibitemOpen
  \bibfield  {author} {\bibinfo {author} {\bibfnamefont {A.}~\bibnamefont
  {Bashyal}} \emph {et~al.} (\bibinfo {collaboration} {MINERvA}),\ }\href
  {https://doi.org/10.1088/1748-0221/16/08/P08068} {\bibfield  {journal}
  {\bibinfo  {journal} {JINST}\ }\textbf {\bibinfo {volume} {16}},\ \bibinfo
  {pages} {P08068}},\ \Eprint {https://arxiv.org/abs/2104.05769}
  {arXiv:2104.05769 [hep-ex]} \BibitemShut {NoStop}%
\bibitem [{\citenamefont {Messerly}\ \emph {et~al.}(2021)\citenamefont
  {Messerly} \emph {et~al.}}]{MINERvA:2021ddh}%
  \BibitemOpen
  \bibfield  {author} {\bibinfo {author} {\bibfnamefont {B.}~\bibnamefont
  {Messerly}} \emph {et~al.} (\bibinfo {collaboration} {MINERvA}),\ }\href
  {https://doi.org/10.1051/epjconf/202125103046} {\bibfield  {journal}
  {\bibinfo  {journal} {EPJ Web Conf.}\ }\textbf {\bibinfo {volume} {251}},\
  \bibinfo {pages} {03046} (\bibinfo {year} {2021})},\ \Eprint
  {https://arxiv.org/abs/2103.08677} {arXiv:2103.08677 [hep-ex]} \BibitemShut
  {NoStop}%
\bibitem [{\citenamefont {Carlson}\ \emph {et~al.}(2009)\citenamefont {Carlson}
  \emph {et~al.}}]{CARLSON20093215}%
  \BibitemOpen
  \bibfield  {author} {\bibinfo {author} {\bibfnamefont {A.~D.}\ \bibnamefont
  {Carlson}} \emph {et~al.},\ }\href
  {https://doi.org/10.1016/j.nds.2009.11.001} {\bibfield  {journal} {\bibinfo
  {journal} {Nucl. Data Sheets}\ }\textbf {\bibinfo {volume} {110}},\ \bibinfo
  {pages} {3215} (\bibinfo {year} {2009})}\BibitemShut {NoStop}%
\bibitem [{\citenamefont {Nieves}\ \emph {et~al.}(2011)\citenamefont {Nieves},
  \citenamefont {Ruiz~Simo},\ and\ \citenamefont
  {Vicente~Vacas}}]{Nieves:2011pp}%
  \BibitemOpen
  \bibfield  {author} {\bibinfo {author} {\bibfnamefont {J.}~\bibnamefont
  {Nieves}}, \bibinfo {author} {\bibfnamefont {I.}~\bibnamefont {Ruiz~Simo}},\
  and\ \bibinfo {author} {\bibfnamefont {M.~J.}\ \bibnamefont
  {Vicente~Vacas}},\ }\href {https://doi.org/10.1103/PhysRevC.83.045501}
  {\bibfield  {journal} {\bibinfo  {journal} {Phys. Rev. C}\ }\textbf {\bibinfo
  {volume} {83}},\ \bibinfo {pages} {045501} (\bibinfo {year} {2011})},\
  \Eprint {https://arxiv.org/abs/1102.2777} {arXiv:1102.2777 [hep-ph]}
  \BibitemShut {NoStop}%
\bibitem [{sup()}]{supplement}%
  \BibitemOpen
  \href@noop {} {}\bibinfo {howpublished} {See Supplemental Material at
  \url{URL_will_be_inserted_by_publisher} for numerical tables and details of
  the fractional uncertainties.}\BibitemShut {Stop}%
\end{thebibliography}%
